\documentclass[epj,nopacs,fleqn]{svjour}
\pdfoutput=1
\usepackage{graphicx,amsmath,amssymb,slashed,cite,verbatim,multirow,url}
\usepackage{lineno}
\modulolinenumbers[5]
\usepackage{color}
\usepackage[utf8]{inputenc}
\newcommand{\nn}{\nonumber}

\newcommand{\MET}{{\rm MET}}

\newcommand {\beq} {\begin{equation}}
\newcommand {\eeq} {\end{equation}}
\newcommand {\bea} {\begin{eqnarray}}
\newcommand {\eea} {\end{eqnarray}}
\newcommand{\GeV}{{\rm\ GeV}}

\newcommand{\TeV}{{\rm\ TeV}}

\begin{document}

\title{Higher-order QCD predictions for dark matter production at the
LHC in simplified models with s-channel mediators} 

\author{
 Mihailo Backovi\'{c}\inst{1}, 
 Michael Kr\"amer\inst{2},
 Fabio Maltoni\inst{1},
 Antony Martini\inst{1},
 Kentarou Mawatari\inst{3},
 Mathieu Pellen\inst{2}
}

\institute{ 
 Centre for Cosmology, Particle Physics and Phenomenology (CP3),
 Universit\'e catholique de Louvain,\\
 B-1348 Louvain-la-Neuve, Belgium
 \and
 Institute for Theoretical Particle Physics and Cosmology,
 RWTH Aachen University,\\
 D-52056 Aachen, Germany
 \and
 Theoretische Natuurkunde and IIHE/ELEM, Vrije Universiteit Brussel,
 and International Solvay Institutes,\\
 Pleinlaan 2, B-1050 Brussels, Belgium
}

\abstract{
Weakly interacting dark matter particles can be pair-produced at colliders and
detected through signatures featuring missing energy in association with
either QCD/EW radiation or heavy quarks.  In order to constrain the mass and the couplings to standard model particles, accurate and precise predictions for
production cross sections and distributions are of prime importance.
In this work, we consider various simplified models with 
$s$-channel mediators. We implement such models in the
{\sc FeynRules}/{\sc MadGraph5\_aMC@NLO} framework, which allows to include 
higher-order QCD corrections in realistic simulations and to study their effect systematically. As a first phenomenological application, we present predictions for dark matter
production in association with jets and with a top-quark pair at the
LHC, at next-to-leading order accuracy in QCD, including
matching/merging to parton showers. Our study shows that higher-order QCD corrections to dark matter production via $s$-channel mediators have a significant impact not only
on total production rates, but also on shapes of distributions. 
We also show that the inclusion of next-to-leading order effects results in a sizeable
reduction of the theoretical uncertainties.
}

\date{}

\titlerunning{
Higher-order QCD predictions for dark matter production at the
LHC in simplified models 
}   
\authorrunning{M.~Backovi\'{c} et al.}

\maketitle

\vspace*{-14cm}
\noindent 
MCnet-15-24, CP3-15-25, TTK-15-19
\vspace*{12cm}

\section{Introduction}

Various astrophysical and cosmological observations provide strong hints
for the existence of dark matter (DM). 
Yet, very little is known about the nature of DM, and constraints on
models from various direct/indirect detection experiments and cosmology
still allow for a wide range of DM masses and couplings to the Standard
Model (SM) particles.  
Various types of DM searches are sensitive to different
regions of the DM model parameter space~\cite{Drees:2012ji,Klasen:2015uma}. 
In order to maximise the chances for discovering -- or at least 
excluding -- DM models,  it is hence imperative to perform both
astrophysical and collider searches for DM.
The most promising way to detect signals of weakly interacting DM particles at the LHC is
through their associated production with jets, EW bosons and heavy
quarks, leading to signatures with missing transverse energy (MET).  
Searches for DM have been performed at LHC Run~I (see \textit{e.g.} Refs.~\cite{Aad:2015zva,CMS:2015jha})
and are one of the central goals of LHC Run~II~\cite{Abercrombie:2015gea}.

While the complementarity of different DM searches is a powerful tool, it is
intrinsically limited in that the comparison of collider and other results
introduces  some degree of model dependence.  
As the nature of DM is still unknown,  there exist a myriad of DM models
and mechanisms to be explored, spanning a wide range of complexity and
ambition.
In this context, it is of utmost importance to follow an approach
where the model dependence is limited, while the salient features that
can provide a useful characterisation of possible signals are kept
intact, $i.e.$, the ``simplified models'' approach~\cite{Alves:2011wf}. 

Many simplified models for DM have been proposed in the past (see Refs.~\cite{Abdallah:2014hon, Malik:2014ggr, Abercrombie:2015gea} and references therein). 
In their simplest form, these models assume DM to be 
a single massive particle which interacts weakly with the SM particles.
The interaction of DM with the SM can be
mediated by a new field, which we dub the mediator. When the mediator
is heavier than the energy scales the experiment can probe, the interaction
becomes point-like and the Lagrangian can be organised in terms of a tower of 
higher-dimensional operators in the framework of an effective field theory (EFT). 
However, when the experimentally accessible energy becomes comparable or higher than the mediator
mass, on-shell effects become important and a properly defined
quantum field theory featuring the mediator state in the spectrum is
needed~\cite{Goodman:2010yf,Bai:2010hh,Fox:2011pm,Shoemaker:2011vi,Busoni:2013lha,Buchmueller:2013dya}.
The LHC can explore a large range of DM and mediator masses,
as well as coupling strengths and possible combinations of DM and
mediator spins. Collider results within the framework of simplified models can then be combined with
direct searches for the mediators, $e.g.$ in Drell--Yan processes,
as well as with
cosmological and astrophysical constraints on DM.  

LHC searches for DM will rely on precision calculations to impose the
most stringent bounds on DM models, and hopefully characterise possible signals.
Higher-order corrections in QCD to DM production processes at the LHC are hence
vital for extracting precise information about DM from the LHC results.

Previous analyses~\cite{Wang:2011sx,Huang:2012hs,Fox:2012ru,
Haisch:2013ata, Mao:2014rga} studied next-to-leading order (NLO) QCD corrections for DM
production in MET $+\,j/\gamma/W$ in the case of EFT,
$i.e.$, in the limit of heavy mediators.%
\footnote{A first discussion of the impact of NLO QCD corrections to DM
production in a specific simplified model can be found in
~\cite{Fox:2012ru} and more recently in~\cite{Abercrombie:2015gea,Haisch:2013ata}.}
In this article, we consider simplified models with $s$-channel mediators. We analyse the impact of the higher-order
corrections to mono-jet signals in various benchmark scenarios with spin-1 (vector and
axial-vector) mediators,
calculate DM production cross sections (both total and differential) at NLO accuracy with up to two
jets, and also merge the corresponding samples via the FxFx procedure~\cite{Frederix:2012ps}. 
To our knowledge, such accuracy has so far not been achieved in the
context of DM simulations/predictions for the LHC Run~II.  

In addition, we consider $t\bar t$ associated production and compute NLO
cross sections and distributions for spin-0 (scalar and pseudo-scalar) 
mediators in representative cases, including those with a light
mediator. Predictions for this class of processes at NLO in QCD  also represent a
novelty in the context of DM computations for LHC Run~II. 

The first goal of this work is to illustrate the feasibility of having a
fully general implementation of DM simplified models in the
{\sc FeynRules}~\cite{Alloul:2013bka}/{\sc MadGraph5\_aMC@NLO} \cite{Alwall:2014hca}
({\sc MG5aMC} henceforth) framework, accurate up to NLO in QCD.
To this aim we start with the simplest (yet non-trivial) case of
$s$-channel mediators (colour singlet, spin 0 and 1 bosons) coupling to
DM and quarks. We assume that DM is a Dirac fermion for concreteness, yet our implementation is not limited by the choice of DM spin or nature (Dirac or Majorana).    We show how predictions and event generation for this
class of models can be achieved at NLO QCD accuracy, {\it in a fully automatic way}, for a wide set of observable/final states, while also employing the most recent
matching/merging multi-jet techniques~\cite{Frederix:2012ps}. 

The second goal of this work is of phenomenological nature, $i.e.$ to investigate the
impact of the NLO corrections on the current and future searches for DM
at the LHC. We consider two examples, among several promising final state signatures:
\begin{align}\label{xxj}
 pp \to X \bar{X} + {\rm jets}
\end{align}
for a spin-1 mediator model, and
\begin{align}\label{xxtt}
 pp \to X \bar{X} + t \bar t
\end{align}
for a spin-0 mediator model, where $X$ is a DM particle. 
We do not only calculate NLO QCD corrections to the overall
production rates, but also study in detail the effects of
higher-order corrections on the differential distributions of relevant
observables.  

Our simulation set-up is based on the {\sc FeynRules}/ {\sc MG5aMC} framework.
The {\sc FeynRules} package provides the relevant Feynman rules starting
from any local Lagrangian~\cite{Degrande:2011ua,deAquino:2011ub,Alloul:2013bka}, as
well as the UV/$R_2$ counterterms~\cite{Ossola:2008xq} necessary for the NLO 
computations via {\sc Nloct}~\cite{Degrande:2014vpa}/{\sc FeynArts}~\cite{Hahn:2000kx}.
Our simplified DM model files are publicly available at the
{\sc FeynRules} repository~\cite{FR-DMsimp:Online}.
With these ingredients, which are only based on the model and are not
process dependent, {\sc MG5aMC} computes tree-level amplitudes,
loop-amplitudes~\cite{Ossola:2007ax,Hirschi:2011pa,Cascioli:2011va} and
subtraction terms for a desired process, as well as their integration
over phase space~\cite{Frederix:2009yq}.
Event generation is obtained by matching short-distance events to the
parton shower employing the MC@NLO method~\cite{Frixione:2002ik}, which
is implemented for {\sc Pythia6}~\cite{Sjostrand:2006za},
{\sc Pythia8}~\cite{Sjostrand:2007gs},
{\sc Herwig6}~\cite{Corcella:2000bw} and
{\sc Herwig++}\cite{Bahr:2008pv}. 
We note that our DM model files can be exported not only to event
generators but also to tools for DM relic abundance as well as 
DM direct and indirect detections such as
{\sc MicrOMEGAs}~\cite{Belanger:2006is,Belanger:2008sj} and
{\sc MadDM}~\cite{Backovic:2013dpa,Backovic:2015cra}, allowing for more
comprehensive DM studies.  

The paper is organised as follows.  
In Sect.~\ref{sec:simplified_models} we introduce simplified
models for DM and specify the relevant interactions and model parameters
of the implementation.
We discuss the impact of the NLO QCD corrections on DM pair production
with jets through spin-1 mediators in
Sect.~\ref{sec:DM_pair_production}, which includes a discussion of
inclusive cross sections, differential distributions as well as a
discussion of the impact of parton showers and the NLO merging of events
with different jet multiplicities. 
NLO QCD predictions for DM production in association with a top-quark
pair through spin-0 mediators are discussed in 
Sect.~\ref{sec:DM_top}.
We provide our conclusions and an outlook in Sect.~\ref{sec:conclusions}.

\section{Simplified dark matter models:\\ the s-channel mediator case}
\label{sec:simplified_models}

We start by defining the particle content and the interactions of the
simplified model, which we dub {\sc DMsimp}. 
We assume that DM is described by a single, massive and weakly interacting
particle, that communicates with the SM through the exchange of a
mediator. For simplicity, we assume that the mediator is not part of
the SM.%
\footnote{So-called portal models where DM interacts with the Higgs
or the $Z/\gamma$ bosons can
be included in our framework as well.}
The first very general classification stems from the class of vertices
that characterise the model: Lagrangians featuring DM--DM--mediator and
SM--SM--mediator type interactions identify models with an $s$-channel
mediator, while  Lagrangians characterised by DM--SM--mediator
interactions define a $t$-channel mediator.  
The former interactions, for example, arise in models featuring an
extra scalar or $Z'$ which couples to a pair of DM particles, while the
latter is common in supersymmetric models.
From the point of view of QCD corrections the two classes are very
different, as an $s$-channel mediator is necessarily a 
colour singlet, while a $t$-channel mediator can be either neutral or coloured.
In this work we focus on the $s$-channel models, leaving the implementation
and validation of $t$-channel models to forthcoming work.
The $s$- and $t$-channel classes can be further organised by the quantum
numbers of the DM particle and the mediator. To start with,  
we focus on the case of Dirac DM with spin-1 or spin-0 mediators coupling
to the matter fields of the SM.  Changing the spin or the nature of the fermion (Dirac or Majorana) of
the DM particle or including a coupling of the mediator to the SM bosons 
is straightforward~\cite{Neubert:2015fka}. On the other hand, extending our analysis to spin-2 mediators, while feasible in principle, entails
dedicated validation work, as such models are, in general, not renormalisable. We defer such an extension to the future. 

\subsection{Spin-1 mediator model}

In the framework of our simplified model, the interaction Lagrangian of
a spin-1 mediator ($Y_1$) with a Dirac fermion DM ($X_D$) is given by 
\begin{align}\label{eq:vector_mediator}
 {\cal L}_{X_D}^{Y_1} = \bar X_D \gamma_{\mu}
  (g^{V}_{X_D}+g^{A}_{X_D}\gamma_5)X_D\,Y_1^{\mu} \,,
\end{align}
and with quarks by
\begin{align}\label{eq:vector_mediator2}
  {\cal L}_{\rm SM}^{Y_1} &= \sum_{i,j} \Big[\bar d_i\gamma_{\mu}
    (g^{V}_{d_{ij}}+g^{A}_{d_{ij}}\gamma_5)d_j \nn\\ 
    &\hspace*{10mm} +\bar u_i\gamma_{\mu}
    (g^{V}_{u_{ij}}+g^{A}_{u_{ij}}\gamma_5)u_j\Big] Y_1^{\mu} \,,
\end{align}
where $d$ and $u$ denote down- and up-type quarks, respectively, 
($i,j$=1,2,3) are flavour indices, and $g^{V/A}$ are the vector/axial-vector
couplings of DM and quarks.
Note that we adopt this notation according to the actual implementation
in {\sc FeynRules}. 
The model file, including an alternative choice for the spin of DM particle
(complex scalar $X_C$), can be downloaded at the {\sc FeynRules} 
repository~\cite{FR-DMsimp:Online}. 

The pure vector and pure axial-vector mediator scenarios are given by
setting the parameters in the Lagrangians~\eqref{eq:vector_mediator} and 
\eqref{eq:vector_mediator2} to
\begin{align}
 &g^V_{X_D} \equiv g_X \quad {\rm and}\quad g^A_{X_D} = 0
 \label{paramX_v} \\
 &g^{V}_{u_{ii}} =  g^{V}_{d_{ii}} \equiv g_{\rm SM} \quad {\rm and}\quad
  g^{A}_{u_{ii}} =  g^{A}_{d_{ii}} = 0
 \label{paramSM_v}
\end{align}
and 
\begin{align}
 &g^V_{X_D} = 0 \quad {\rm and}\quad g^A_{X_D} \equiv g_X 
 \label{paramX_a}\\
 &g^{V}_{u_{ii}} = g^{V}_{d_{ii}} = 0 \quad {\rm and}\quad
  g^{A}_{u_{ii}} = g^{A}_{d_{ii}} \equiv g_{\rm SM}\,,
 \label{paramSM_a}
\end{align}
respectively, where we assume quark couplings to the mediator to be
flavour universal and set all flavour off-diagonal couplings to zero.  
With this simplification of a single universal coupling for the SM-$Y_1$
interactions, the model has only four independent parameters, $i.e.$ two couplings
and two masses:  
\begin{equation}\label{param}
 \{g_{\rm SM},\,g_X,\,m_X,\,m_Y\} \,.
\end{equation}
We note that the mediator width is calculated from the above
parameters. 

Finding a signal of DM in this parameter space (or to constrain these
parameters) is the primary goal of the DM searches at the 
LHC Run~II~\cite{Abercrombie:2015gea}, and the most important signature
in this model is mono-jet plus MET. 
The di-jet final state via the $Y_1$ Drell--Yan process can be an important 
complementary channel.

\subsection{Spin-0 mediator model}

Similarly, in the case of a spin-0 mediator ($Y_0$) interacting with the
Dirac fermion DM and the SM particles, we define the interaction part of
the Lagrangians as
\begin{align}\label{eq:scalar_mediator}
 {\cal L}_{X_D}^{Y_0} = \bar X_D
 (g^{S}_{X_D}+ig^{P}_{X_D}\gamma_5)X_D\, Y_0 \,,
\end{align}
and
\begin{align}\label{eq:scalar_mediator2}
  {\cal L}_{\rm SM}^{Y_0} & =
   \sum_{i,j} \Big[\bar d_i \frac{y_{ij}^d}{\sqrt{2}}
       (g^{S}_{d_{ij}}+ig^{P}_{d_{ij}}\gamma_5)d_j \nn\\
   &\hspace*{10mm} + \bar u_i \frac{y_{ij}^u}{\sqrt{2}}
       (g^{S}_{u_{ij}}+ig^{P}_{u_{ij}}\gamma_5)u_j\Big] Y_0 \,,
\end{align}
where $g^{S/P}$ are the scalar/pseudo-scalar couplings of DM and
quarks.  Assuming a UV complete description of the scalar theory with 
the couplings of the mediator to the SM particles proportional to the
particle masses, we normalise these couplings to the SM Yukawa
couplings, $y_{ii}^f = \sqrt{2}m_f/v$, and set all flavour off-diagonal couplings to zero. This implies that, in a five-flavour scheme with massless bottom quarks, 
only top quarks are relevant for DM production in this model. Extension to a four-flavour
scheme with massive bottom quarks is possible.
The model file for the spin-0 mediator case, including other
 choices for the spin of the DM particle (real scalar $X_R$ and complex
 scalar $X_C$), is also available at the {\sc FeynRules}
 repository~\cite{FR-DMsimp:Online}.  

The pure scalar and pure pseudo-scalar mediator scenarios are given by 
setting the parameters in the Lagrangians~\eqref{eq:scalar_mediator} and
\eqref{eq:scalar_mediator2} 
to 
\begin{align}
 &g^S_{X_D} \equiv g_X \quad {\rm and}\quad g^P_{X_D} = 0
 \label{paramX_s} \\
 &g^{S}_{u_{33}} \equiv g_{\rm SM} \quad {\rm and}\quad
  g^{P}_{u_{33}} = 0
 \label{paramSM_s}
\end{align}
and
\begin{align}
 &g^S_{X_D} = 0 \quad {\rm and}\quad g^P_{X_D} \equiv g_X 
 \label{paramX_p}\\
 &g^{S}_{u_{33}} = 0 \quad {\rm and}\quad
  g^{P}_{u_{33}} \equiv g_{\rm SM}\,,
 \label{paramSM_p}
\end{align}
respectively.
All the other $g^{S/P}_{u_{ij}}$ and $g^{S/P}_{d_{ij}}$ parameters are
irrelevant. 
Similar to the spin-1 case, the model has only four independent parameters as
in~\eqref{param}. 

In the spin-0 mediator model with Yukawa-type couplings, the most relevant tree-level process at the LHC is DM
pair production associated with a top-quark pair. On the other hand similarly to Higgs production, at one loop, gluon fusion can give rise to MET + jets signatures which are in general phenomenologically relevant. 
For the heavy mediator case, the four-top final state can be also relevant.

At this stage, we do not see compelling reasons to introduce couplings of the mediator to leptons,
even though it is straightforward to do. We do not include effective gluon--gluon--$Y_0$ interactions either, 
for several reasons. The first is that this operator is higher dimensional (dim=5) and therefore
might lead to unitarity-violating effects that need to be studied on a model and benchmark basis.  The second is that a simplified model assumes no other new physics particle
beyond the dark matter particle $X$ and mediator $Y$ at the weak scale. If such new particle effects decouple with their mass,  the main contribution to the gluon--gluon--$Y_0$ coupling would then be due to loop of SM quarks.  Depending on the masses involved
(that of the DM, the mediator and the quarks contributing in the loop) and the momentum transfer of the process, such interactions might be considered point-like. This is,  however, very much model and process dependent. To be accurate one should first calculate the loop-induced processes exactly and study the range of applicability of such an effective interaction by explicit comparison.
This is possible in {\sc MG5aMC}~\cite{Hirschi:2015iia} and has been
considered with the same DM implementation as presented here in
Ref.~\cite{Mattelaer:2015haa}. 
Other studies of the loop-induced process for mono-jet + MET can be
found in
Refs.~\cite{Haisch:2012kf,Buckley:2014fba,Harris:2014hga,Haisch:2015ioa}.  
We also note that couplings of the mediator to the SM gauge
bosons can be introduced easily~\cite{Neubert:2015fka}.

\section{Dark matter production with jets}\label{sec:DM_pair_production}

In this section, we consider a spin-1 mediator scenario and discuss the
impact of the NLO-QCD corrections on DM pair production with jets,
$i.e.$, 
\begin{align}
 pp \to X \bar X + j(j) \,.
\end{align}
In {\sc MG5aMC} the code and events for the above process can be
automatically generated by issuing the following commands:
\begin{verbatim}
 > import model DMsimp_s_spin1
 > generate p p > xd xd~ j [QCD]
 > add process p p > xd xd~ j j [QCD]
 > output
 > launch
\end{verbatim}
We have checked that our model can reproduce the SM predictions for
$pp\to Zj(j)\to\tau^+\tau^-j(j)$ by adjusting the corresponding coupling and mass
parameters.

To illustrate the effect of the higher-order corrections, we consider
pure vector, Eqs.~\eqref{paramX_v} and \eqref{paramSM_v}, or pure
axial-vector, Eqs.~\eqref{paramX_a} and \eqref{paramSM_a}, couplings 
with a simplified flavour structure.
We take 
\begin{align}
 g_X=1 \quad {\rm and}\quad g_{\rm SM}=0.25
 \label{param_spin1}
\end{align}
as our benchmark for the spin-1 mediator scenario.

We assume that the mediator can only decay into the DM particle and the
SM quarks (if kinematically allowed) through the interactions specified
in Eqs.~\eqref{eq:vector_mediator} and \eqref{eq:vector_mediator2}, so that
the mediator width, $\Gamma_{Y}$, is determined by the particle masses
and the couplings $g_{X}$ and $g_{\rm SM}$.  
In our framework, the width is automatically computed by using the {\sc MadWidth} module~\cite{Alwall:2014bza} for each parameter point.
The above benchmark coupling strength in~\eqref{param_spin1} 
 leads to $\Gamma_Y/m_Y\sim 0.05$ for $m_Y>2m_X$ and 
$\Gamma_Y/m_Y\sim 0.025$ for $m_Y<2m_X$, both for the vector and
axial-vector cases.
Note that, if we take $g_{\rm SM}=1\ (0.5)$ with $g_X=1$, the $Y_1$
width becomes very large as $\Gamma_Y/m_Y\sim 0.5\ (0.15)$ for
$m_Y>2m_X$.

We provide LO and NLO QCD predictions for $pp\to X\bar X+j(j)$ at the
center-of-mass energy $\sqrt{s}=13$~TeV. 
The central value $\mu_0$ for the renormalisation ($\mu_R$) and
factorisation ($\mu_F$) scales is set to $H_T/2$, where $H_T$ is the sum
of the transverse momenta of all jets in the event and the missing
transverse energy. 
The scale uncertainty is estimated by varying the scales $\mu_R$ and $\mu_F$,
independently, by a factor two around $\mu_0$.
We adopt the five-flavour scheme and the LO and NLO {\sc NNPDF}2.3
set~\cite{Botje:2011sn} through the LHAPDF
interface~\cite{Whalley:2005nh}, with the corresponding values of
$\alpha_{s}^{\rm LO}(M_Z) = 0.130$ and 
$\alpha_{s}^{\rm NLO}(M_Z) = 0.118$, for the LO and NLO predictions, 
respectively.
The PDF uncertainties are computed automatically~\cite{Frederix:2011ss}, following the prescription summarised in~\cite{Alekhin:2011sk}. 

Where relevant, we apply a parton shower to the events using
{\sc Pythia8}~\cite{Sjostrand:2007gs}. 
We then define jets using the anti-$k_T$
algorithm~\cite{Cacciari:2008gp} as implemented in 
{\sc FastJet}~\cite{Cacciari:2011ma} with the jet cone radius $R=0.4$,
where we require $p_{T}(j)>30$~GeV and $|\eta(j)|<4.5$ for all jets in
the event.

\subsection{Total cross sections}\label{sec:DM_pair_cross_sections}

\begin{figure}
\center 
\includegraphics[width=1.\columnwidth]{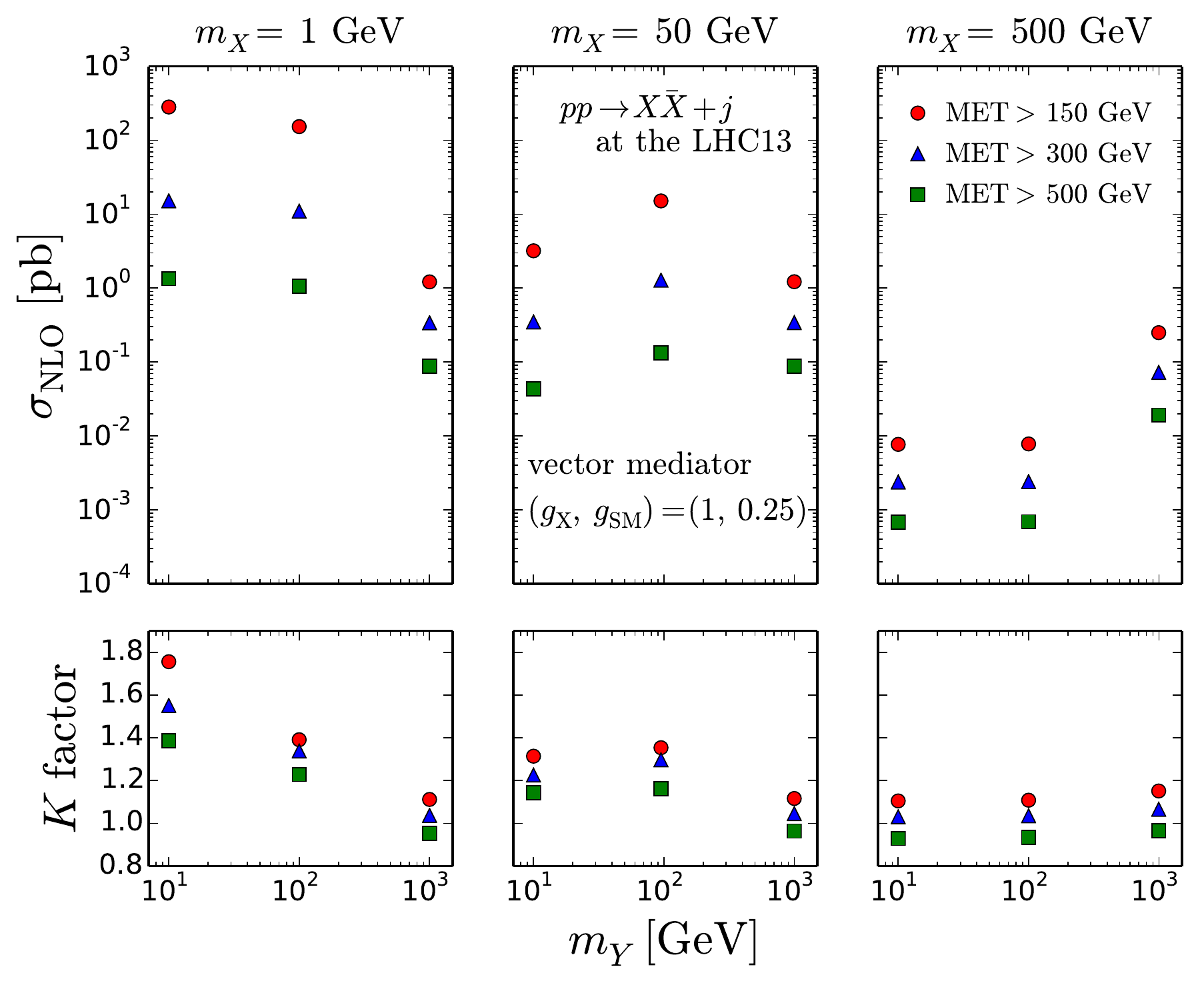}
\caption{Summary plot of NLO cross sections and corresponding $K$
 factors in Table~\ref{tab:xxj}.}
\label{fig:summary_xsec}
\end{figure}

\begin{table*}
\begin{footnotesize}
\center
\begin{tabular}{l|l|r|l|l|l}
\hline\rule{0pt}{3ex}
 &&&\multicolumn{3}{c}{vector}\\[1mm]
 $(m_{Y},m_{X})$ [GeV] & & & 
 MET $>150$ GeV & MET $>300$ GeV & MET $>500$ GeV \\[1mm] 
\hline\hline\rule{0pt}{3ex}
 & & $\sigma_{{\rm LO}}$ [pb]  
  & ${2.923\times10^{2}\,}^{+10.7}_{-8.9}${\scriptsize\,${\pm1.6\,\%}$} 
  & ${1.734\times10^{1}\,}^{+14.2}_{-11.9}${\scriptsize\,${\pm1.1\,\%}$} 
  & ${1.695\times10^{0}\,}^{+17.4}_{-14.0}${\scriptsize\,${\pm1.8\,\%}$} \\ 
 $10$  & undecayed &$\sigma_{{\rm NLO}}$ [pb]  
  & ${5.093\times10^{2}\,}^{+10.3}_{-8.2}${\scriptsize\,${\pm0.5\,\%}$} 
  & ${2.689\times10^{1}\,}^{+10.4}_{-9.1}${\scriptsize\,${\pm0.6\,\%}$} 
  & ${2.433\times10^{0}\,}^{+11.1}_{-10.0}${\scriptsize\,${\pm1.1\,\%}$}  \\ 
 & &$K$ factor 
  & 1.74   
  & 1.55 
  & 1.44 \\ 
\hline\rule{0pt}{3ex}
& & $\sigma_{{\rm LO}}$ [pb]  
  &$ {1.605\times10^{2}\,}^{+10.7}_{-8.9}${\scriptsize\,${\pm1.6\,\%}$}&$ {0.978\times10^{1}\,}^{+14.3}_{-12.0}${\scriptsize\,${\pm1.1\,\%}$}&$ {0.970\times10^{0}\,}^{+17.4}_{-14.1}${\scriptsize\,${\pm2.0\,\%}$} \\ 
$(10,1)$ & $m_{Y}\!>\!2m_{X}$ &$\sigma_{{\rm NLO}}$ [pb]  
  &$ {2.818\times10^{2}\,}^{+10.1}_{-8.1}${\scriptsize\,${\pm0.5\,\%}$}&$ {1.517\times10^{1}\,}^{+10.0}_{-8.9}${\scriptsize\,${\pm0.6\,\%}$}&$ {1.345\times10^{0}\,}^{+10.5}_{-9.6}${\scriptsize\,${\pm1.1\,\%}$} \\ 
 & &$K$ factor & $1.76$ & $1.55$ & $1.39$ \\ 
\hline\rule{0pt}{3ex}
& &$\sigma_{{\rm LO}}$ [pb]  &$ {2.434\times10^{0}\,}^{+11.8}_{-10.1}${\scriptsize\,${\pm1.5\,\%}$}&$ {2.843\times10^{-1}\,}^{+15.0}_{-12.5}${\scriptsize\,${\pm1.2\,\%}$}&$ {3.786\times10^{-2}\,}^{+18.0}_{-14.5}${\scriptsize\,${\pm2.4\,\%}$} \\ 
$(10,50)$ & $m_{Y}\!<\!2m_{X}$ &$\sigma_{{\rm NLO}}$ [pb]  &$ {3.198\times10^{0}\,}^{+5.6}_{-5.4}${\scriptsize\,${\pm0.5\,\%}$}&$ {3.485\times10^{-1}\,}^{+5.9}_{-6.3}${\scriptsize\,${\pm0.7\,\%}$}&$ {4.325\times10^{-2}\,}^{+7.3}_{-7.8}${\scriptsize\,${\pm1.3\,\%}$} \\ 
 & &$K$ factor & $1.31$ & $1.23$ & $1.14$ \\ 
\hline\rule{0pt}{3ex}
& &$\sigma_{{\rm LO}}$ [pb]  &$ {6.968\times10^{-3}\,}^{+17.4}_{-14.0}${\scriptsize\,${\pm4.3\,\%}$}&$ {2.314\times10^{-3}\,}^{+18.9}_{-15.0}${\scriptsize\,${\pm4.6\,\%}$}&$ {7.317\times10^{-4}\,}^{+20.6}_{-16.1}${\scriptsize\,${\pm5.6\,\%}$} \\ 
$(10,500)$ &$m_{Y}\!<\!2m_{X}$ &$\sigma_{{\rm NLO}}$ [pb]  &$ {7.698\times10^{-3}\,}^{+5.4}_{-6.4}${\scriptsize\,${\pm2.2\,\%}$}&$ {2.385\times10^{-3}\,}^{+5.7}_{-6.9}${\scriptsize\,${\pm2.3\,\%}$}&$ {6.800\times10^{-4}\,}^{+5.5}_{-7.1}${\scriptsize\,${\pm2.6\,\%}$} \\ 
 & &$K$ factor & $1.10$ & $1.03$ & $0.93$ \\ 
\hline\hline\rule{0pt}{3ex}
& & $\sigma_{{\rm LO}}$ [pb]  
  & ${2.148\times10^{2}\,}^{+10.6}_{-9.3}${\scriptsize\,${\pm1.5\,\%}$} 
  & ${1.616\times10^{1}\,}^{+14.4}_{-12.0}${\scriptsize\,${\pm1.0\,\%}$} 
  & ${1.644\times10^{0}\,}^{+17.4}_{-14.1}${\scriptsize\,${\pm1.9\,\%}$} \\ 
 $100$  & undecayed &$\sigma_{{\rm NLO}}$ [pb]  
  & ${3.011\times10^{2}\,}^{+6.6}_{-5.9}${\scriptsize\,${\pm0.5\,\%}$}
  & ${2.121\times10^{1}\,}^{+7.3}_{-7.1}${\scriptsize\,${\pm0.6\,\%}$} 
  & ${1.955\times10^{0}\,}^{+8.1}_{-8.2}${\scriptsize\,${\pm1.2\,\%}$}  \\ 
 & &$K$ factor 
  & 1.40  
  & 1.31  
  & 1.19 \\ 
\hline\rule{0pt}{3ex}
& &$\sigma_{{\rm LO}}$ [pb]   &$ {1.100\times10^{2}\,}^{+10.6}_{-9.3}${\scriptsize\,${\pm1.5\,\%}$}&$ {0.822\times10^{1}\,}^{+14.4}_{-12.0}${\scriptsize\,${\pm1.1\,\%}$}&$ {0.862\times10^{0}\,}^{+17.4}_{-14.1}${\scriptsize\,${\pm1.9\,\%}$} \\ 
$(100,1)$ &$m_{Y}\!>\!2m_{X}$ &$\sigma_{{\rm NLO}}$ [pb]  &$ {1.530\times10^{2}\,}^{+6.5}_{-5.7}${\scriptsize\,${\pm0.5\,\%}$}&$ {1.100\times10^{1}\,}^{+7.4}_{-7.2}${\scriptsize\,${\pm0.6\,\%}$}&$ {1.059\times10^{0}\,}^{+8.0}_{-8.1}${\scriptsize\,${\pm1.2\,\%}$} \\ 
 & &$K$ factor & $1.39$ & $1.34$ & $1.23$ \\ 
\hline\rule{0pt}{3ex}
& &$\sigma_{{\rm LO}}$ [pb]  &$ {1.117\times10^{1}\,}^{+11.0}_{-9.6}${\scriptsize\,${\pm1.5\,\%}$}&$ {0.988\times10^{0}\,}^{+14.7}_{-12.2}${\scriptsize\,${\pm1.1\,\%}$}&$ {1.140\times10^{-1}\,}^{+17.6}_{-14.2}${\scriptsize\,${\pm2.0\,\%}$} \\ 
$(95,50)$ &$m_{Y}\!\lesssim\!2m_{X}$ &$\sigma_{{\rm NLO}}$ [pb]  &$ {1.512\times10^{1}\,}^{+6.0}_{-5.5}${\scriptsize\,${\pm0.5\,\%}$}&$ {1.281\times10^{0}\,}^{+6.8}_{-6.8}${\scriptsize\,${\pm0.6\,\%}$}&$ {1.325\times10^{-1}\,}^{+7.2}_{-7.6}${\scriptsize\,${\pm1.2\,\%}$} \\ 
 & &$K$ factor & $1.35$ & $1.30$ & $1.16$ \\ 
\hline\rule{0pt}{3ex}
& &$\sigma_{{\rm LO}}$ [pb]  &$ {7.043\times10^{-3}\,}^{+17.4}_{-14.0}${\scriptsize\,${\pm4.3\,\%}$}&$ {2.329\times10^{-3}\,}^{+18.9}_{-15.0}${\scriptsize\,${\pm4.6\,\%}$}&$ {7.395\times10^{-4}\,}^{+20.6}_{-16.1}${\scriptsize\,${\pm5.6\,\%}$} \\ 
$(100,500)$ &$m_{Y}\!<\!2m_{X}$ &$\sigma_{{\rm NLO}}$ [pb]  &$ {7.804\times10^{-3}\,}^{+5.3}_{-6.4}${\scriptsize\,${\pm2.2\,\%}$}&$ {2.411\times10^{-3}\,}^{+5.5}_{-6.8}${\scriptsize\,${\pm2.3\,\%}$}&$ {6.908\times10^{-4}\,}^{+5.5}_{-7.1}${\scriptsize\,${\pm2.6\,\%}$} \\ 
 & &$K$ factor & $1.11$ & $1.04$ & $0.93$ \\ 
\hline\hline\rule{0pt}{3ex}
& & $\sigma_{{\rm LO}}$ [pb]  
  & ${2.248\times10^{0}\,}^{+16.1}_{-13.2}${\scriptsize\,${\pm3.2\,\%}$}  
  & ${6.865\times10^{-1}\,}^{+17.7}_{-14.3}${\scriptsize\,${\pm3.3\,\%}$}  
  & ${1.979\times10^{-1}\,}^{+19.6}_{-15.5}${\scriptsize\,${\pm4.1\,\%}$} \\ 
 $1000$ & undecayed &$\sigma_{{\rm NLO}}$ [pb]
  & ${2.601\times10^{0}\,}^{+5.1}_{-6.0}${\scriptsize\,${\pm1.7\,\%}$} 
  & ${7.393\times10^{-1}\,}^{+5.2}_{-6.4}${\scriptsize\,${\pm1.8\,\%}$}
  & ${1.909\times10^{-1}\,}^{+5.3}_{-6.8}${\scriptsize\,${\pm2.1\,\%}$} \\ 
 & &$K$ factor 
  & 1.16  
  & 1.08  
  & 0.96 \\ 
\hline\rule{0pt}{3ex}
& &$\sigma_{{\rm LO}}$ [pb]  &$ {1.093\times10^{0}\,}^{+16.4}_{-13.3}${\scriptsize\,${\pm3.1\,\%}$}&$ {3.278\times10^{-1}\,}^{+18.0}_{-14.4}${\scriptsize\,${\pm3.3\,\%}$}&$ {9.182\times10^{-2}\,}^{+19.7}_{-15.6}${\scriptsize\,${\pm4.1\,\%}$} \\ 
$(1000,1)$ &$m_{Y}\!>\!2m_{X}$ &$\sigma_{{\rm NLO}}$ [pb]  &$ {1.215\times10^{0}\,}^{+4.2}_{-5.5}${\scriptsize\,${\pm1.7\,\%}$}&$ {3.399\times10^{-1}\,}^{+4.5}_{-6.0}${\scriptsize\,${\pm1.7\,\%}$}&$ {8.743\times10^{-2}\,}^{+4.8}_{-6.5}${\scriptsize\,${\pm2.0\,\%}$} \\ 
 & &$K$ factor & $1.11$ & $1.04$ & $0.95$ \\ 
\hline\rule{0pt}{3ex}
& &$\sigma_{{\rm LO}}$ [pb]  &$ {1.094\times10^{0}\,}^{+16.4}_{-13.3}${\scriptsize\,${\pm3.1\,\%}$}&$ {3.268\times10^{-1}\,}^{+18.0}_{-14.4}${\scriptsize\,${\pm3.3\,\%}$}&$ {9.137\times10^{-2}\,}^{+19.7}_{-15.6}${\scriptsize\,${\pm4.1\,\%}$} \\ 
$(1000,50)$ &$m_{Y}\!>\!2m_{X}$ &$\sigma_{{\rm NLO}}$ [pb]  &$ {1.221\times10^{0}\,}^{+4.3}_{-5.6}${\scriptsize\,${\pm1.7\,\%}$}&$ {3.416\times10^{-1}\,}^{+4.6}_{-6.0}${\scriptsize\,${\pm1.7\,\%}$}&$ {8.807\times10^{-2}\,}^{+4.9}_{-6.6}${\scriptsize\,${\pm2.0\,\%}$} \\ 
 & &$K$ factor & $1.12$ & $1.05$ & $0.96$ \\ 
\hline\rule{0pt}{3ex}
& &$\sigma_{{\rm LO}}$ [pb]  &$ {2.169\times10^{-1}\,}^{+16.4}_{-13.3}${\scriptsize\,${\pm3.4\,\%}$}&$ {6.777\times10^{-2}\,}^{+18.0}_{-14.4}${\scriptsize\,${\pm3.6\,\%}$}&$ {1.981\times10^{-2}\,}^{+19.7}_{-15.6}${\scriptsize\,${\pm4.4\,\%}$} \\ 
$(995,500)$ &$m_{Y}\!\lesssim\!2m_{X}$ &$\sigma_{{\rm NLO}}$ [pb]  &$
	     {2.497\times10^{-1}\,}^{+5.3}_{-6.2}${\scriptsize\,${\pm1.8\,\%}$}&$
		 {7.223\times10^{-2}\,}^{+5.5}_{-6.6}${\scriptsize\,${\pm1.9\,\%}$}&$  {1.914\times10^{-2}\,}^{+5.3}_{-6.8}${\scriptsize\,${\pm2.1\,\%}$} \\ 
 & &$K$ factor & $1.15$ & $1.07$ & $0.97$ \\ 
\hline\hline\rule{0pt}{3ex}
& &$\sigma_{{\rm LO}}$ [pb]  & $ {8.487\times10^{-6}\,}^{+18.0}_{-14.3}${\scriptsize\,${\pm4.3\,\%}$}& $ {2.666\times10^{-6}\,}^{+20.0}_{-15.7}${\scriptsize\,${\pm5.5\,\%}$}&  $ {8.238\times10^{-7}\,}^{+22.0}_{-17.0}${\scriptsize\,${\pm7.3\,\%}$}\\ 
$(10000,1)$ & $m_{Y}\!\gg\!\sqrt{\hat s}$  &$\sigma_{{\rm NLO}}$ [pb]  & $ {8.835\times10^{-6}\,}^{+3.1}_{-5.1}${\scriptsize\,${\pm2.5\,\%}$}& $ {2.579\times10^{-6}\,}^{+3.1}_{-5.5}${\scriptsize\,${\pm3\,\%}$}& $ {7.148\times10^{-7}\,}^{+5.0}_{-7.0}${\scriptsize\,${\pm4.4\,\%}$} \\ 
 & &$K$ factor & $1.04 $ & $0.97 $ & $0.87 $ \\ \hline
\end{tabular}
\caption{LO and NLO cross sections and corresponding $K$ factors for
 DM pair production in association with a jet for the vector mediator
 scenario at the 13-TeV LHC, where different MET cuts are imposed. 
 The uncertainties represent the scale and PDF uncertainties in per cent,
 respectively.  
 We show several benchmark model points for the mediator and
 DM masses with the coupling parameters $g_{X}=1$ and 
 $g_{\rm SM}=0.25$.}
\label{tab:xxj} 
\end{footnotesize}
\end{table*}

\begin{figure*}
\center 
\includegraphics[width=0.47\textwidth]{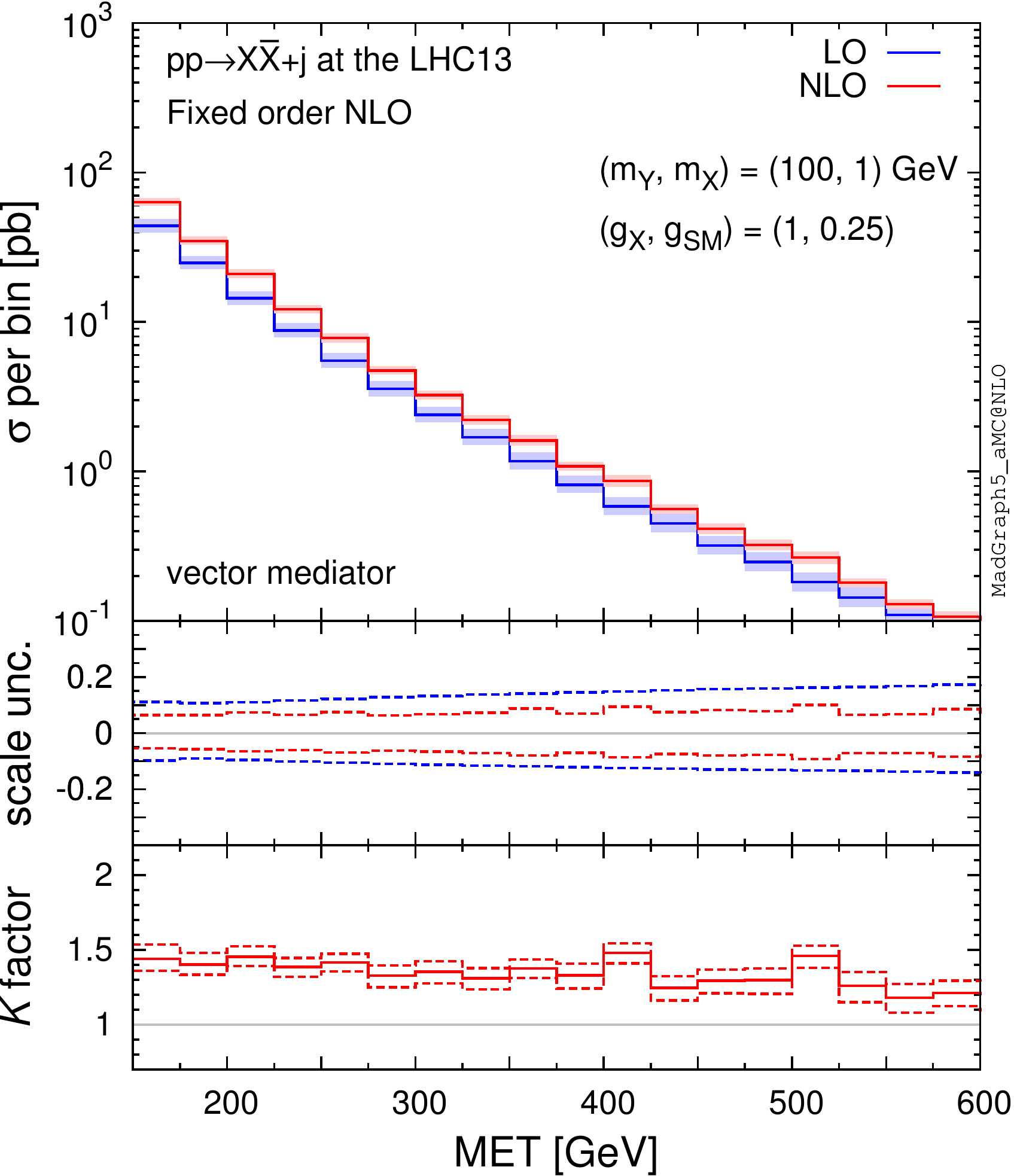}\qquad
\includegraphics[width=0.47\textwidth]{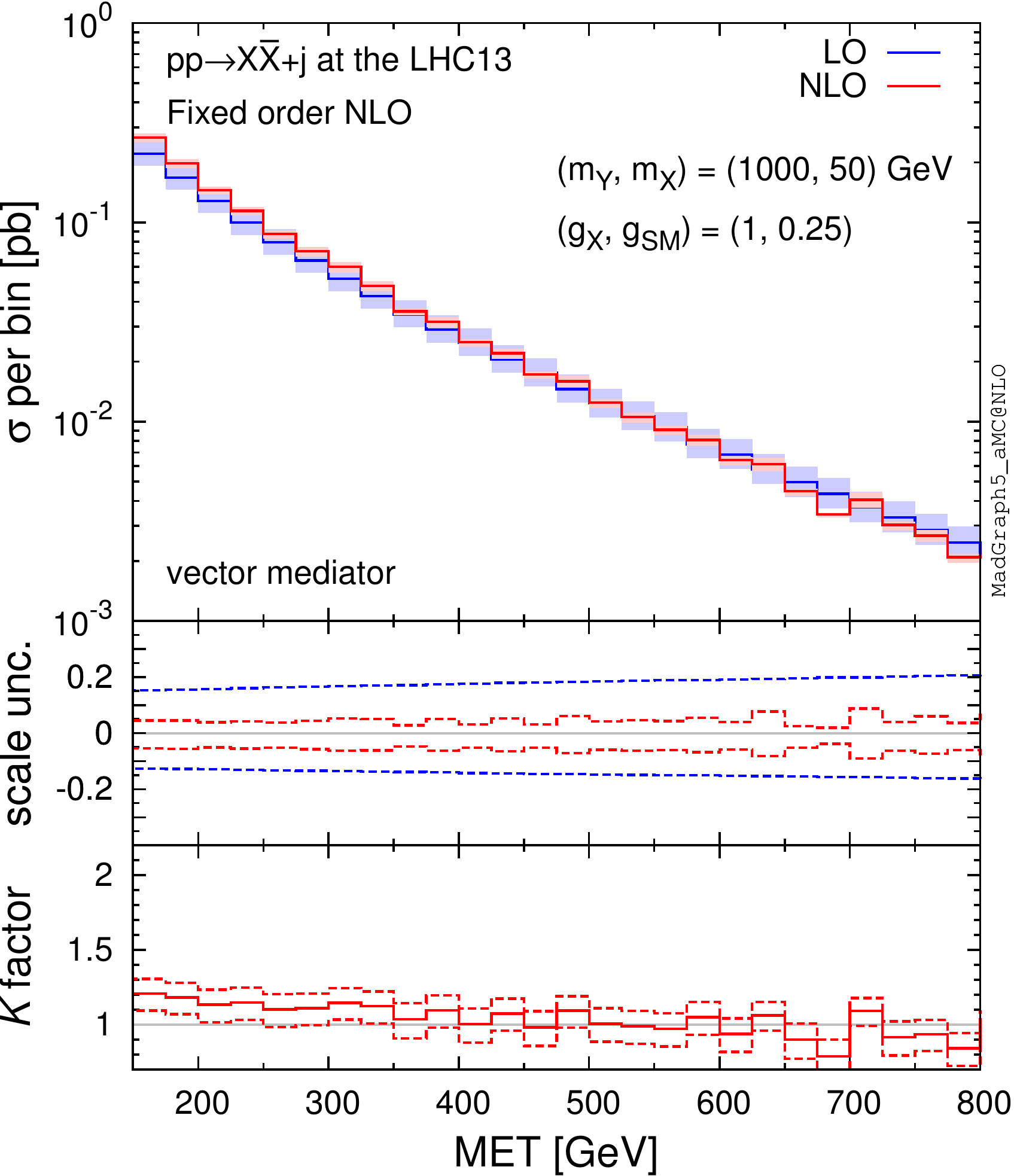}
\caption{
 MET distributions at FO (N)LO accuracy for $pp\to X\bar X+j$ at the
 13-TeV LHC for $(m_Y,m_X)=(100,1)$ and $(1000,50)$~GeV, where we
 assume a pure vector mediator and Dirac DM. 
 The middle and bottom panels show the differential scale uncertainties
 and $K$ factors, respectively.}
\label{fig:met_fo}
\end{figure*}

In Table~\ref{tab:xxj} we present LO
and NLO cross sections (in pb) for DM pair production in association
with a jet at fixed order (FO) in perturbation theory.
We show results for a pure vector mediator by fixing the parameters as
in Eqs.~\eqref{paramX_v}, \eqref{paramSM_v} and \eqref{param_spin1}. 
We cover various benchmark points suggested by the ATLAS/CMS DM
forum~\cite{Abercrombie:2015gea} in the $m_{Y}-m_{X}$ plane,
representing four different cases:
on-shell ($m_{Y}>2m_{X}$) and off-shell ($m_{Y}<2m_{X}$) production of
the mediator, in the threshold regime ($m_{Y}\lesssim 2m_{X}$) and in
the EFT limit ($m_{Y}\gg\sqrt{\hat s}$). 
We also present scale and PDF uncertainties in \% as well as $K$ factors
which we define as the ratio of the central values of the NLO and LO
cross sections.
We compute the table entries with different MET cuts: 150, 300, and
500~GeV.
For convenience, we also show a graphical summary of our results
in Fig.~\ref{fig:summary_xsec}. 
As a reference, the cross sections for $pp\to Y_1+j$ are also shown,
where the vector mediator $Y_1$ is produced on-shell and does not
decay.
For $m_{Y}>2m_{X}$, the mono-jet rate is given by 
$\sigma (pp\to X\bar X+j)\sim\sigma (pp\to Y_1+j)\times B(Y_1\to X\bar X)$ 
in the narrow width approximation. 

The production rate strongly depends on the both masses as well as on
the kinematic cuts, and varies by orders of magnitude in the parameter
scan.  
On the other hand, the $K$ factors, $i.e.$ higher-order effects, are not
so sensitive to the mass spectra;
$K\sim1.1$ for the heavy-mediator and/or heavy-DM cases, while
$K\sim1.3-1.4$ for the $\sim 100 \GeV$ mediator with light DM,  assuming the
$\MET>150$~GeV cut.
We find that in the case of a relatively light mediator with a very
light DM, $(m_Y,m_X)=(10,1)$~GeV, the $K$ factor can reach a value as
large as $1.8$. 

Different benchmark points probe different Bjorken-$x$ regions of the parton distribution functions. As heavy mediators/DM are produced from very high-$x$ partons, the
dominant contribution comes from the $q\bar{q}$ initial state, as
the gluon PDF is sub-dominant in the high-$x$ region. 
For light mediators with light DM, on the other hand, a large contribution arises from the $qg$ initial state. For instance, we find that the ratio of production
cross sections via $u\bar{u}$ and $ug$ initial states, $\sigma(u\bar{u}) / \sigma(ug)$, is 1.4 in the case of 
$(m_Y, m_X)=(1000, 50) \GeV $ while 0.2 for 
$(m_Y, m_X)=(100, 1) \GeV$, at LO.

\begin{table*}
\begin{footnotesize}
\center
\begin{tabular}{l|l|r|l|l|l}
\hline\rule{0pt}{3ex}
 &&&\multicolumn{3}{c}{axial-vector}\\[1mm]
 $(m_{Y},m_{X})$ [GeV] & & & 
 MET $>150$ GeV & MET $>300$ GeV & MET $>500$ GeV \\[1mm] 
\hline\hline\rule{0pt}{3ex}
& & $\sigma_{{\rm LO}}$ [pb]  
  & ${2.130\times10^{2}\,}^{+10.6}_{-9.3}${\scriptsize\,${\pm1.6\,\%}$} 
  & ${1.573\times10^{1}\,}^{+14.4}_{-12.0}${\scriptsize\,${\pm1.1\,\%}$} 
  & ${1.633\times10^{0}\,}^{+17.3}_{-14.0}${\scriptsize\,${\pm1.9\,\%}$} \\ 
 $100$  & undecayed &$\sigma_{{\rm NLO}}$ [pb]   
  & ${3.063\times10^{2}\,}^{+6.9}_{-6.1}${\scriptsize\,${\pm0.5\,\%}$}
  & ${2.153\times10^{1}\,}^{+7.7}_{-7.4}${\scriptsize\,${\pm0.6\,\%}$} 
  & ${2.055\times10^{0}\,}^{+8.4}_{-8.3}${\scriptsize\,${\pm1.6\,\%}$}  \\ 
 & &$K$ factor 
  & 1.44  
  & 1.37  
  & 1.26 \\ 
\hline\rule{0pt}{3ex}
& &$\sigma_{{\rm LO}}$ [pb]   &$ {1.101\times10^{2}\,}^{+10.6}_{-9.3}${\scriptsize\,${\pm1.6\,\%}$}&$ {0.825\times10^{1}\,}^{+14.4}_{-12.1}${\scriptsize\,${\pm1.1\,\%}$}&$ {0.854\times10^{0}\,}^{+17.4}_{-14.1}${\scriptsize\,${\pm2\,\%}$} \\ 
$(100,1)$ &$m_{Y}\!>\!2m_{X}$ &$\sigma_{{\rm NLO}}$ [pb]  &$ {1.549\times10^{2}\,}^{+6.8}_{-6.0}${\scriptsize\,${\pm0.5\,\%}$}&$ {1.127\times10^{1}\,}^{+7.4}_{-7.2}${\scriptsize\,${\pm0.6\,\%}$}&$ {1.063\times10^{0}\,}^{+8.2}_{-8.2}${\scriptsize\,${\pm1.2\,\%}$} \\ 
 & &$K$ factor & $1.41$ & $1.37$ & $1.24$ \\ 
\hline\rule{0pt}{3ex}
& &$\sigma_{{\rm LO}}$ [pb]  &$ {3.070\times10^{0}\,}^{+11.6}_{-10.0}${\scriptsize\,${\pm1.5\,\%}$}&$ {3.359\times10^{-1}\,}^{+14.9}_{-12.4}${\scriptsize\,${\pm1.2\,\%}$}&$ {4.457\times10^{-2}\,}^{+17.7}_{-14.3}${\scriptsize\,${\pm1.8\,\%}$} \\ 
$(95,50)$ &$m_{Y}\!\lesssim\!2m_{X}$ &$\sigma_{{\rm NLO}}$ [pb]  &$ {4.093\times10^{0}\,}^{+6.0}_{-5.7}${\scriptsize\,${\pm0.5\,\%}$}&$ {4.302\times10^{-1}\,}^{+6.7}_{-6.9}${\scriptsize\,${\pm0.7\,\%}$}&$ {5.079\times10^{-2}\,}^{+6.9}_{-7.4}${\scriptsize\,${\pm1.3\,\%}$} \\ 
 & &$K$ factor & $1.33$ & $1.28$ & $1.14$ \\ 
\hline\rule{0pt}{3ex}
& &$\sigma_{{\rm LO}}$ [pb]  &$ {2.298\times10^{-3}\,}^{+18.1}_{-14.5}${\scriptsize\,${\pm5\,\%}$}&$ {7.839\times10^{-4}\,}^{+19.5}_{-15.4}${\scriptsize\,${\pm5.3\,\%}$}&$ {2.558\times10^{-4}\,}^{+21.2}_{-16.5}${\scriptsize\,${\pm6.3\,\%}$} \\ 
$(100,500)$ &$m_{Y}\!<\!2m_{X}$ &$\sigma_{{\rm NLO}}$ [pb]  &$ {2.502\times10^{-3}\,}^{+5.9}_{-6.8}${\scriptsize\,${\pm2.5\,\%}$}&$ {7.972\times10^{-4}\,}^{+6.2}_{-7.3}${\scriptsize\,${\pm2.6\,\%}$}&$ {2.383\times10^{-4}\,}^{+6.1}_{-7.5}${\scriptsize\,${\pm3.0\,\%}$} \\ 
 & &$K$ factor & $1.09$ & $1.02$ & $0.93$ \\ 
\hline
\end{tabular}
\caption{
 Same as Table~\ref{tab:xxj}, but for the axial-vector mediator scenario.
}
\label{tab:xxj_axial} 
\end{footnotesize}
\end{table*}

As expected, most of the results at NLO accuracy display significantly smaller
scale uncertainties compared to the LO calculations. An  exception is provided by the
$(m_Y,m_X)=(10,1)$~GeV case, which we discuss in detail at the end of
this subsection. 
The PDF uncertainties are sub-leading in both the LO and NLO results and reduced by going from LO to NLO.
Furthermore, the scale and PDF uncertainties increase when the mass
scale of the mediator and/or DM increases.

The higher $\MET$ cut leads to smaller $K$ factors and to larger
scale and PDF uncertainties, which one can clearly see in the MET
distributions in Fig.~\ref{fig:met_fo}.

\begin{figure*}
\center 
\includegraphics[width=0.47\textwidth]{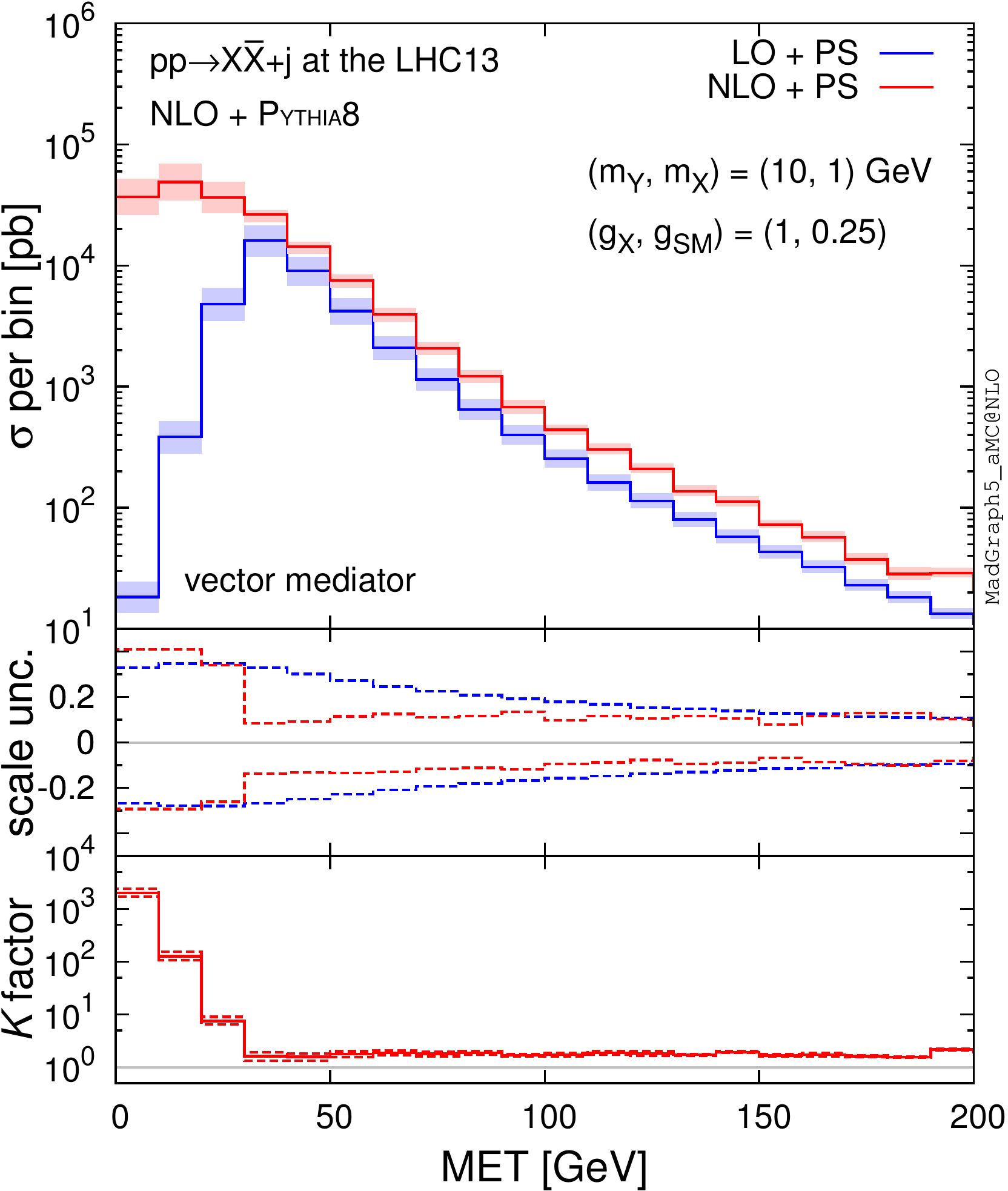}\qquad
\includegraphics[width=0.47\textwidth]{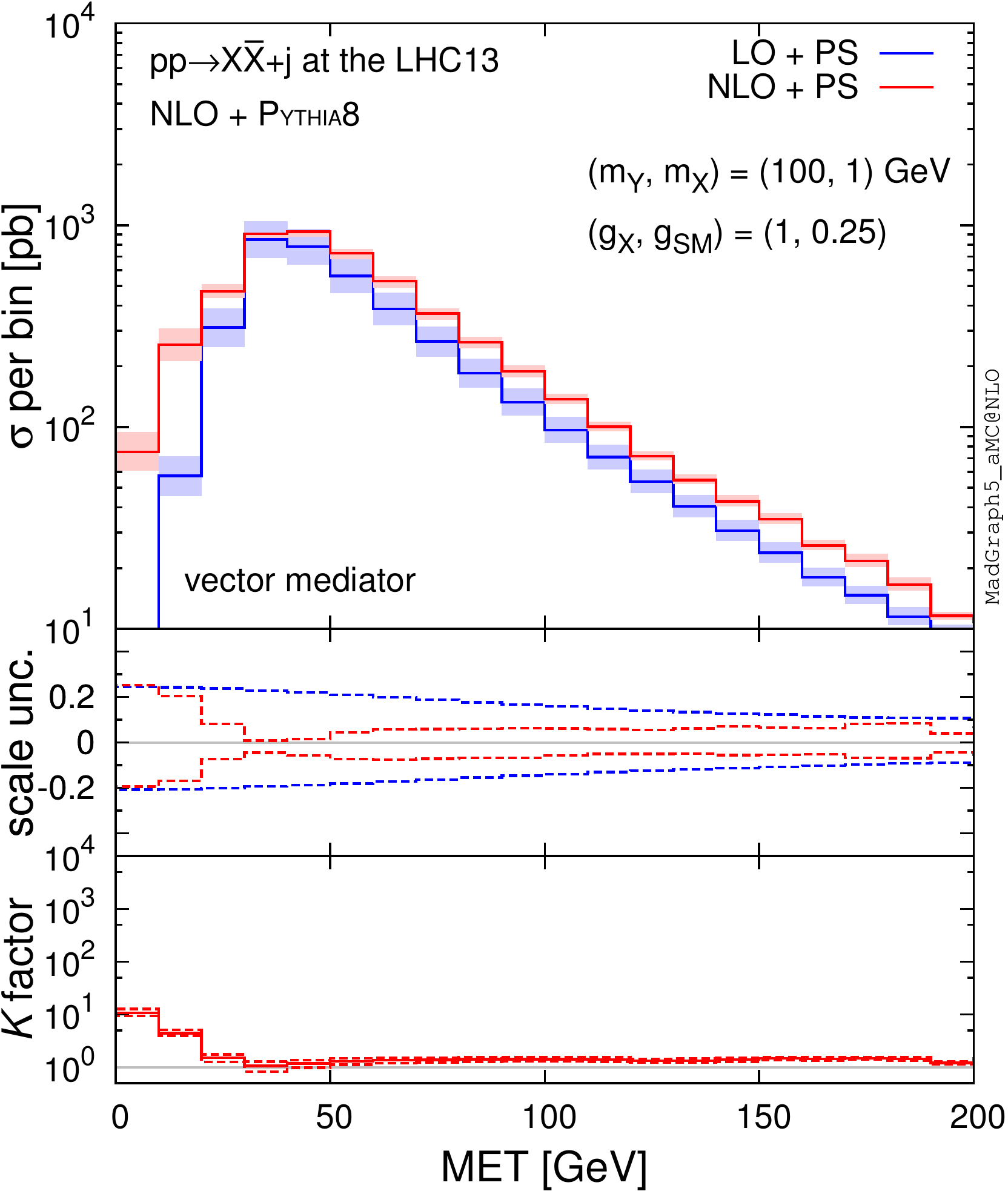}
\caption{
 MET distribution at (N)LO+PS accuracy for $pp\to X\bar X+j$ at the
 13-TeV LHC for $(m_Y,m_X)=(10,1) \GeV$ (left) and
 $(m_Y,m_X)=(100,1) \GeV$ (right). 
 The lower panels provide information on the differential scale
 uncertainty and $K$ factor.}  
\label{fig:met10}
\end{figure*}

In Table~\ref{tab:xxj_axial}, we present the pure axial-vector mediator
case by fixing the parameters as in Eqs.~\eqref{paramX_a},
\eqref{paramSM_a} and \eqref{param_spin1}. 
The resulting cross sections are very similar compared to the pure
vector case for $m_Y>2m_X$, while in the off-shell regime, we
find that the cross sections are suppressed compared to the production
via pure vector mediators.  In the off-shell situation the DM pair is
produced at threshold.
A pair of (Dirac) DM originating from a decay of a spin-1 mediator will
be in a $~^{2S+1}L_J$ state with $J=1$.
If the coupling is vector-like the DM pair can be in a $^3S_1$ state,
while if it is axial-like it will be in a $^3P_1$ state, {\it i.e.}
suppressed at threshold.
The similar argument holds in case of $gg\to Y_0+t\bar t$ case, as we
show in Sect.~\ref{sec:DM_top}.

The NLO effects are very similar to the vector mediator
scenario for all the mass combinations as well as the $\MET$ cuts.
Although we do not show the mixed scenario of vector and axial-vector,
one can easily compute such scenarios by changing the coupling
parameters in our simplified model. 

The parameter point $(m_Y,m_X)=(10,1)$~GeV warrants special attention,
as it illustrates a case of large NLO corrections  
(so-called ``giant $K$ factors''~\cite{Rubin:2010xp}), which might arise
in the limit where $p_T^j \gg m_{Y},m_{X}$. 
The giant $K$ factors in the $p p \to Y_1+j$ process occur due to the opening of the 
$pp\to Y_1+jj$ channel at NLO. This process can lead to a di-jet event
topology with a soft, collinear emission of $Y_1$.
In the regime of $p_T^j \gg m_Y$, the $Y_1$ emission behaves similar to
an emission of a massless gauge boson, where the diagrams with dijet
topologies contribute factors of
$\alpha_{s}^2 g_X^2\mathrm{log}^2(p_T^j/m_Y)$, and hence NLO $K$ factors
which scale as $\sim \alpha_{s} \mathrm{log}^2(p_T^j/m_Y)$.
Similar features commonly appear in calculations of SM $W/Z$+jets
processes at high jet $p_T$~\cite{Rubin:2010xp}. 

Topologies leading to giant $K$ factors are naturally suppressed in the case of DM production
by the cut on MET. This restricts the
calculation to regions of phase space which are insensitive to the soft
and collinear double-logs of di-jet event topologies with a soft $Y_1$
emission.
Figure~\ref{fig:met10} (left) illustrates the effect in case of
$m_Y=10$~GeV and $m_{X}=1$~GeV.
The region of low missing energy displays a two to three orders of
magnitude difference in rate between the LO and NLO calculations,
whereas we see that above $\MET> 50 \GeV$, the $K$ factor is
drastically reduced.
On the other hand, we see that already for  
$(m_Y,m_X)=(100,1)$~GeV in Fig.~\ref{fig:met10} (right), such logarithmic enhancements
are only very weak.

\subsection{Differential distributions}\label{sec:DM_distributions}

\begin{figure*}
\center 
\includegraphics[width=0.48\textwidth]{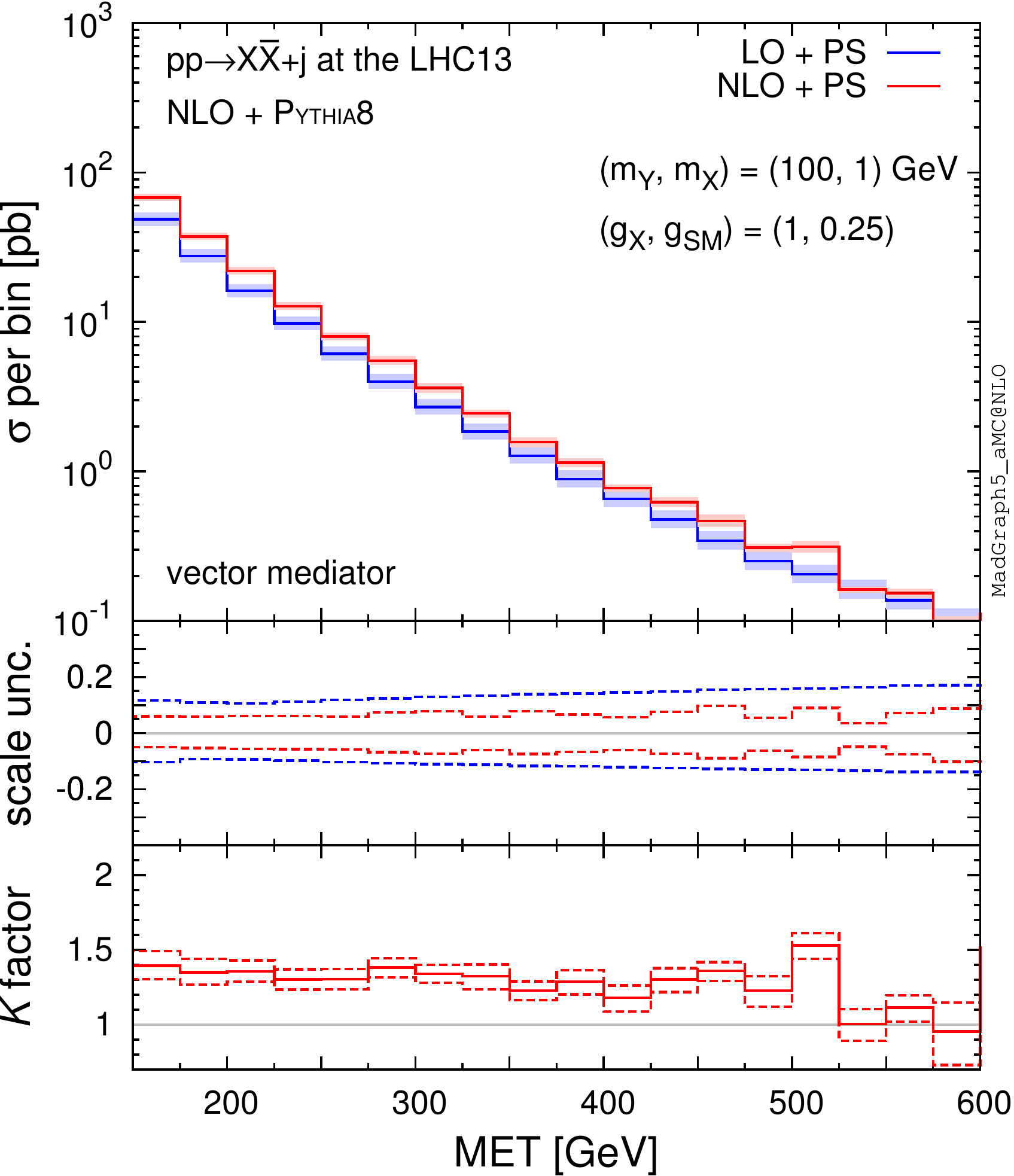}\qquad
\includegraphics[width=0.48\textwidth]{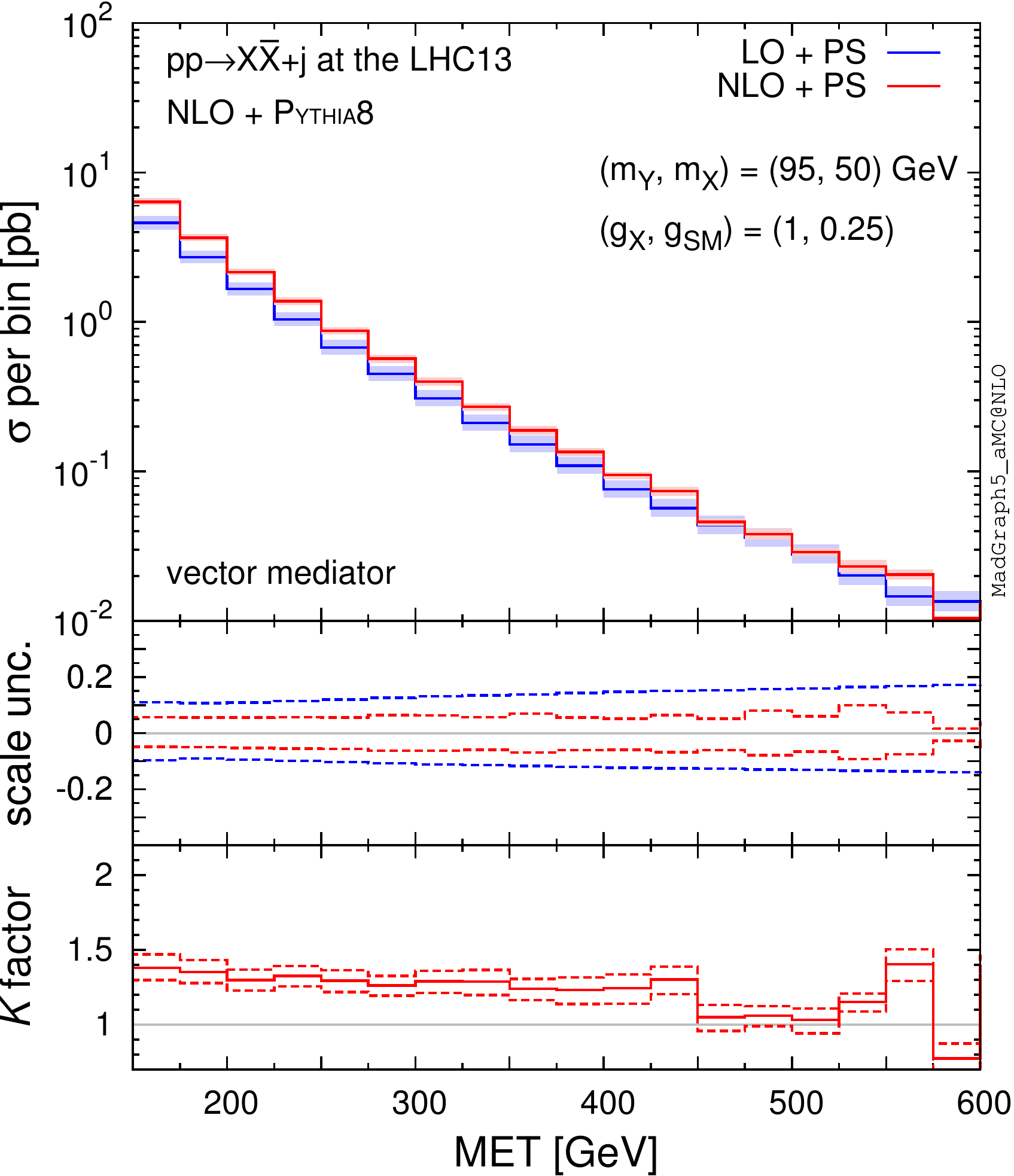}\\[5mm]
\includegraphics[width=0.48\textwidth]{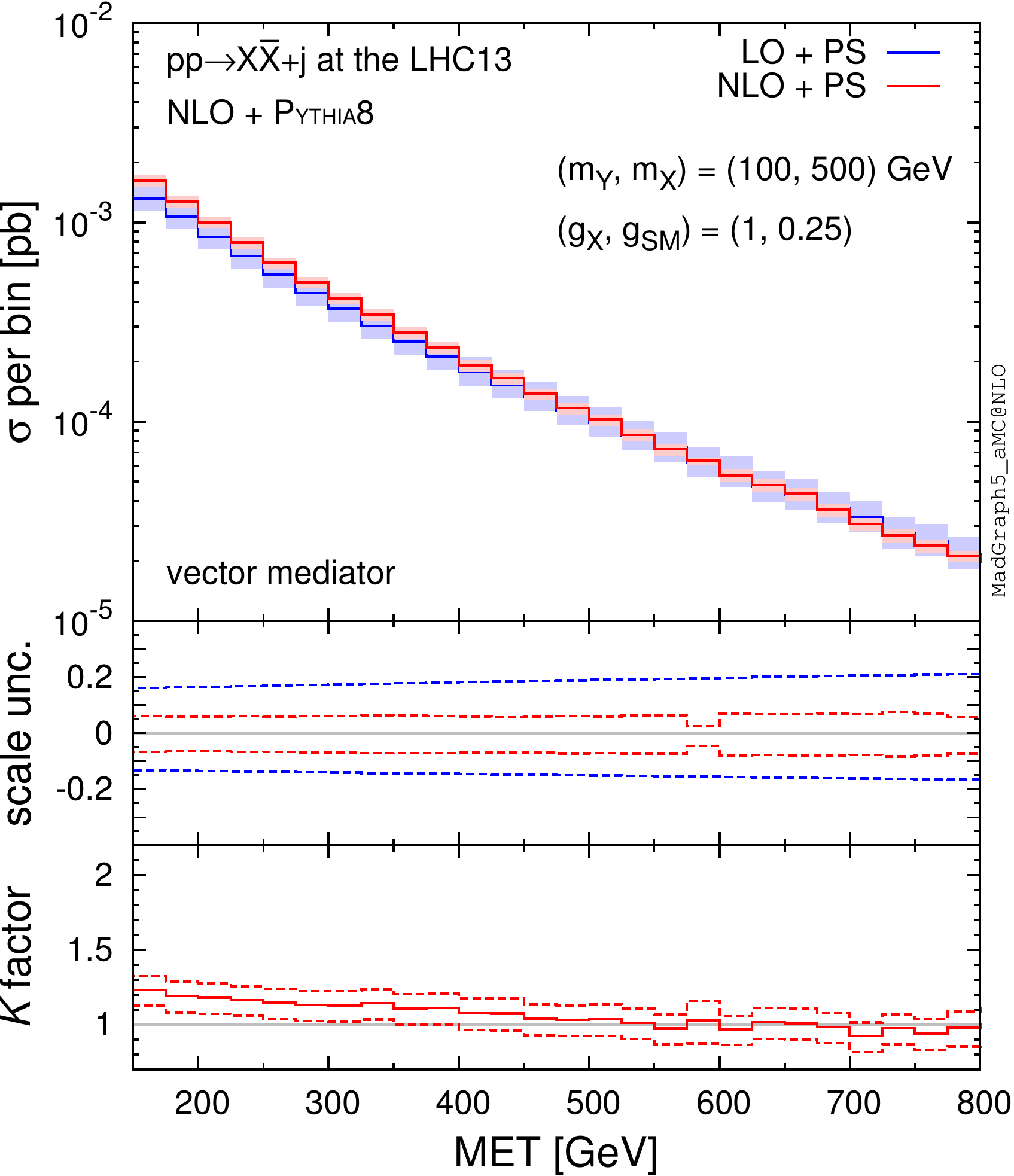}\qquad
\includegraphics[width=0.48\textwidth]{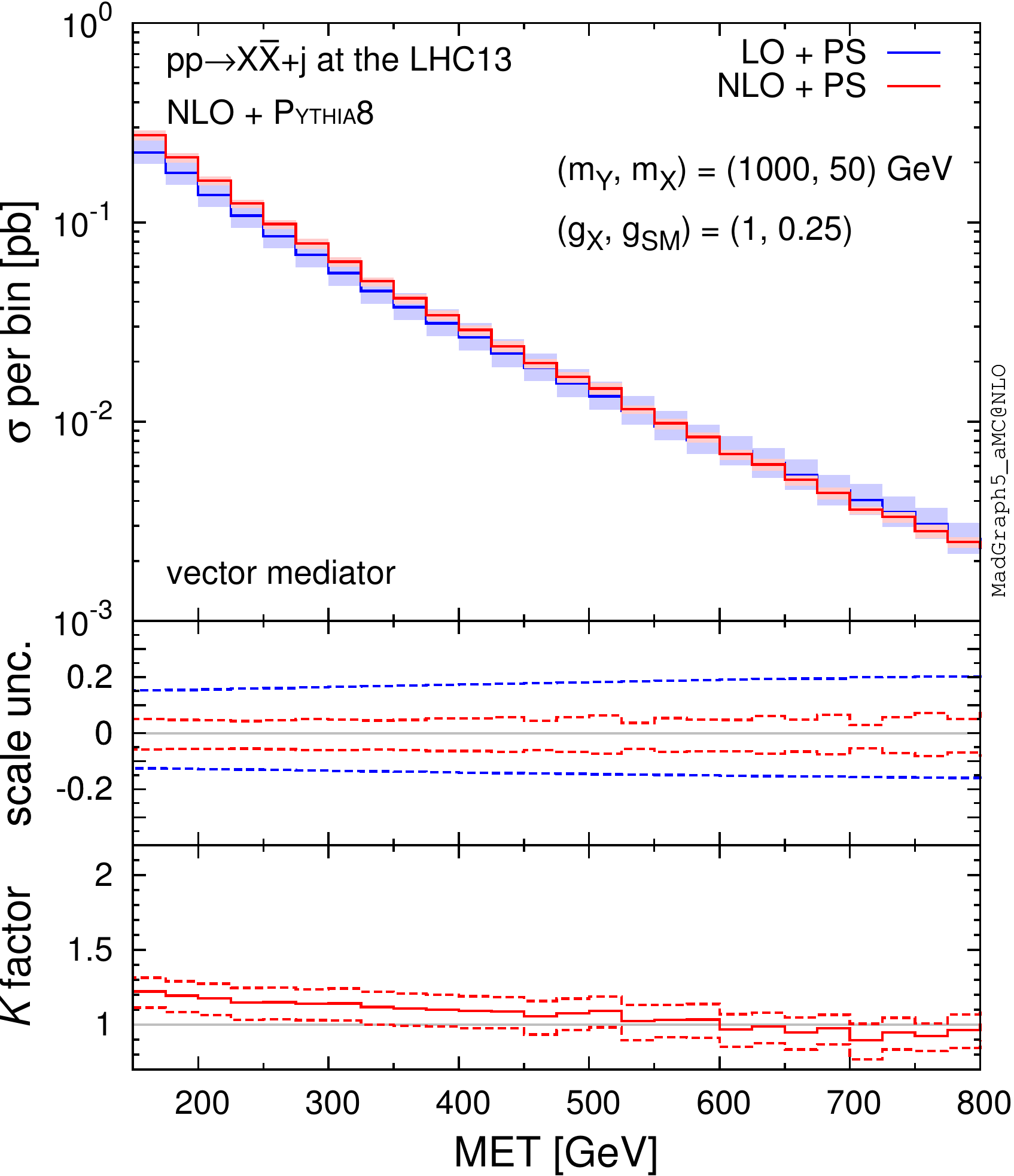}
\caption{
 MET distributions for $pp\to X\bar X+j$ at the
 13-TeV LHC for four benchmark points specified by ($m_Y,m_X$), where we
 assume a pure vector mediator and Dirac DM. 
 The middle and bottom panels show the differential scale uncertainties
 and $K$ factors, respectively.}
\label{fig:met}
\end{figure*}

\begin{figure*}
\center 
\includegraphics[width=0.48\textwidth]{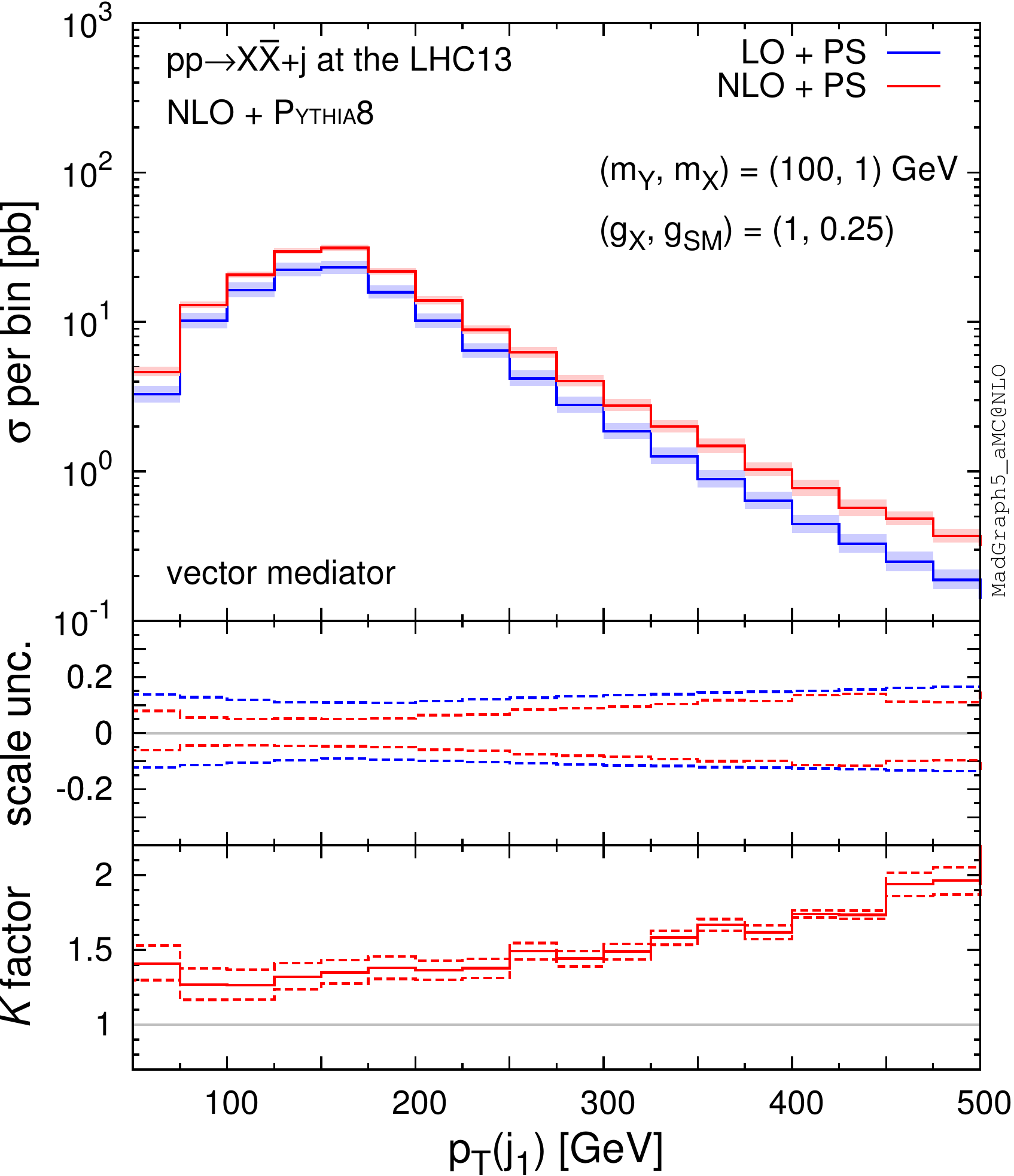}\qquad
\includegraphics[width=0.48\textwidth]{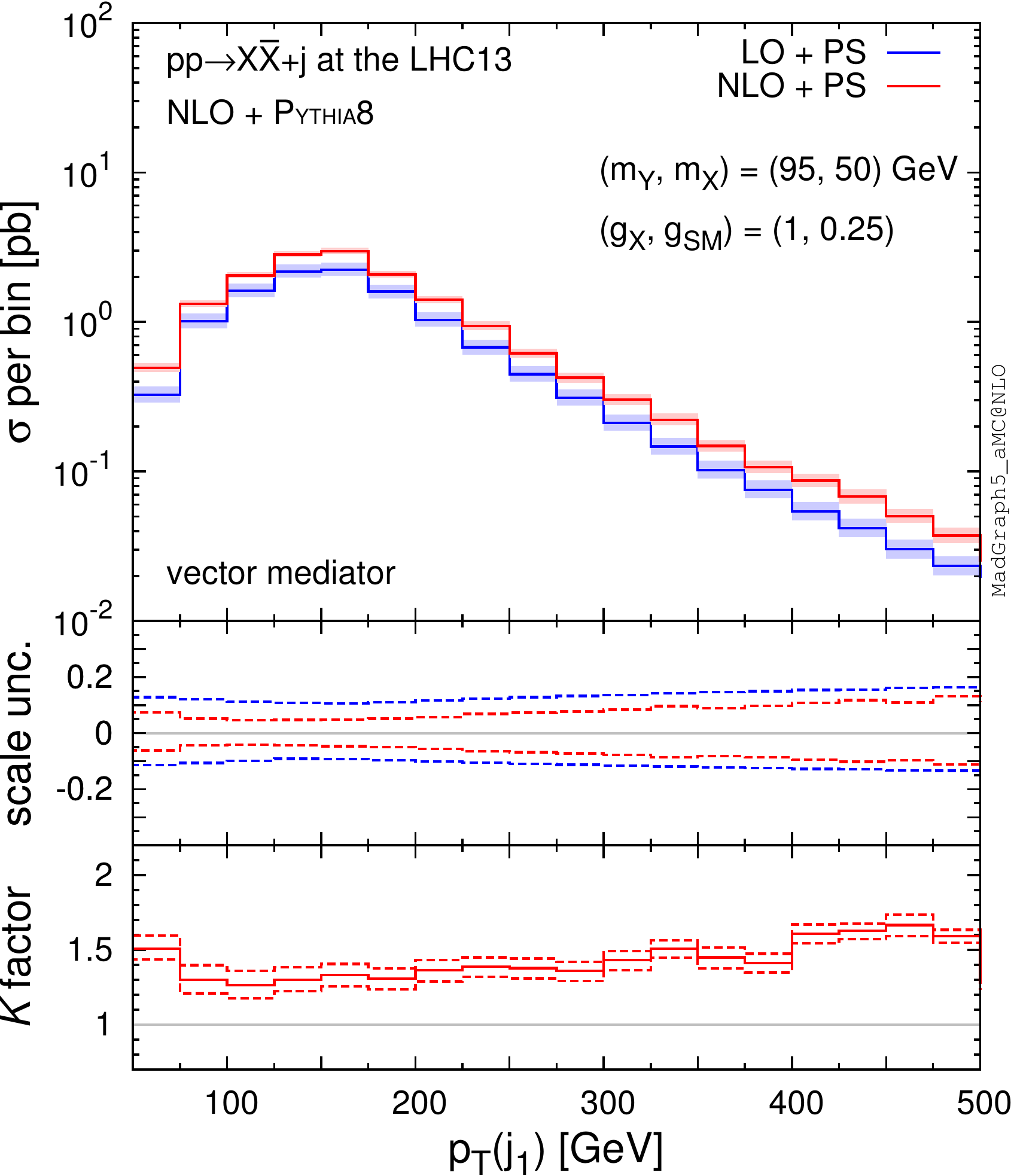}\\[5mm]
\includegraphics[width=0.48\textwidth]{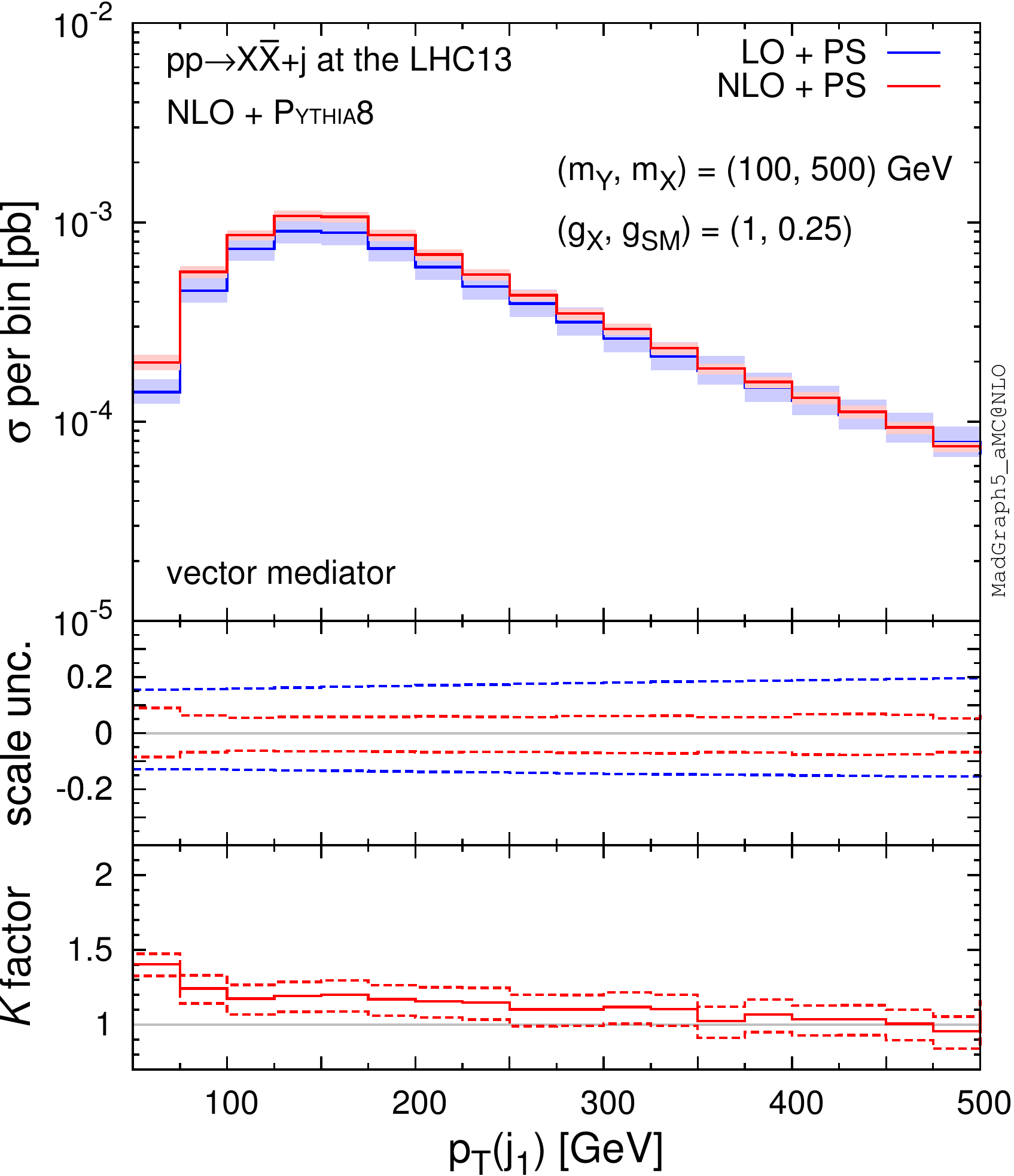}\qquad
\includegraphics[width=0.48\textwidth]{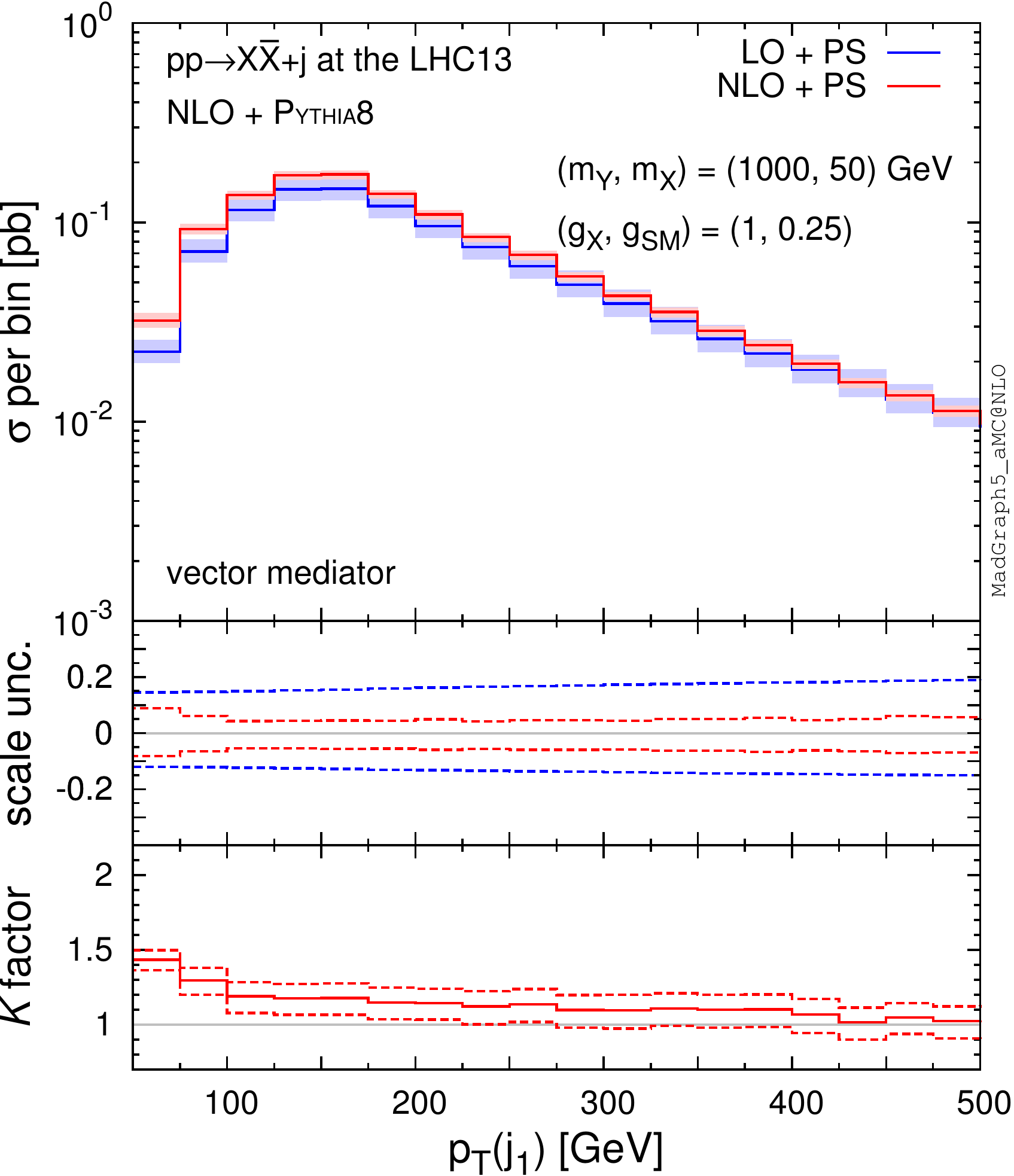}
\caption{
 $p_T$ distributions of the hardest jet for
 $pp\to X\bar X+j$ at the 13-TeV LHC for four benchmark points specified by
 ($m_Y,m_X$), where we assume a pure vector mediator and Dirac DM and
 the $\MET > 150 \GeV$ cut is imposed.  
 The middle and bottom panels show the differential scale uncertainties
 and $K$ factors, respectively.}
\label{fig:ptj1}
\end{figure*}

\begin{figure*}
\center 
\includegraphics[width=0.48\textwidth]{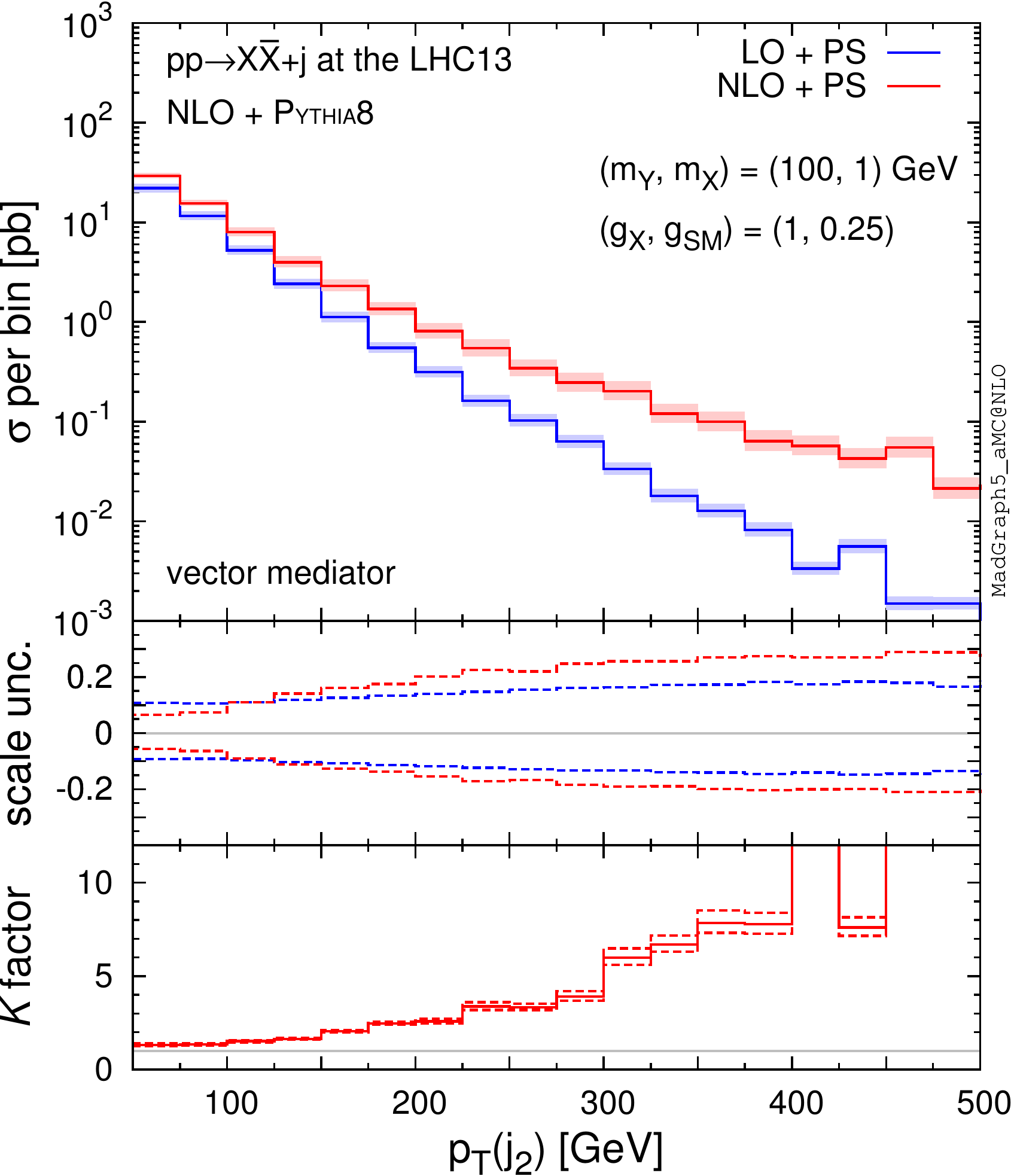}\qquad
\includegraphics[width=0.48\textwidth]{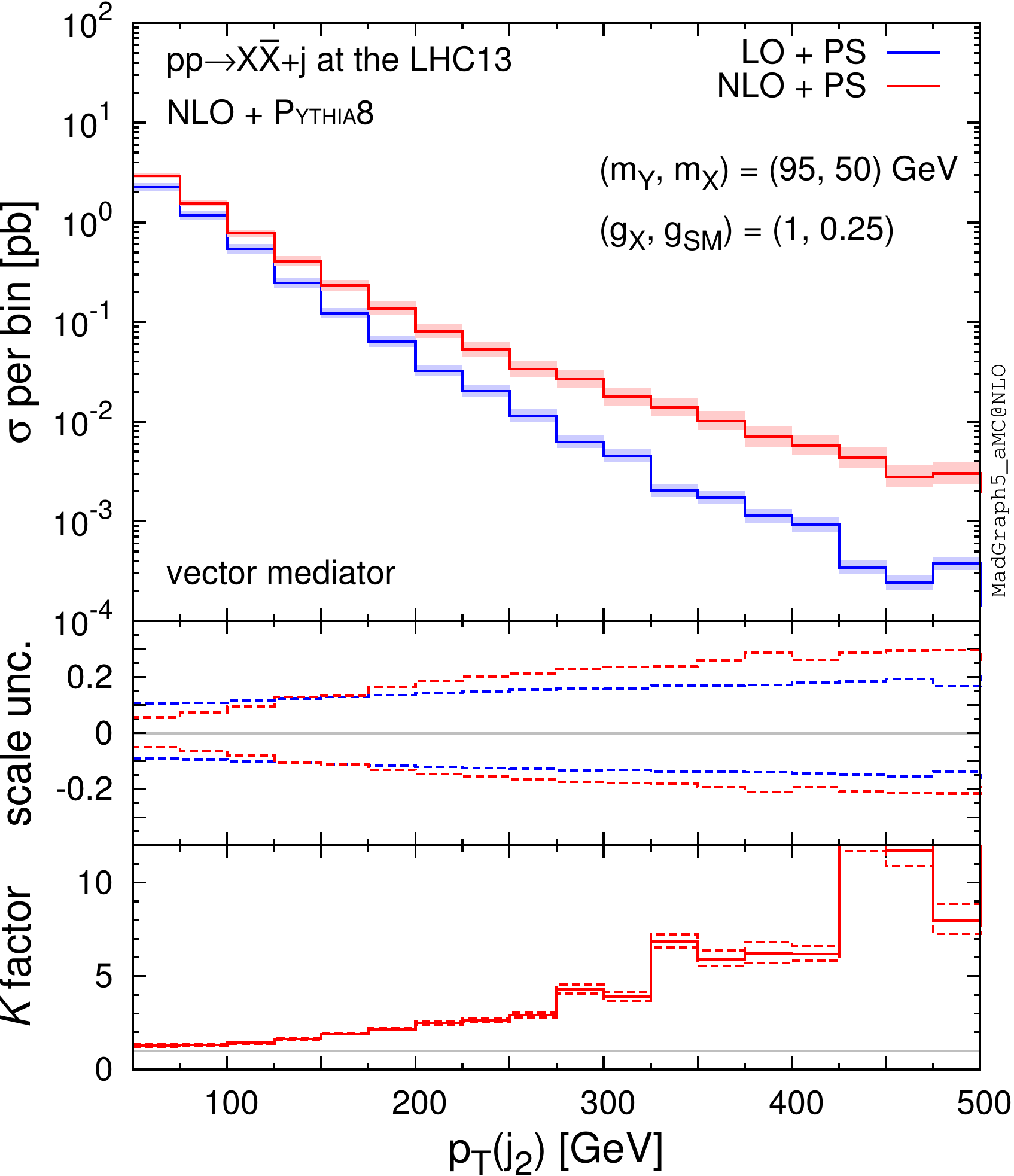}\\[5mm]
\includegraphics[width=0.48\textwidth]{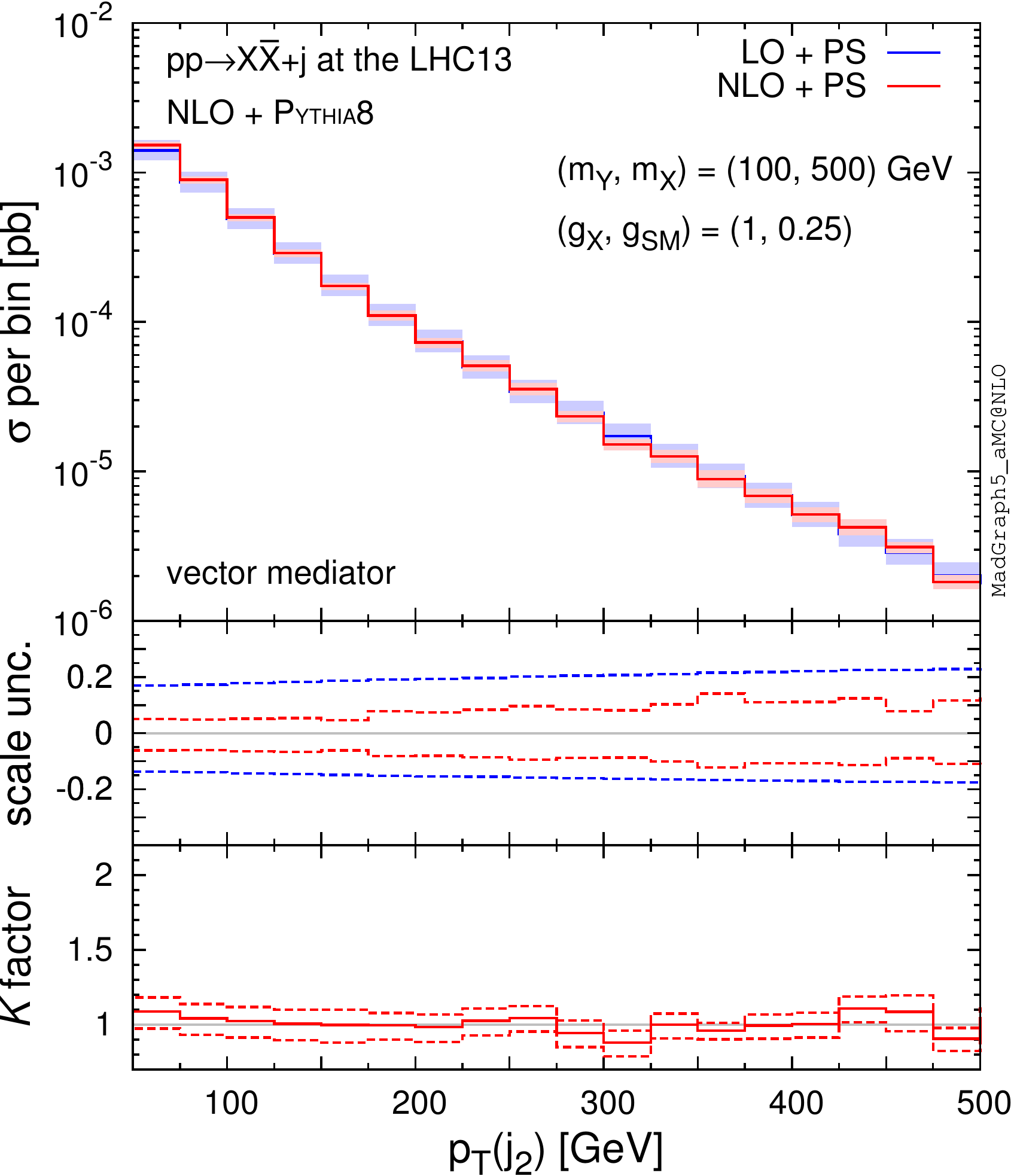}\qquad
\includegraphics[width=0.48\textwidth]{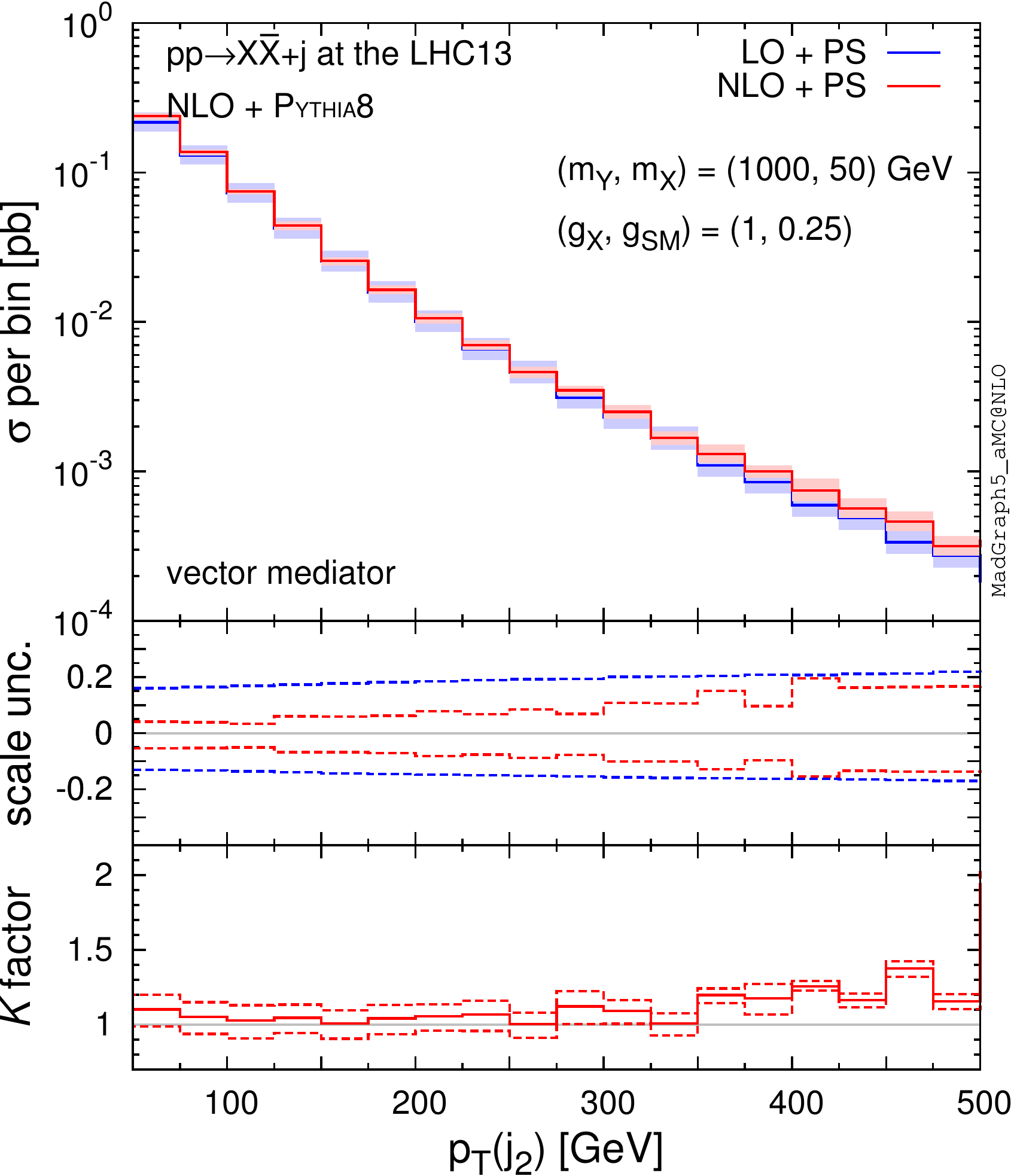}
\caption{Same as Fig.~\ref{fig:ptj1}, but for the second hardest jet.}
\label{fig:ptj2}
\end{figure*}

\begin{figure*}
\center 
\includegraphics[width=0.48\textwidth]{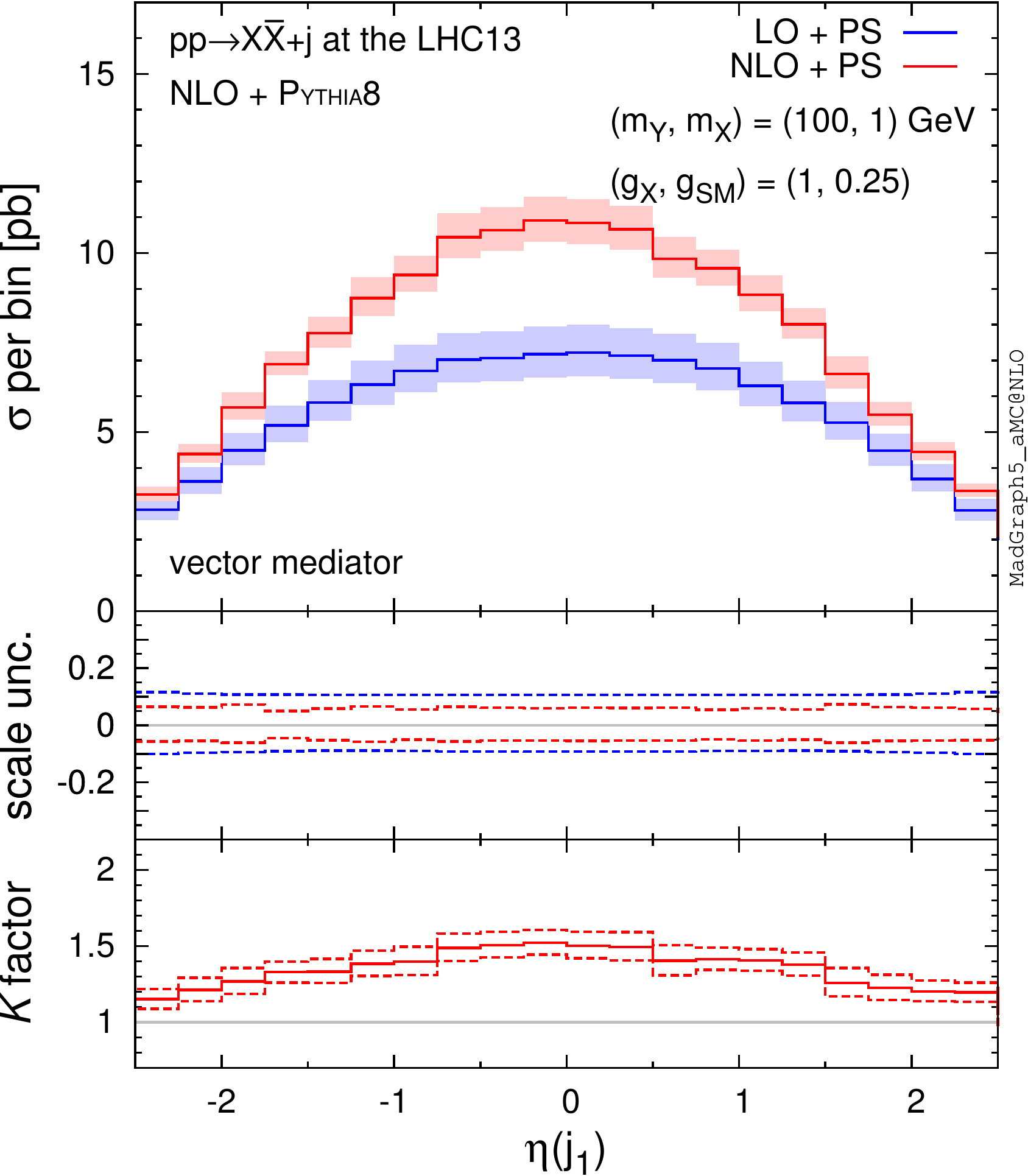}\qquad
\includegraphics[width=0.48\textwidth]{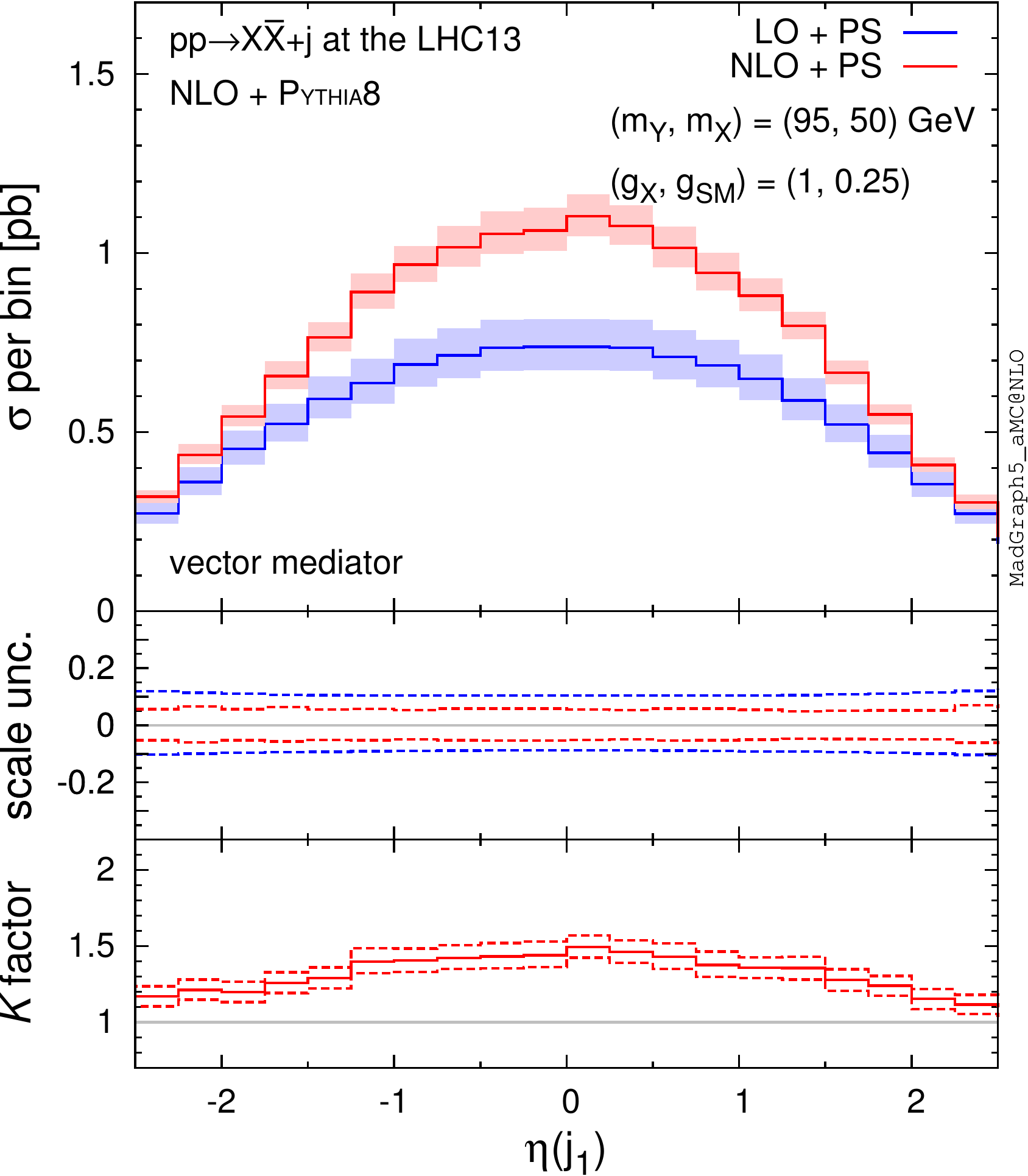}\\[5mm]
\includegraphics[width=0.48\textwidth]{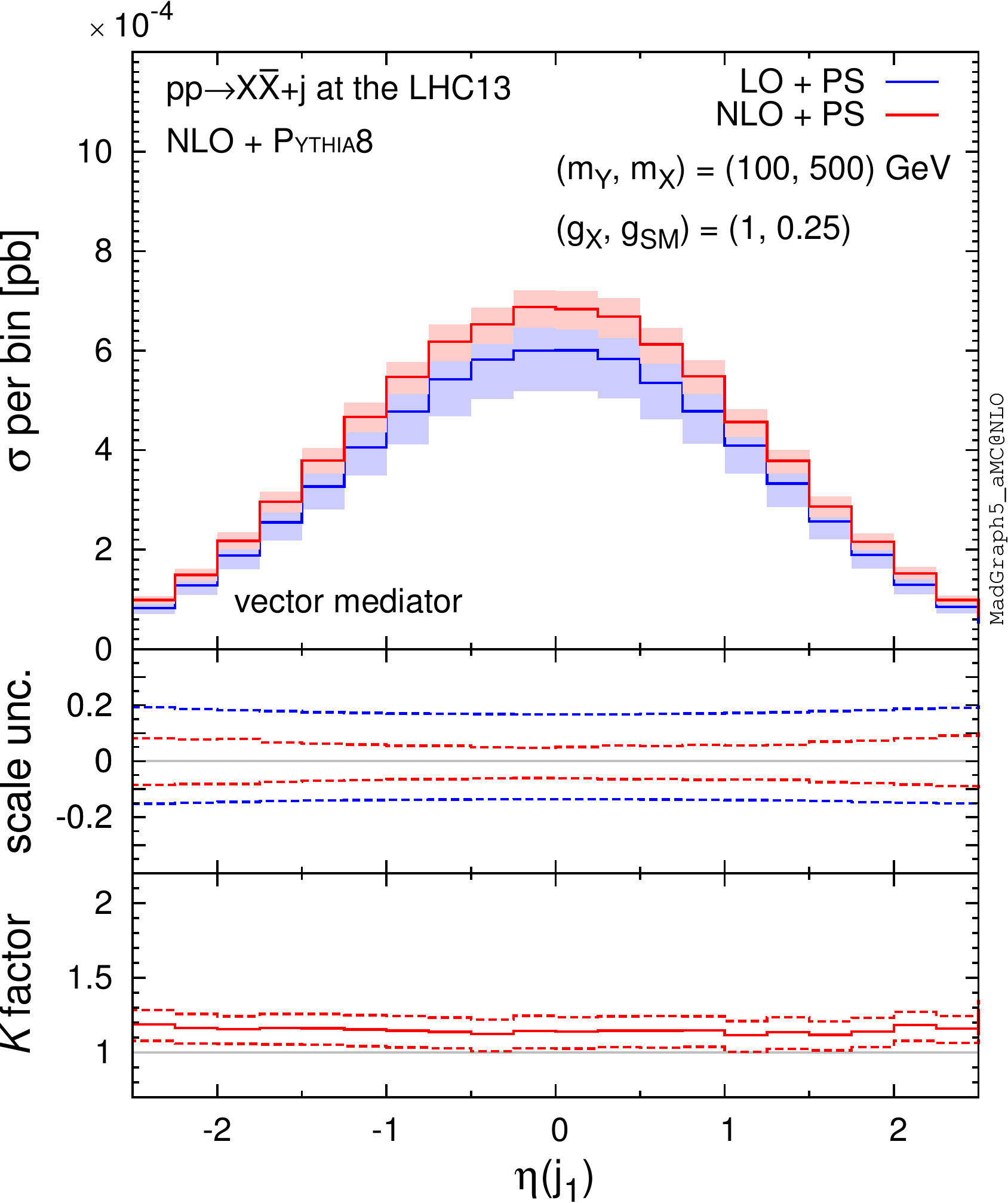}\qquad
\includegraphics[width=0.48\textwidth]{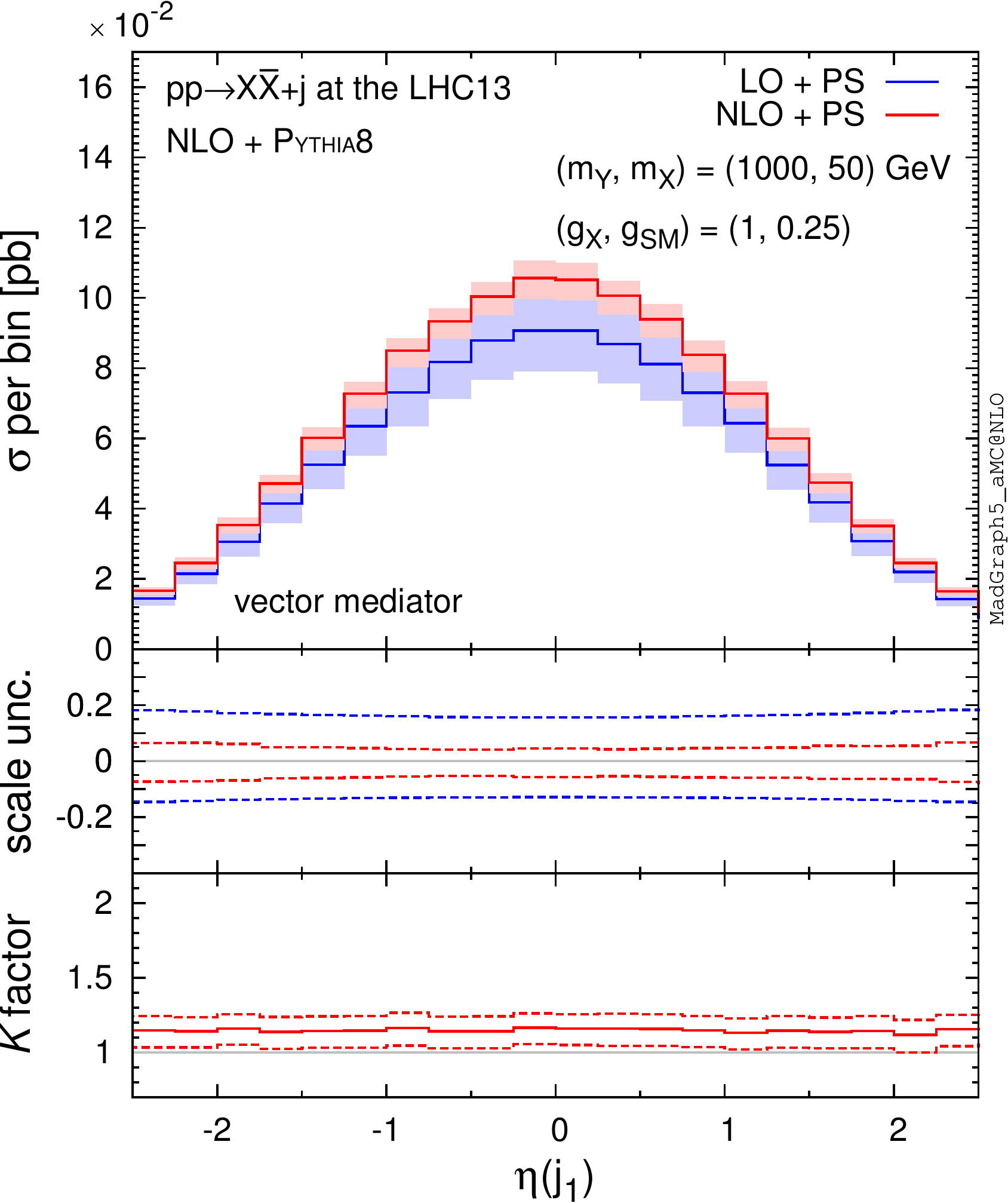}
\caption{$\eta$ distributions of the hardest jet for
 $pp\to X\bar X+j$ at the 13-TeV LHC for four benchmark points specified by
 ($m_Y,m_X$), where we assume a pure vector mediator and Dirac DM and
 the $\MET > 150 \GeV$ cut is imposed.  
 The middle and bottom panels show the differential scale uncertainties
 and $K$ factors, respectively.}
\label{fig:etaj1}
\end{figure*}

\begin{figure*}
\center 
\includegraphics[width=0.48\textwidth]{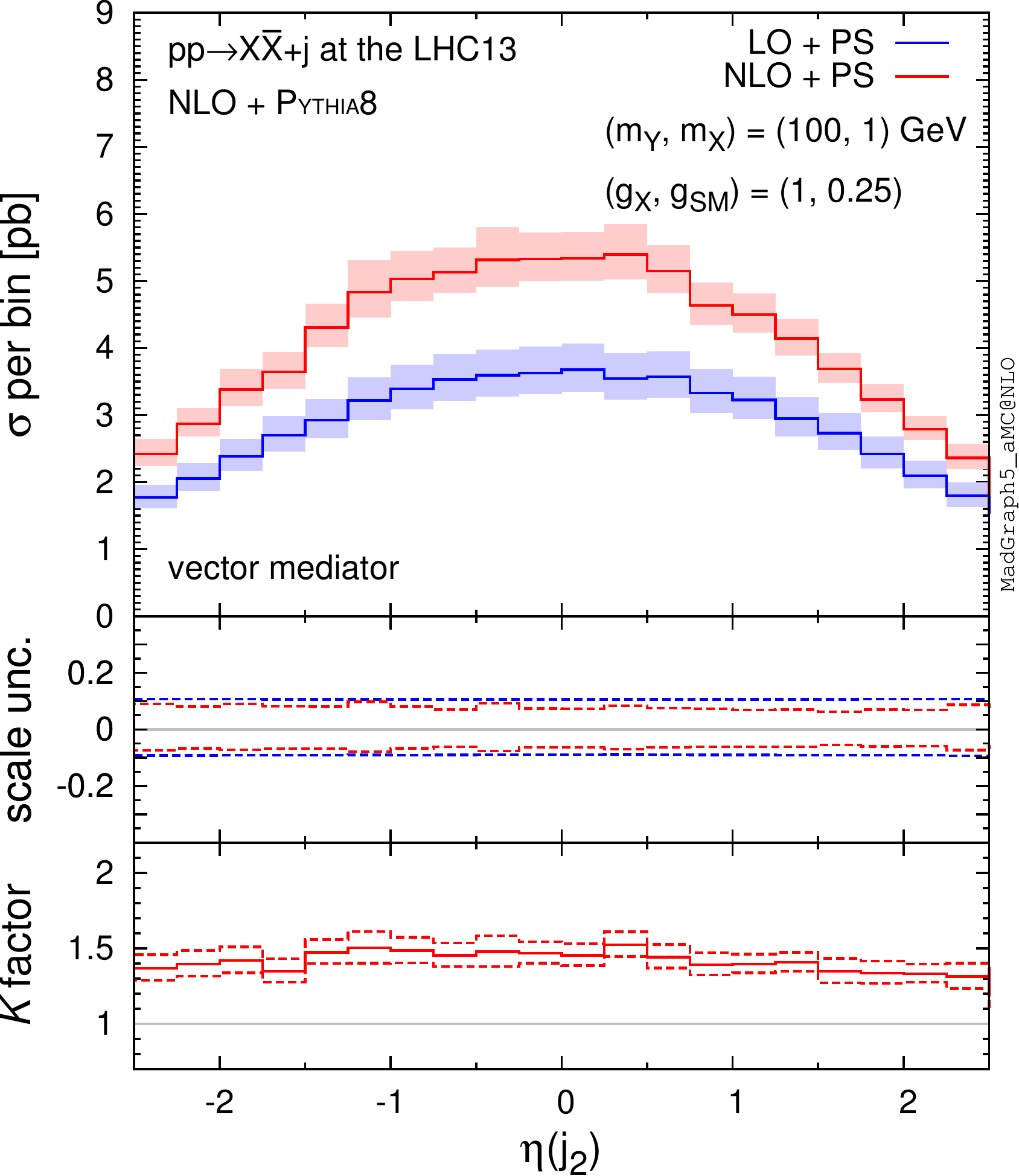}\qquad
\includegraphics[width=0.48\textwidth]{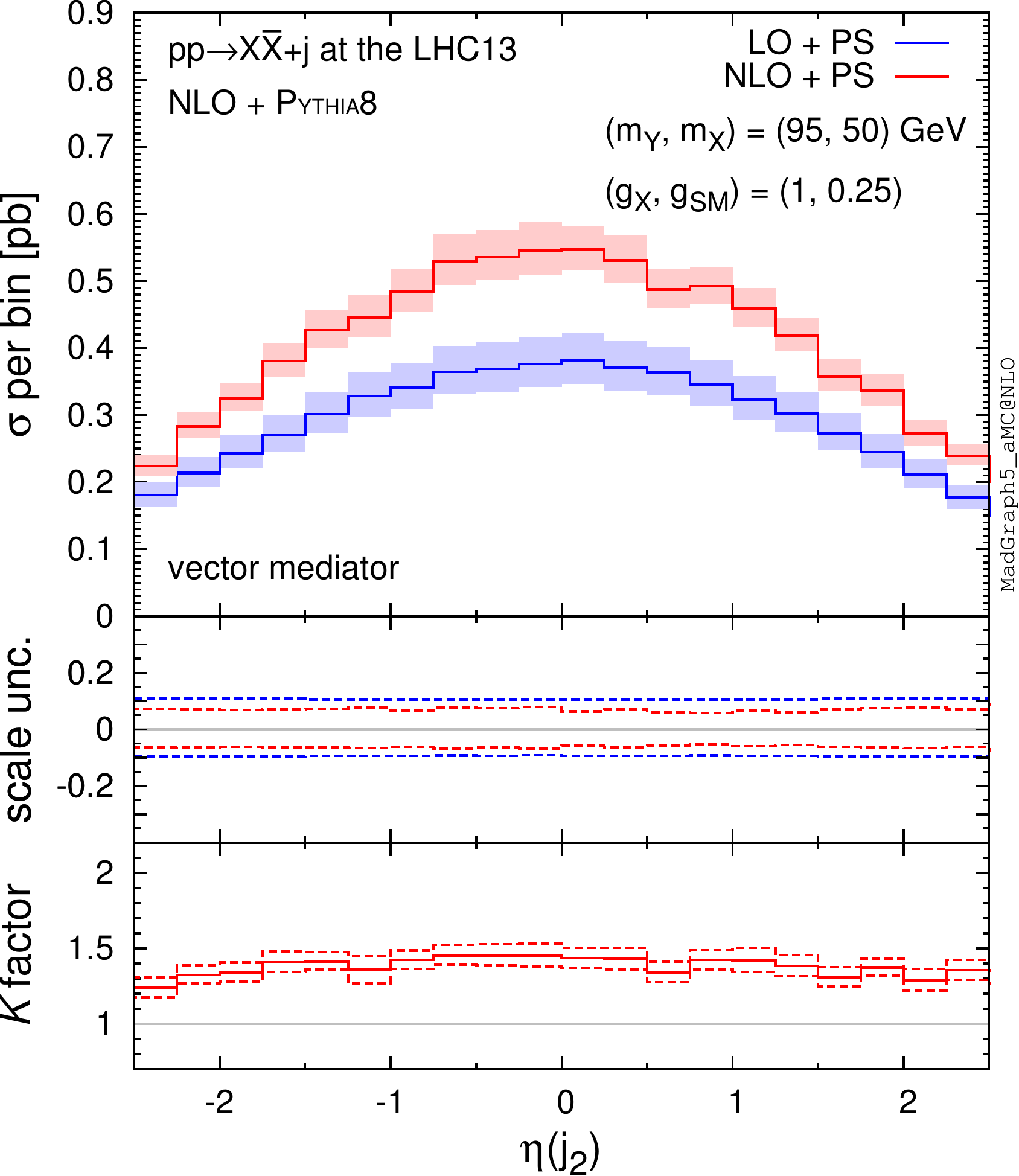}\\[5mm]
\includegraphics[width=0.48\textwidth]{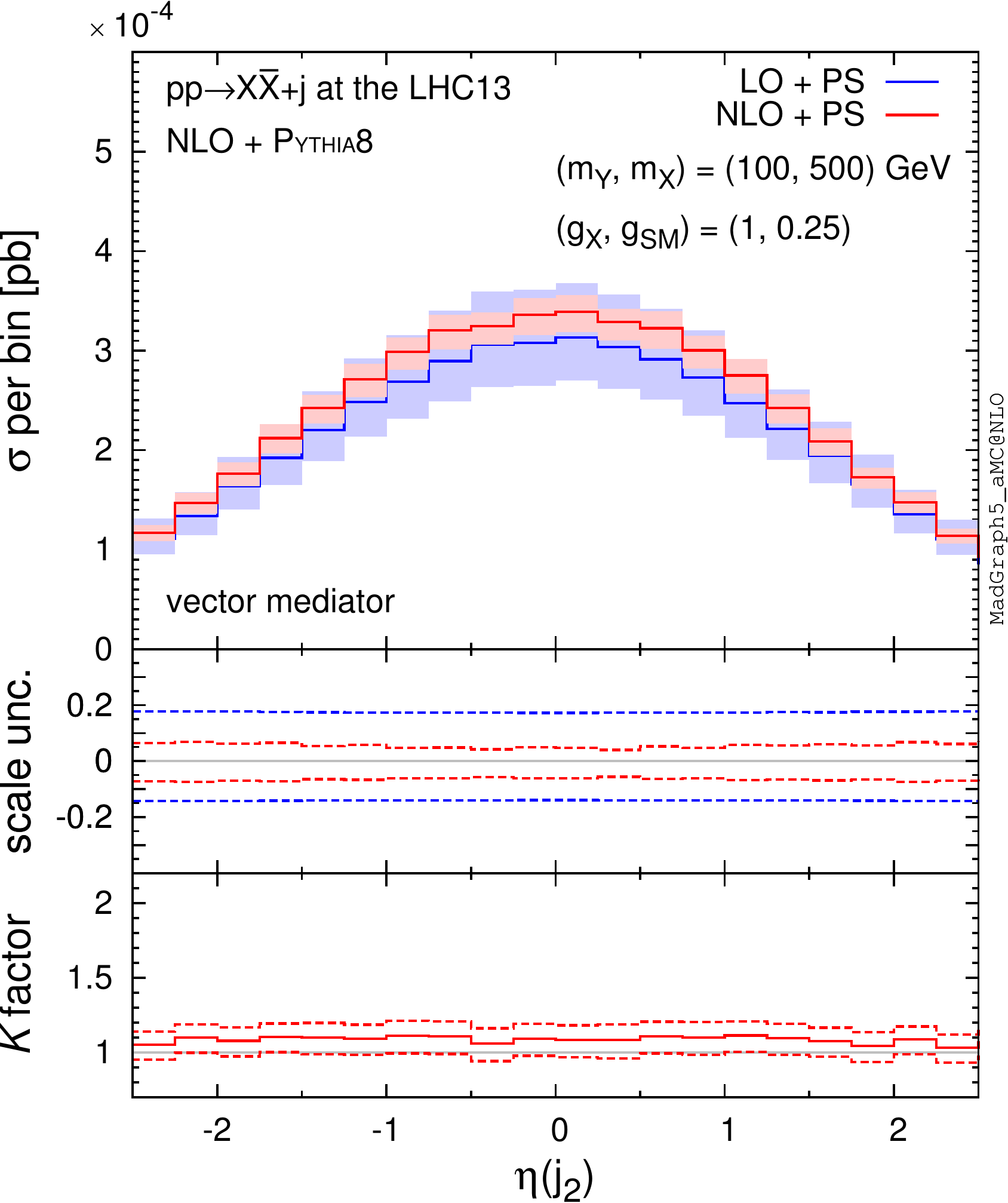}\qquad
\includegraphics[width=0.48\textwidth]{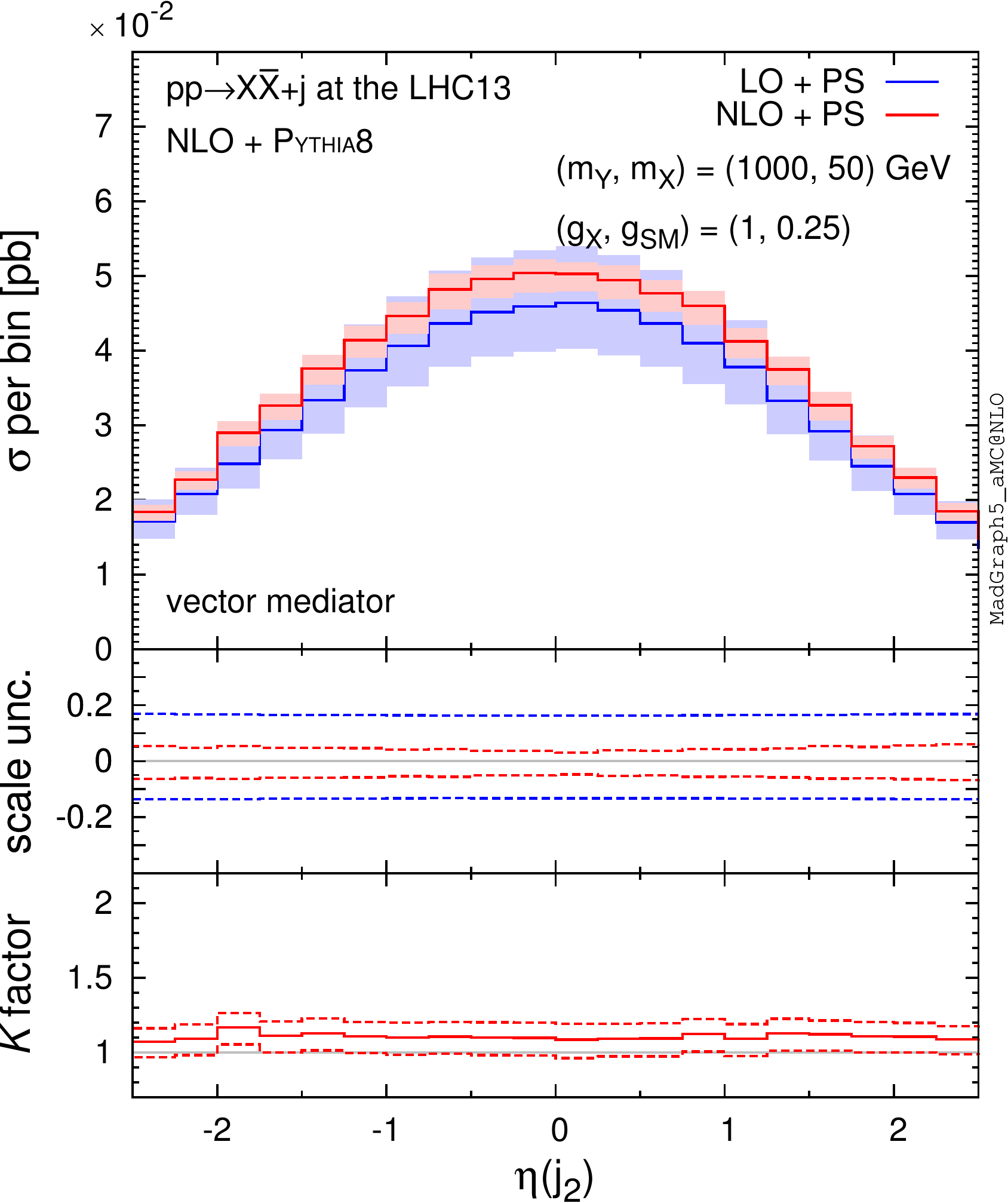}
\caption{Same as Fig.~\ref{fig:etaj1}, but for the second hardest jet.}
\label{fig:etaj2}
\end{figure*}

We proceed with the discussion of the features of the differential
distributions relevant for DM studies.
We begin with Fig.~\ref{fig:met} which shows the $\MET$ distributions at
LO and NLO for four benchmark points of the simplified model,
assuming a pure vector mediator and Dirac fermion DM.
As seen in the total rates, the NLO effects in the distributions do not
depend on the mass relation between the mediator and the DM, $i.e.$
on-shell or off-shell, but do depend on the energy scale of the final
state. 
In the top panels, the energy scale is ${\cal O}(100)$~GeV for $m_Y$ or
$2m_X$. We find that the two benchmark points display striking
similarities in the shape of the $\MET$ distributions, while the rate of
the latter is suppressed due to off-shell $Y_1$ production.
The largest effects of NLO corrections are in the low $\MET$ regions,
where NLO corrections reach $K$ factors of about 1.4 for
$\MET \sim 150 \GeV$, with a steady decrease with increasing $\MET$. 
We observe similar features also in the high-scale benchmark points of
${\cal O}(1)$~TeV for $m_Y$ or $2m_X$ (bottom panels of
Fig.~\ref{fig:met}), where the largest $K$ factors are about 1.2 for
$\MET \sim 150 \GeV$.
Comparing with the FO distributions in Fig.~\ref{fig:met_fo}, we observe
that the parton shower does not affect the MET distribution.
Note that the NLO corrections are different for different $\MET$
regions, with the largest NLO corrections occurring in the lower MET
regions where the rate is the highest.  
Hence the careful estimation of NLO effects is very important for
accurate LHC studies of DM in each signal region. 

Next, we study the features of jet kinematic distributions produced in
association with DM.  Figures~\ref{fig:ptj1}, \ref{fig:ptj2}, \ref{fig:etaj1} and
\ref{fig:etaj2} show example $p_T$ and $\eta$ distributions of the
hardest and second hardest jets for the four benchmark points as in
Fig.~\ref{fig:met}, assuming  
\begin{align}\label{metcut}
 \MET>150\ {\rm GeV} \,.
\end{align}
Distributions of the hardest jet transverse momentum show very
interesting features. 
In Fig.~\ref{fig:ptj1} we find that, in all benchmark points, the LO
distributions match the NLO predictions at the peak, $i.e.$ 
$p_T(j_1)\sim150$~GeV, to a very good degree.
The agreement can be attributed to the imposed $\MET$ cut
in~\eqref{metcut}, which forces the events into a back-to-back
configuration of 
the leading jet and the $Y_1$ mediator (on average).
We also note that the NLO scale uncertainty in the peak region becomes
very small compared to the LO estimates. 

The NLO corrections to $p_T (j_1)$ distributions affect not only the
overall rate, but the shape of the distribution as well. 
In the low-$p_T$ region, $K$ factors are about $1.2-1.5$.
In the high $p_T$ region, we find significant NLO effects again for the 
$(m_Y,m_X)=(100,1)$ and (95, 50)~GeV cases (top panels), but not for the
$(m_Y,m_X)=(100, 500)$ and (1000, 50)~GeV cases (bottom panels).
We note that the scale uncertainty does not significantly reduce at
NLO in the $p_T$ regions away from the peak, especially for the light
mediator and DM case (top panels). 
Significant differences in NLO contributions and theoretical
uncertainties in different regions of the
$p_T(j_1)$ spectrum suggest that the proper modelling of the hardest jet
differential distributions has to go beyond the simple scaling by a
constant $K$ factor.

Apart from the highest $p_T$ jet which is modelled by the hard matrix
element, all other jets in the LO simulation are generated by the parton
shower. 
By contrast, the NLO corrections include real emission diagrams which can
contain two hard and well-separated partons in the final state as well
as virtual corrections to one parton emission.
One could expect significant differences between LO and NLO in the
kinematic distributions of the second highest $p_T$ jet. 
For the $(m_Y,m_X)=(100,1)$ and (95,50)~GeV cases (top panels), we
observe giant $K$ factors in the high-$p_T$ tails of the distributions.
The large difference between LO and NLO computations is a consequence of
the inadequacy of the parton shower to accurately model high-$p_T$
emissions. 
In Fig.~\ref{fig:ptj2} (bottom panels), on the other hand, we find no
significant differences between LO and NLO for the overall rate and
shape of the second jet emission in case of very heavy mediators
($i.e.$ $m_Y = 1 \TeV$) or heavy DM ($i.e.$ $m_X = 500 \GeV$),
suggesting that the second hardest jet is described  very well by the parton shower.  
This is because the scale of the shower is very high and therefore extra
parton emission from the parton shower can be sufficiently hard. 

Features similar to those observed in $p_T(j_{1,2})$ 
also occur in distributions of the hardest/second-hardest jet
pseudo-rapidity ($\eta(j_{1,2})$), shown in Figs.~\ref{fig:etaj1} and
\ref{fig:etaj2}. 
For the light mediator/DM (top panels), we observe that the
rate at which the hardest jet is emitted at NLO in the low rapidity
region is enhanced by a factor about $1.5$, with the corrections
falling off with the increase in rapidity.
However, even though the overall rate for the second-hardest jet
increases by a factor of roughly $\sim 1.5$ the shape of the $\eta(j_2)$
distribution is affected only mildly. 
In the case of heavy mediator/DM, the hardest jet is emitted
at a lower rapidity (on average) at a significantly higher rate compared
to light mediators as illustrated by the width of the $\eta$
distributions in Fig.~\ref{fig:etaj1}. 
As the hardest jet typically recoils against $\MET$, this explains why
the $\MET$ spectrum falls off more quickly for lighter mediators than
for the heavier ones.

\subsection{Merging samples at NLO accuracy}\label{sec:DM_merging}

In addition to total and differential production cross sections for the
$pp\to X\bar X+j$ process, we study NLO effects for different jet
multiplicities in the final state.
For this purpose we utilise the FxFx merging
procedure~\cite{Frederix:2012ps} within the framework of {\sc MG5aMC},
and 
consider
\begin{align}
 pp\to X\bar X +0,1,2\ \mathrm{jets}\,.
\end{align}
We take the merging scales at 45 and 30~GeV for 2- and 1-jet merged
samples, respectively. 

\begin{figure}
 \center
 \includegraphics[width=1.\columnwidth]{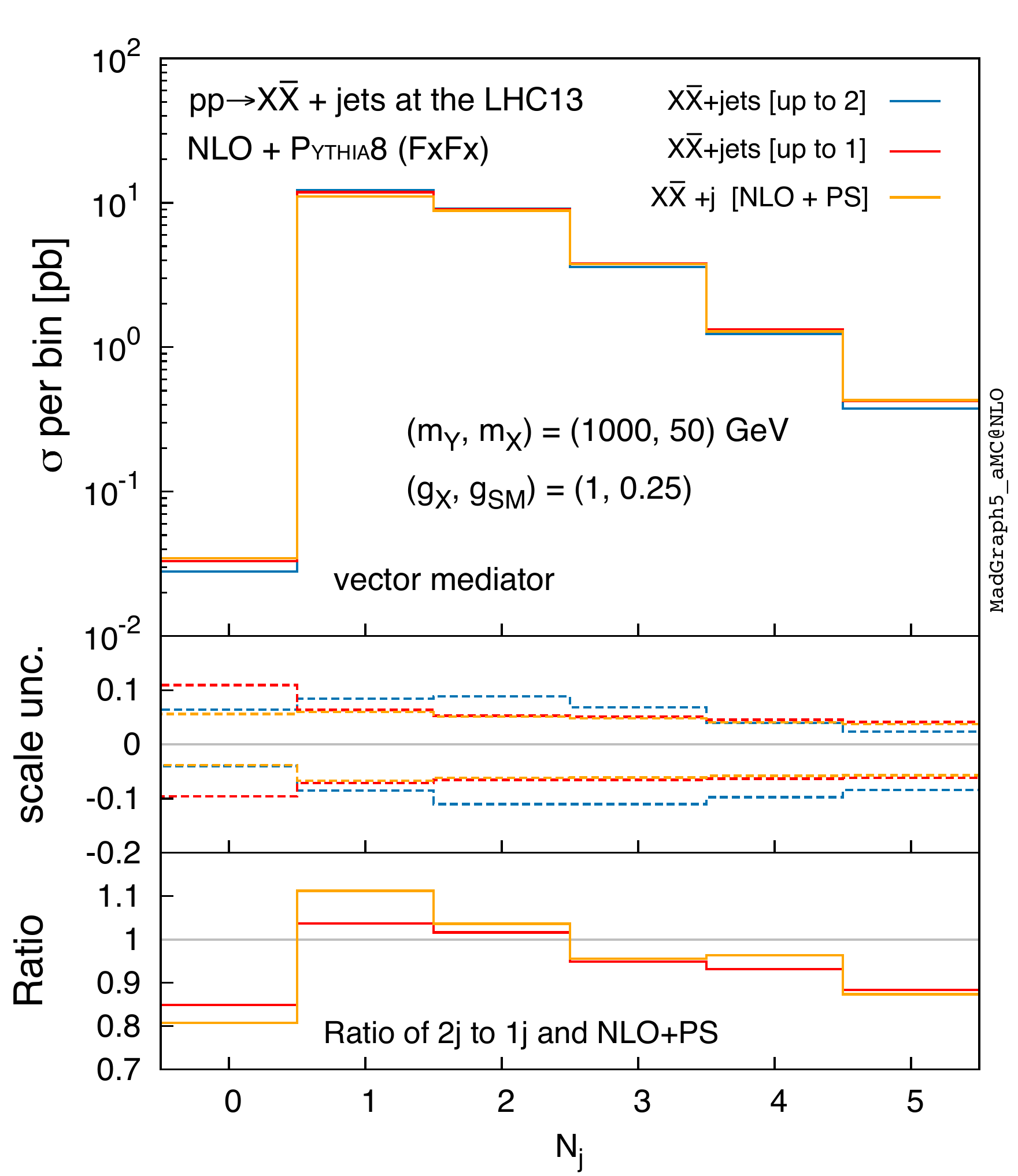}
 \caption{
 Jet multiplicity for $pp\to X\bar X+$ jets at the 13-TeV LHC with FxFx
 merging in case of $m_Y=1$~TeV with $m_X=50$~GeV, where
 NLO+PS samples are merged up to one (red) and two (blue) jets.
 The 1-jet NLO+PS sample without merging (orange) is also shown for
 comparison.  
 The middle panel shows the relative scale uncertainties,
 while the bottom panel presents the ratio of the 2$j$ merged sample
 to the 1$j$ merged one and to the non-merged one.
 We assume $\MET>100 \GeV$ and jets with $p_T > 30 \GeV$ and $|\eta| <
 4.5$ for the purpose of illustration.  
}\label{fig:fxfx}
\end{figure}

\begin{figure}
  \center
  \includegraphics[width=1.\columnwidth]{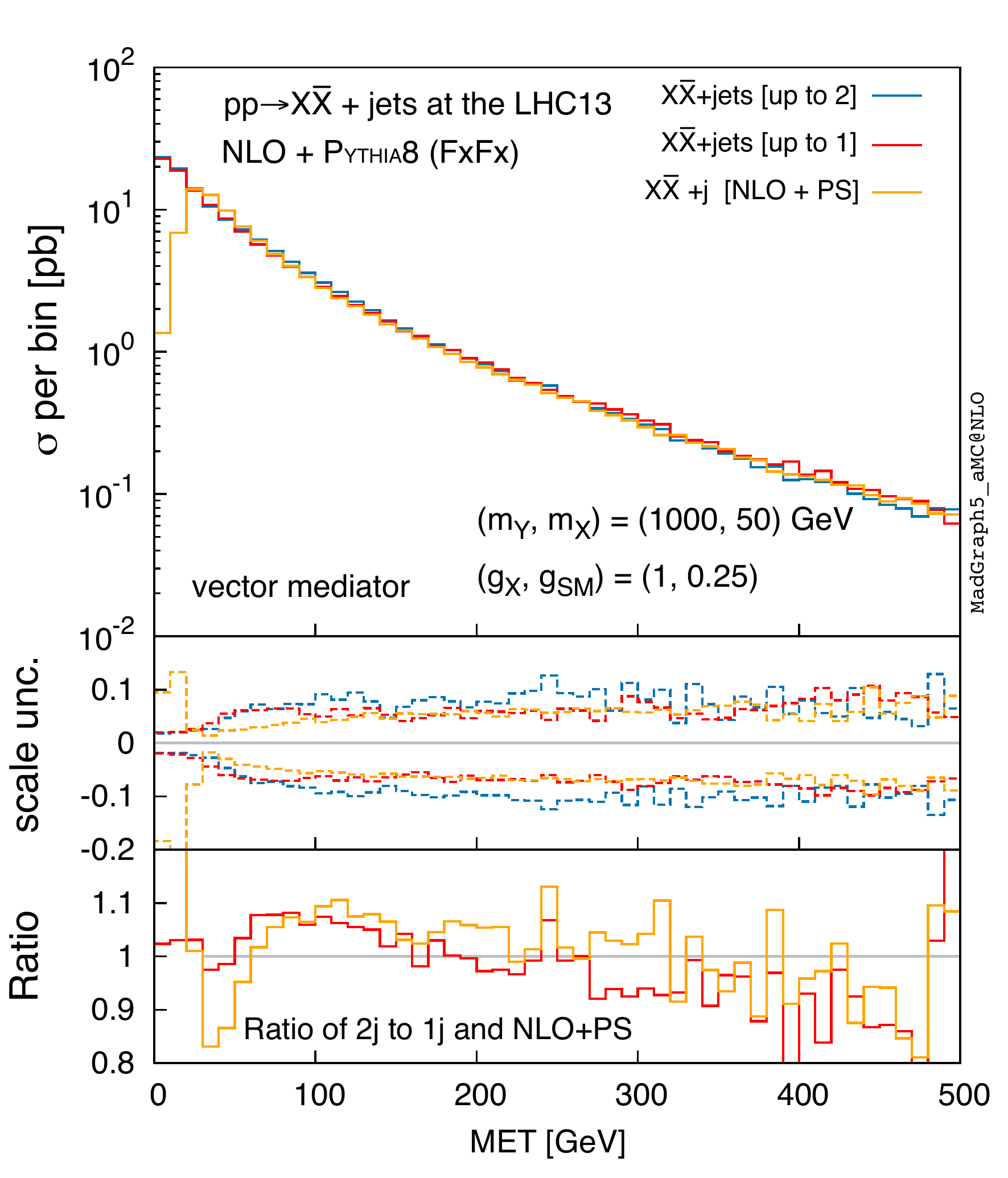}
  \caption{Same as Fig.~\ref{fig:fxfx}, but for the MET distribution.
}\label{fig:fxfxMET}
\end{figure}

Figure~\ref{fig:fxfx} shows the number of jets in the final state for
NLO merged samples in case of $m_{Y} = 1$~TeV and $m_{X} = 50$~GeV.
The red and blue curves show the results of merging up to 1 and 2 jets,
respectively, while the orange curve shows the $pp\to X\bar X+1j$
process at NLO+PS without merging for comparison.
An inspection  of the three samples in the lowest panel of
Fig.~\ref{fig:fxfx} shows that the effects of NLO merging are mild.
The non-merged NLO sample over-estimates the production rate in the $0j$
and $> 2j$ bins by 20 and 10\,\% respectively, and underestimates the rate in the $1j-2j$
bins by $<10\,\%$.
The differences are even milder between the samples merged to $1j$ and
to $2j$.
As the $0j$ bin is phenomenologically irrelevant, we can conclude that
the effects of jet merging at NLO are within $10\,\%$. 

We show effects of jet merging on the MET distribution in
Fig.~\ref{fig:fxfxMET}.
Except in the low MET region, we find that the effects of NLO merging
are again mild and within 10\,\%.

\subsection{Comparison of signal distributions to the Standard Model}\label{sec:comparison}

In discussions of NLO corrections to DM production, it is important to
consider how the possible signal events at NLO look in the midst of
large SM backgrounds. For the purpose of illustration, we consider only
the largest background in mono-jet searches for DM, {\it i.e.} $Z$+jets
and simulate it  to NLO merged up to 2 extra jets via the FxFx method. 

\begin{figure*}
\center 
\includegraphics[width=0.48\textwidth]{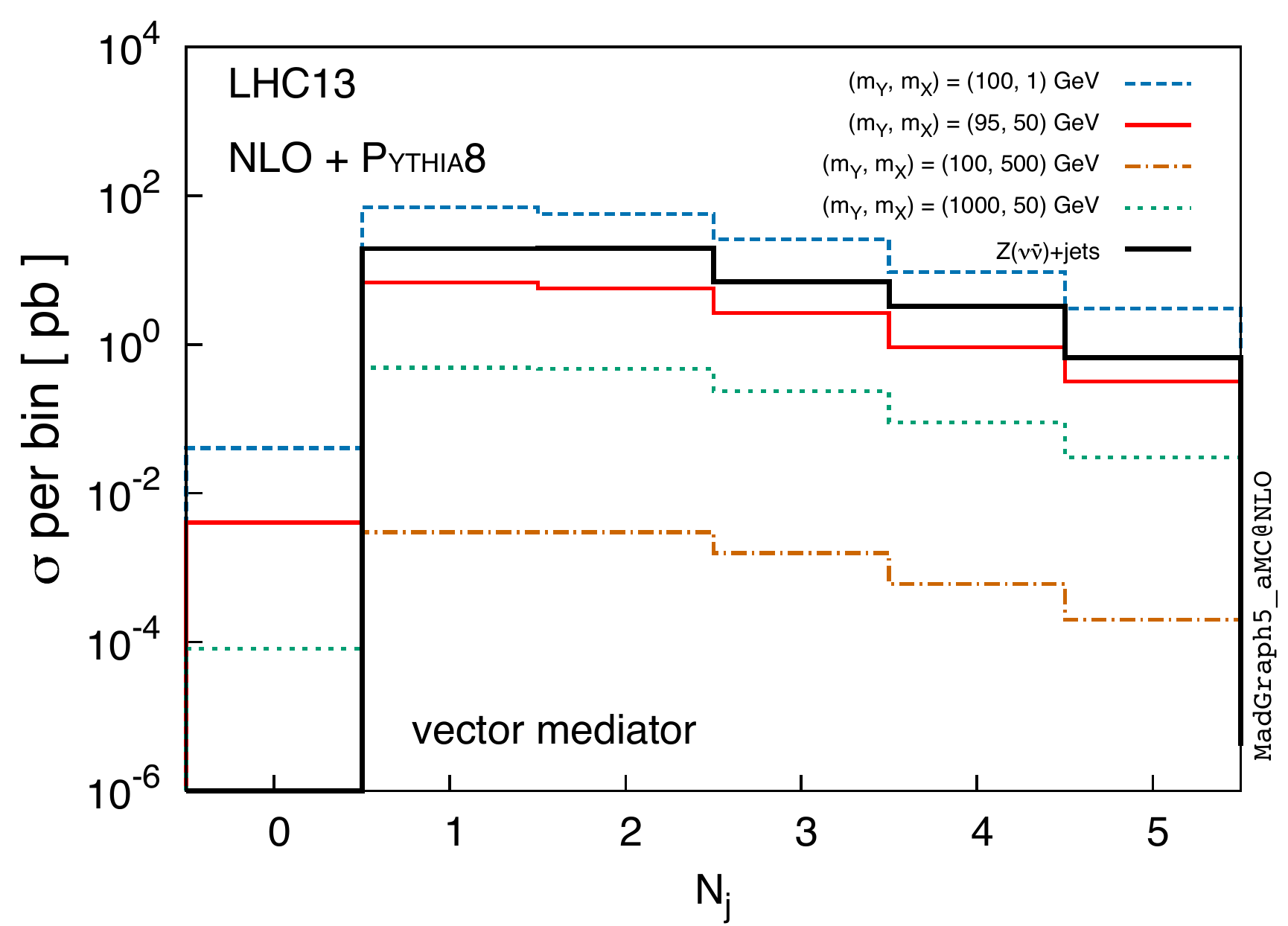}\qquad
\includegraphics[width=0.48\textwidth]{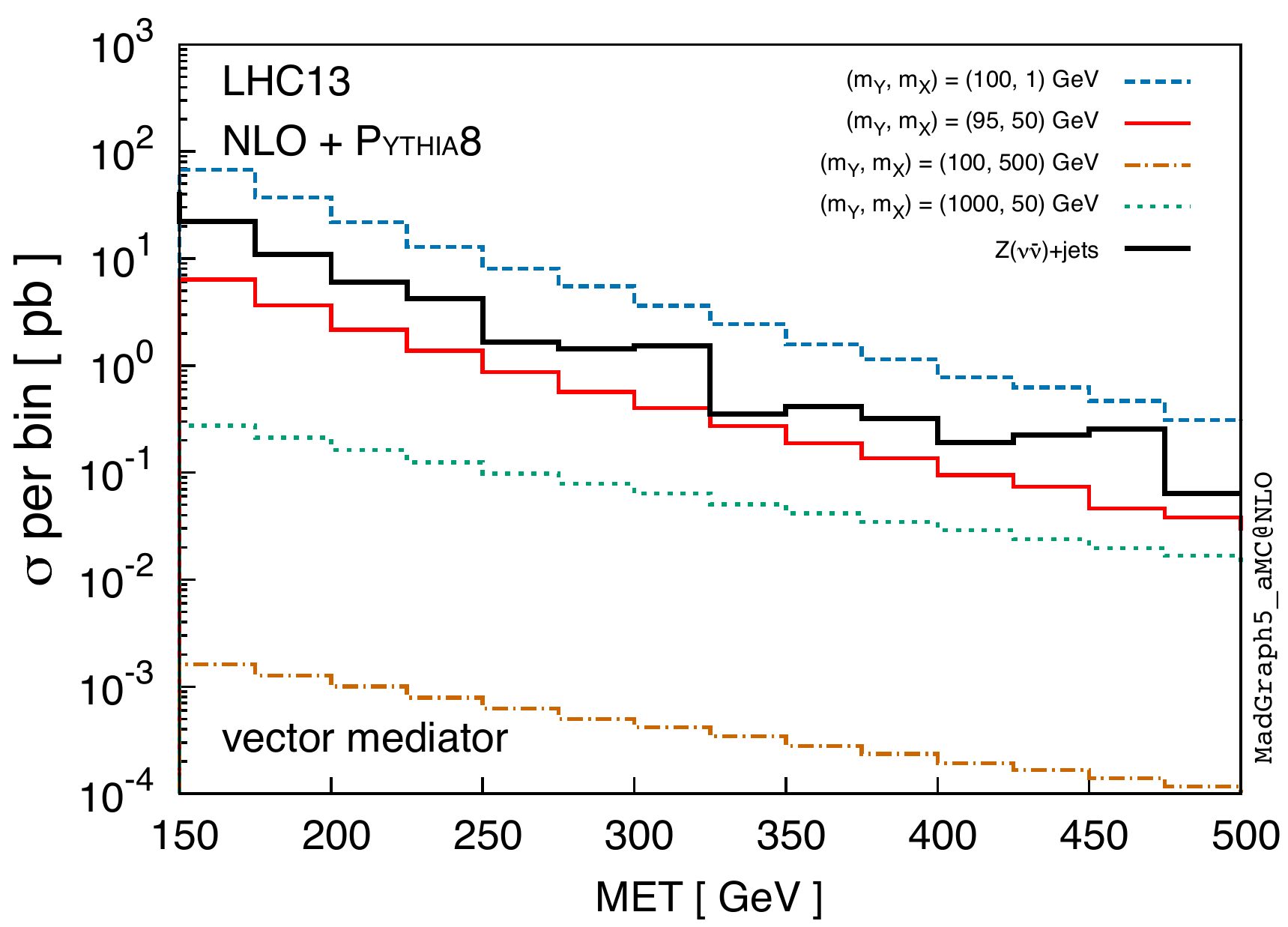}\\
\includegraphics[width=0.48\textwidth]{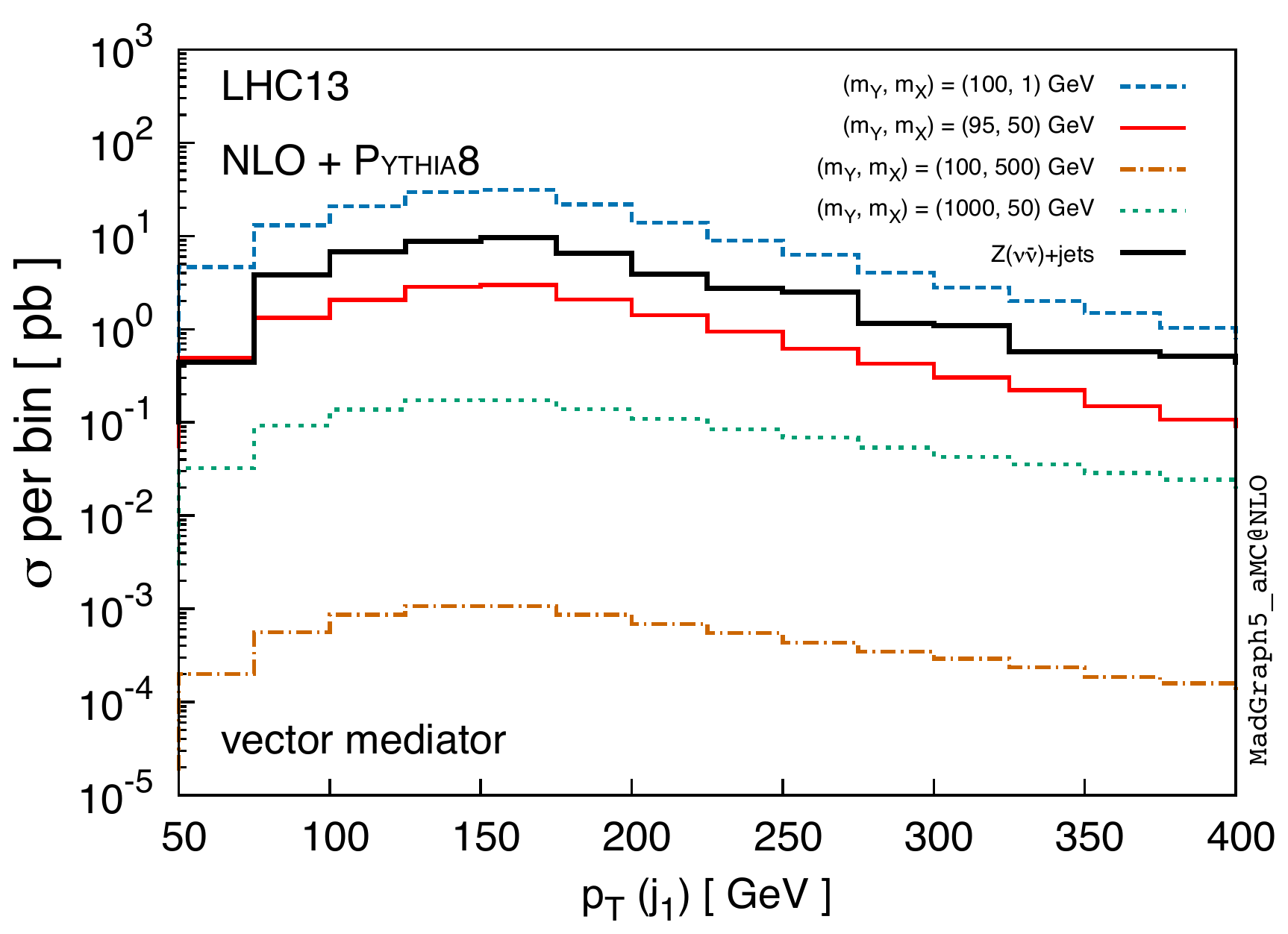}\qquad
\includegraphics[width=0.48\textwidth]{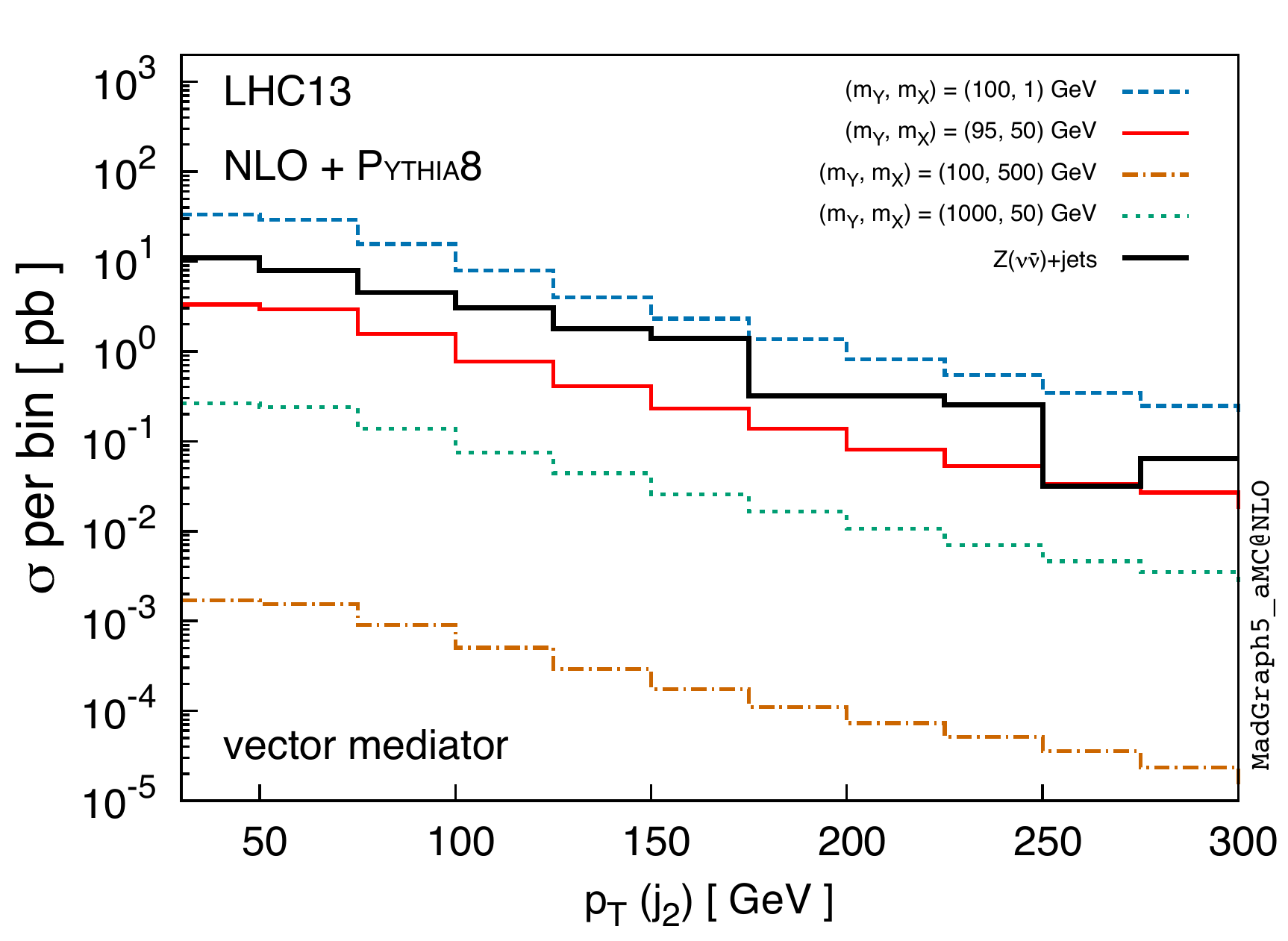}
\caption{Distributions for various signal benchmark points and 
 $Z(\nu\bar{\nu})+$ jets, where 
 $\MET > 150 \GeV$ and jets with $p_T > 30 \GeV$ and 
 $|\eta| < 4.5 $ are considered. 
 We assume a pure vector mediator and Dirac DM with the coupling
 parameters  $g_X=1$ and $g_{\rm SM}=0.25$.}  
\label{fig:sigbgd}
\end{figure*}

Figure~\ref{fig:sigbgd} shows an example comparison of the $Z$+jets channel to several benchmark points of the simplified model discussed in previous sections.
The shape of signal jet multiplicity distributions (upper left panel)
resembles the $Z$+jets distribution to a good degree, while the overall
rate varies wildly depending on the model point. Note that events
containing one jet are produced at almost an identical rate to $2j$
events and comparable to $3j$ events, both in $Z$+jets and all of the
benchmark model points we considered. The production rate for different
jet multiplicities implies that it could be beneficial to consider DM
searches beyond mono-jet, either inclusive, or at fixed jet
multiplicities. 

Next, the MET (as well as the hardest jet $p_T$) distributions for
mediators of mass $\sim 100 \GeV$ with light DM naturally resemble the
$Z$+jets distribution in shape, as the kinematics of the MET and hardest
jet are determined by the mass scale of the heavy object ($i.e.$ the $Z$
boson or the mediator).
For heavy mediators, the MET and $p_T(j_1)$ distributions fall off with
a milder slope, suggesting that the signal to background ratio ($S/B$)
in searches for DM could be improved by requiring a higher
$p_T^{\rm min}$ for the hardest jet. 

The $p_T$ distribution of the second hardest jet, on the other hand,
seems to display a similar shape in all model points, as well as the
$Z$+jets background channel. Requiring a high $p_T$ on the second jet
would hence not improve neither $S/B$ nor the signal significance,
suggesting that in searches which exploit the presence of a second jet,
only minimum cuts on $p_T(j_2)$ should be applied. Note that the modelling
of the second jet from LO calculations grossly underestimates the overall
rate in case of lighter mediators for $p_T \gtrsim 100 \GeV$, as
illustrated in Fig.~\ref{fig:ptj2}.

\section{Dark matter production with a top-quark pair}\label{sec:DM_top}

In the spin-0 mediator model, due to the normalisation of the Yukawa
couplings in the Lagrangian~\eqref{eq:scalar_mediator2}, the most
relevant tree-level process at the LHC is 
\begin{align}
 pp\to X\bar X + t\bar t\,.
\end{align}
Such models have in the past been studied in the context of 
EFT~\cite{Cheung:2010zf,Lin:2013sca,Artoni:2013zba} and 
simplified models~\cite{Buckley:2014fba,Haisch:2015ioa}, and searched for at the LHC Run~I~\cite{Aad:2014vea,Khachatryan:2015nua}. 
Past work on DM interactions with the top quarks has mainly focused on
LO estimates, with only a few analyses including NLO corrections.
Here we present a comprehensive study of NLO effects of DM interaction
with top quarks in the framework of the simplified model.   
We note that a wide class of so called ``top-philic DM'' models exist
where the LO production is via top loops. Reference~\cite{Greiner:2014qna}
studied such a scenario in a minimally model-dependent framework, while
more recently, Ref.~\cite{Mattelaer:2015haa} presented concrete
predictions for loop-induced DM production for the current LHC13 run, using the same simulation framework as in this work.    

\begin{table*}
\begin{footnotesize}
\center
\begin{tabular}{l|l|r|l|l}
\hline\rule{0pt}{3ex}
 $(m_{Y},m_{X})$ [GeV] & & & 
 scalar & pseudo-scalar\\[1mm] 
\hline\hline\rule{0pt}{3ex}
 & & $\sigma_{{\rm LO}}$ [pb]  
  &  ${2.278\times10^{1}\,}^{+28.0}_{-20.4}${\scriptsize\,${\pm4.2\,\%}$}   
  &  ${5.202\times10^{-1}\,}^{+30.8}_{-22.0}${\scriptsize\,${\pm6.0\,\%}$}\\ 
10  & undecayed &$\sigma_{{\rm NLO}}$ [pb]  & ${2.435\times10^{1}\,}^{+5.4}_{-8.5}${\scriptsize\,${\pm1.8\,\%}$}& ${5.431\times10^{-1}\,}^{+7.4}_{-10.2}${\scriptsize\,${\pm2.6\,\%}$} \\ 
 & &$K$ factor &1.07   &1.04   \\
\hline\rule{0pt}{3ex}
 & & $\sigma_{{\rm LO}}$ [pb]  
  &  ${2.294\times10^{1}\,}^{+28.0}_{-20.5}${\scriptsize\,${\pm4.2\,\%}$}   
  &  ${5.500\times10^{-1}\,}^{+30.8}_{-22.1}${\scriptsize\,${\pm6.0\,\%}$}\\ 
(10, 1)  & $m_{Y}\!>\!2m_{X}$ &$\sigma_{{\rm NLO}}$ [pb]  & ${2.460\times10^{1}\,}^{+5.4}_{-8.5}${\scriptsize\,${\pm1.8\,\%}$}& ${5.739\times10^{-1}\,}^{+7.4}_{-10.2}${\scriptsize\,${\pm2.6\,\%}$} \\ 
 & &$K$ factor &1.07   &1.04   \\  
\hline\rule{0pt}{3ex}
& & $\sigma_{{\rm LO}}$ [pb]  
  & ${2.415\times10^{-3}\,}^{+30.5}_{-21.8}${\scriptsize\,${\pm5.8\,\%}$}
  & ${3.329\times10^{-3}\,}^{+33.9}_{-23.8}${\scriptsize\,${\pm8.7\,\%}$}  \\ 
(10, 50)  & $m_{Y}\!<\!2m_{X}$ &$\sigma_{{\rm NLO}}$ [pb]  & ${2.340\times10^{-3}\,}^{+5.8}_{-9.1}${\scriptsize\,${\pm2.8\,\%}$}&  ${3.133\times10^{-3}\,}^{+7.5}_{-11.0}${\scriptsize\,${\pm3.9\,\%}$} \\ 
 & &$K$ factor &0.97   &0.94    \\ 
\hline\hline\rule{0pt}{3ex}
 & & $\sigma_{{\rm LO}}$ [pb]  
  & ${8.226\times10^{-1}\,}^{+28.7}_{-20.9}${\scriptsize\,${\pm4.4\,\%}$} 
  & ${2.442\times10^{-1}\,}^{+32.2}_{-22.9}${\scriptsize\,${\pm7.2\,\%}$} \\ 
100  & undecayed &$\sigma_{{\rm NLO}}$ [pb]  & ${8.391\times10^{-1}\,}^{+5.3}_{-8.6}${\scriptsize\,${\pm2.1\,\%}$} & ${2.431\times10^{-1}\,}^{+7.6}_{-10.7}${\scriptsize\,${\pm3.2\,\%}$}  \\ 
 & &$K$ factor & 1.02  & 1.00     \\ 
\hline\rule{0pt}{3ex}
 & & $\sigma_{{\rm LO}}$ [pb]  
  & ${8.135\times10^{-1}\,}^{+28.8}_{-20.9}${\scriptsize\,${\pm4.4\,\%}$} 
  & ${2.464\times10^{-1}\,}^{+32.4}_{-23.0}${\scriptsize\,${\pm7.2\,\%}$} \\ 
(100, 1)  & $m_{Y}\!>\!2m_{X}$ &$\sigma_{{\rm NLO}}$ [pb]  & ${8.207\times10^{-1}\,}^{+4.8}_{-8.3}${\scriptsize\,${\pm2.1\,\%}$} & ${2.427\times10^{-1}\,}^{+7.0}_{-10.4}${\scriptsize\,${\pm3.2\,\%}$}  \\ 
 & &$K$ factor & 1.01  & 0.98     \\ 
\hline\rule{0pt}{3ex}
& & $\sigma_{{\rm LO}}$ [pb]  
  & ${7.986\times10^{-3}\,}^{+29.5}_{-21.3}${\scriptsize\,${\pm5.0\,\%}$}
  & ${1.404\times10^{-2}\,}^{+32.9}_{-23.3}${\scriptsize\,${\pm7.8\,\%}$} \\ 
(95, 50)  & $m_{Y}\!\lesssim\!2m_{X}$ &$\sigma_{{\rm NLO}}$ [pb]  & ${7.897\times10^{-3}\,}^{+5.5}_{-8.8}${\scriptsize\,${\pm2.4\,\%}$}& ${1.362\times10^{-2}\,}^{+7.4}_{-10.8}${\scriptsize\,${\pm3.5\,\%}$}  \\
 & &$K$ factor & 0.99  & 0.97     \\
 \hline\hline\rule{0pt}{3ex}
 & & $\sigma_{{\rm LO}}$ [pb]  
  & ${1.571\times10^{-3}\,}^{+40.2}_{-27.0}${\scriptsize\,${\pm17.1\,\%}$}
  & ${1.827\times10^{-3}\,}^{+40.4}_{-27.1}${\scriptsize\,${\pm17.4\,\%}$}\\ 
1000  & undecayed &$\sigma_{{\rm NLO}}$ [pb]  &${1.127\times10^{-3}\,}^{+10.9}_{-13.6}${\scriptsize\,${\pm8.7\,\%}$} &${1.297\times10^{-3}\,}^{+10.8}_{-13.7}${\scriptsize\,${\pm8.8\,\%}$} \\ 
 & &$K$ factor & 0.72  & 0.71     \\
\hline\rule{0pt}{3ex}
 & & $\sigma_{{\rm LO}}$ [pb]  
  & ${7.499\times10^{-4}\,}^{+40.8}_{-27.2}${\scriptsize\,${\pm16.8\,\%}$}
  & ${8.174\times10^{-4}\,}^{+40.9}_{-27.3}${\scriptsize\,${\pm17.0\,\%}$}\\ 
(1000, 1)  & $m_{Y}\!>\!2m_{X}$ &$\sigma_{{\rm NLO}}$ [pb]  &${5.201\times10^{-4}\,}^{+8.4}_{-12.7}${\scriptsize\,${\pm8.4\,\%}$} &${5.675\times10^{-4}\,}^{+8.6}_{-12.9}${\scriptsize\,${\pm8.5\,\%}$} \\ 
 & &$K$ factor & 0.69  & 0.69     \\  
\hline\rule{0pt}{3ex}
& & $\sigma_{{\rm LO}}$ [pb]  
  &  ${7.354\times10^{-4}\,}^{+40.7}_{-27.2}${\scriptsize\,${\pm16.8\,\%}$}
  & ${8.137\times10^{-4}\,}^{+41.0}_{-27.3}${\scriptsize\,${\pm17.0\,\%}$} \\ 
(1000, 50)  & $m_{Y}\!>\!2m_{X}$ &$\sigma_{{\rm NLO}}$ [pb]  & ${5.125\times10^{-4}\,}^{+8.6}_{-12.8}${\scriptsize\,${\pm8.5\,\%}$}& ${5.595\times10^{-4}\,}^{+8.3}_{-12.7}${\scriptsize\,${\pm8.5\,\%}$}  \\
 & &$K$ factor & 0.69   & 0.69     \\ \hline
\end{tabular}
\caption{LO and NLO cross sections and corresponding $K$ factors for 
 DM pair production in association with a top-quark pair for the scalar
 and pseudo-scalar mediator scenario at the 13-TeV LHC. 
 The uncertainties represent the scale and PDF uncertainties in per cent,
 respectively.   
 We show several benchmark model points for the mediator and DM
 masses with the coupling parameters $g_{X}=1$ and $g_{\rm SM}=1$. 
}\label{tab:ttbarNLO}
\end{footnotesize}
\end{table*}

The code and events for the above process can be automatically generated
by issuing the following commands in {\sc MG5aMC}:
\begin{verbatim}
 > import model DMsimp_s_spin0
 > generate p p > xd xd~ t t~ [QCD]
 > output
 > launch
\end{verbatim}
We have checked that our model can reproduce the SM predictions for 
$pp\to ht\bar t\to\tau^+\tau^-t\bar t$ by adjusting the coupling and
mass parameters.
Note that we use the on-shell renormalisation for the NLO model
construction. 
The top-quark decays can be subsequently performed by
{\sc MadSpin}~\cite{Artoisenet:2012st}, which keeps production and decay
spin correlations.

To illustrate the NLO effects, we consider 
pure scalar, Eqs.~\eqref{paramX_s} and \eqref{paramSM_s}, or pure
pseudo-scalar, Eqs.~\eqref{paramX_p} and \eqref{paramSM_p}, couplings, 
and take
\begin{align}
 g_X=1 \quad {\rm and}\quad g_{\rm SM}=1
 \label{param_spin0}
\end{align}
as the default couplings.
With these values, the scalar and pseudo-scalar mediator
width is 
$\Gamma_Y/m_Y\sim 0.06-0.1$ for $m_Y>2m_{X},2m_t$, while  
$\Gamma_Y/m_Y\sim 0.04$ for $m_Y>2m_X$ and $m_Y<2m_t$.
The width for scalar is slightly smaller due to the additional $\beta^2$ factor where
$\beta=(1-4m_{X,t}^2/m_Y^2)^{1/2}$.

\subsection{Total cross sections}\label{sec:DM_top_cross_sections}
 
We start by showing total production rates of $pp\to X\bar X+t\bar{t}$,
where the top quark is considered stable.
Table~\ref{tab:ttbarNLO} shows the LO and NLO cross sections (in pb) for
the scalar and pseudo-scalar mediator scenarios, where we use
$m_t=172$~GeV. 
The central renormalisation and factorisation scales are set to half the
sum of the transverse mass of the top quarks and the missing transverse
energy.  We also present scale and PDF uncertainties in \% as well as $K$ factors.

\begin{figure*}
\center 
\includegraphics[width=0.32\textwidth]{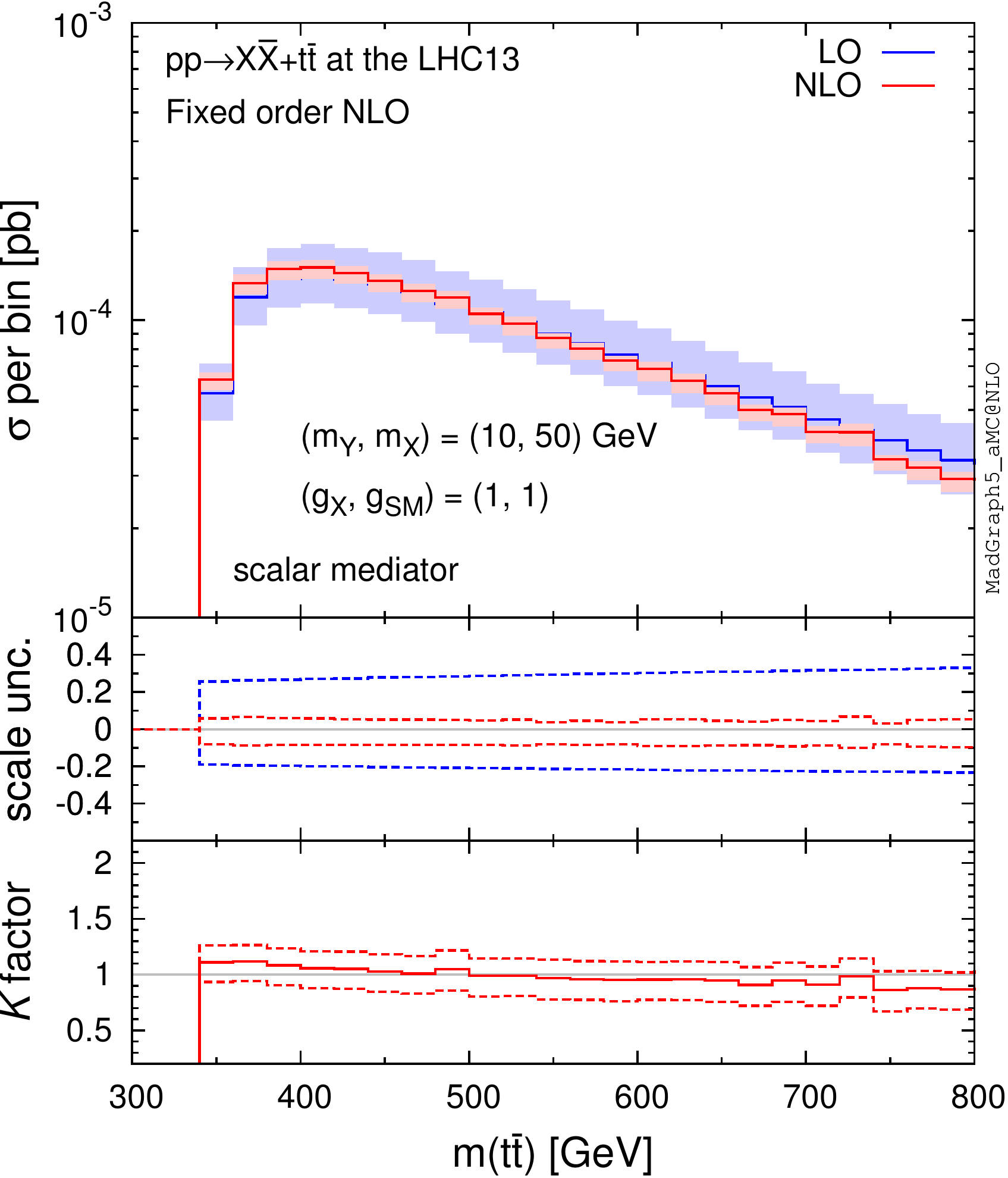}\quad
\includegraphics[width=0.32\textwidth]{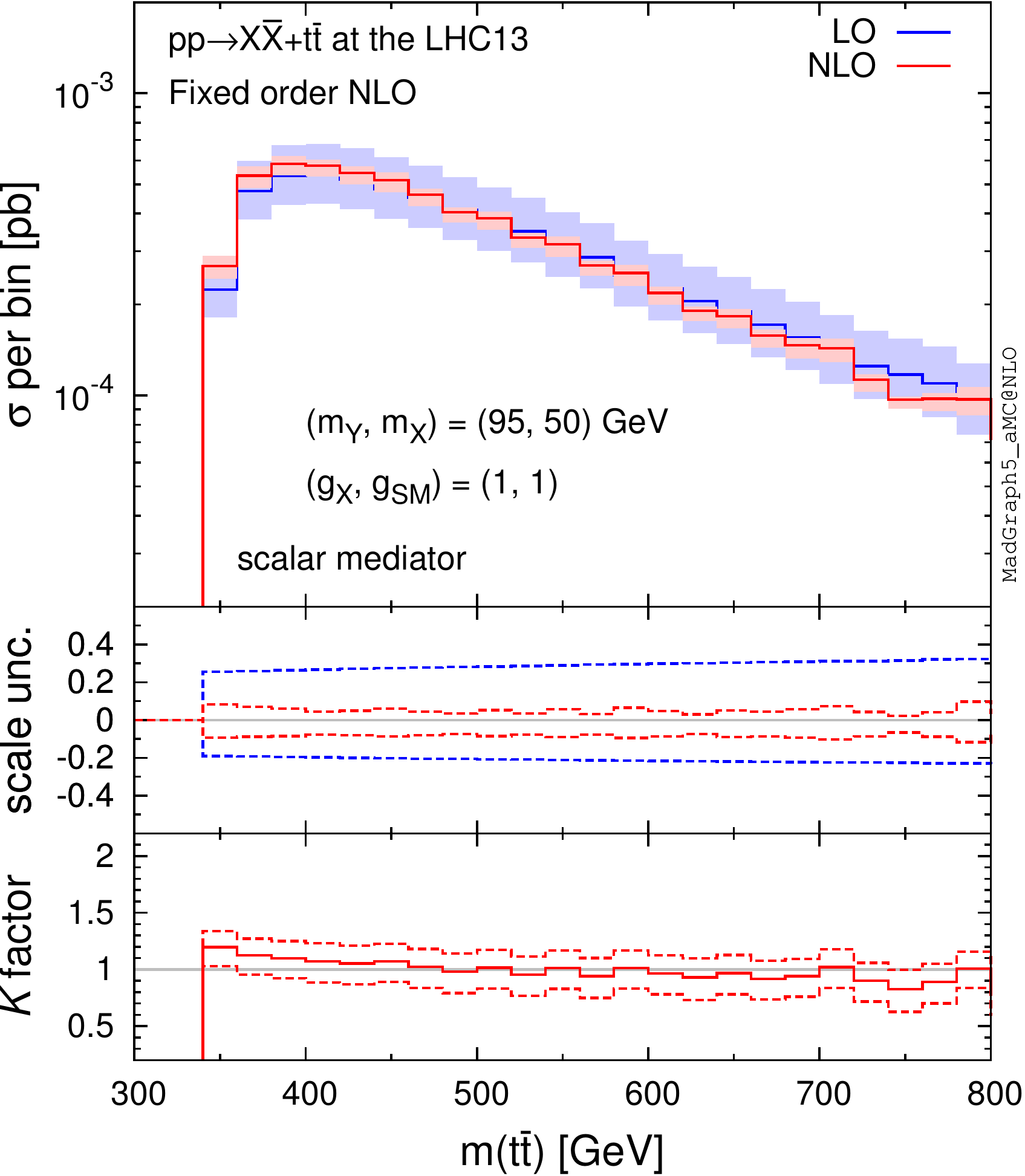}\quad
\includegraphics[width=0.32\textwidth]{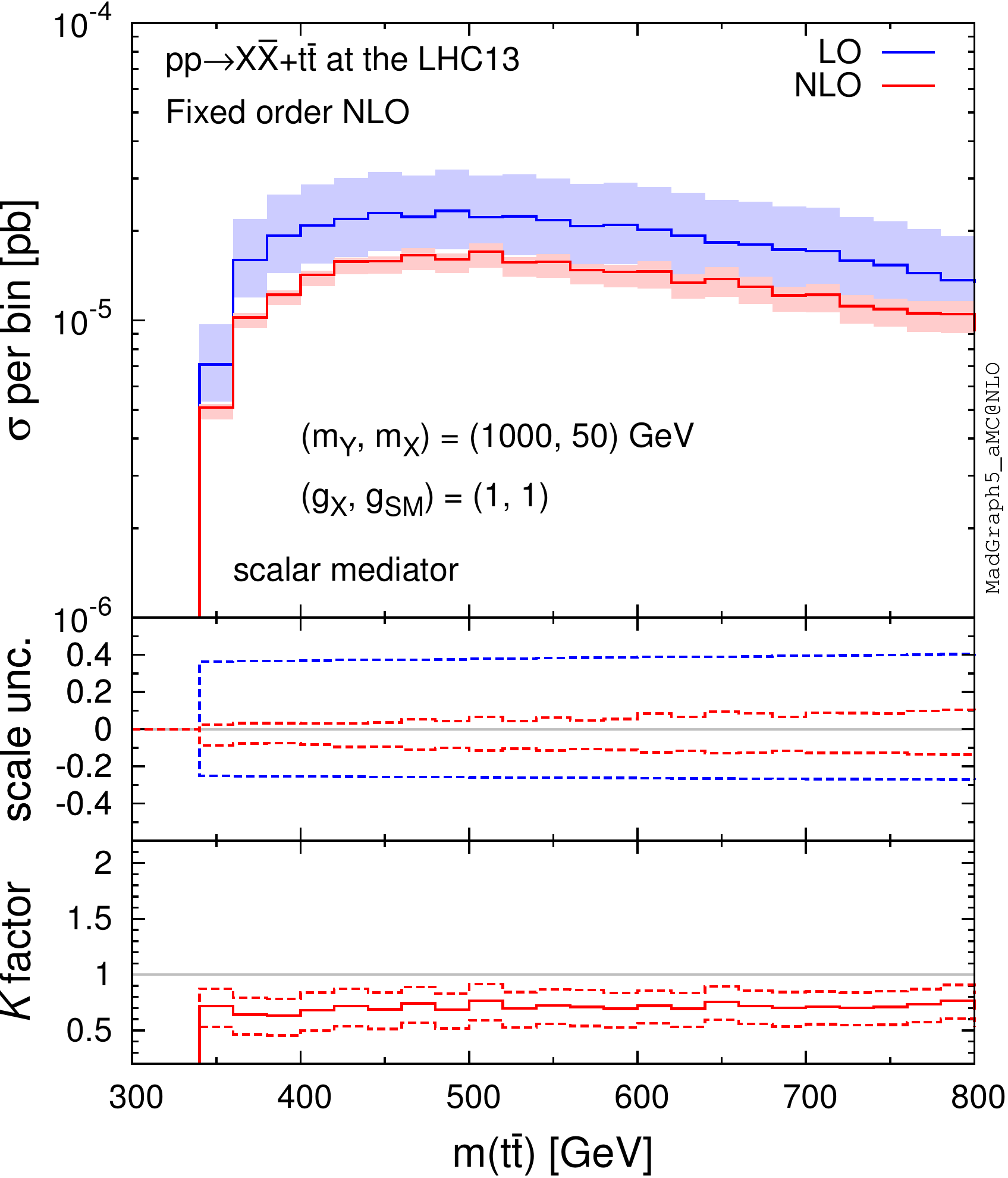}
\caption{Distribution of the invariant mass of the top-quark pair for 
 $pp\to X\bar X+t\bar t$ at the 13-TeV LHC for different mediator masses
 with the DM mass fixed at 50~GeV, where we assume a pure scalar
 mediator and Dirac DM. 
 The middle and bottom panels show the differential scale uncertainties
 and $K$ factors, respectively.} 
\label{fig:tt1}
\end{figure*}

\begin{figure*}
\center 
\includegraphics[width=0.32\textwidth]{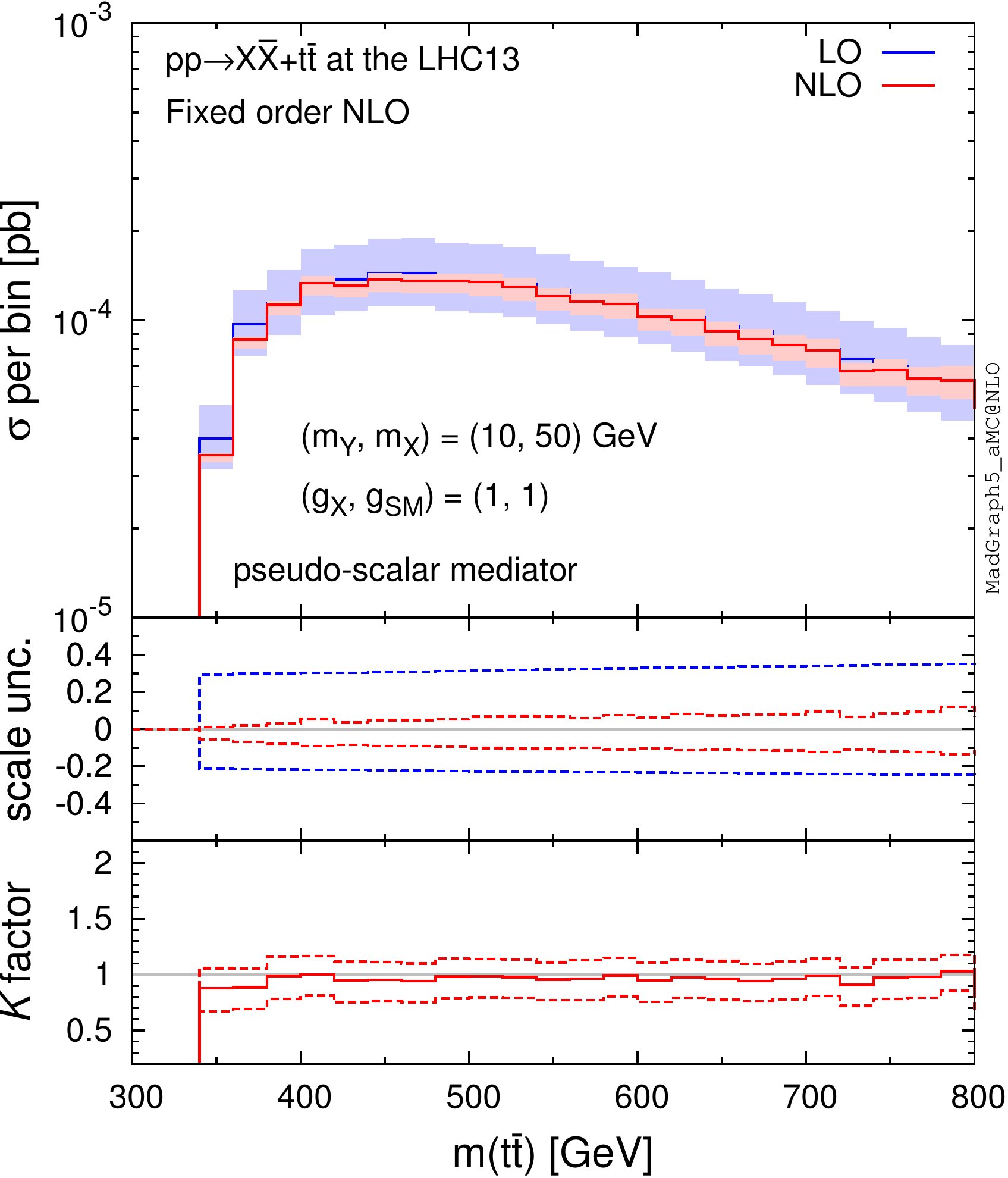}\quad
\includegraphics[width=0.32\textwidth]{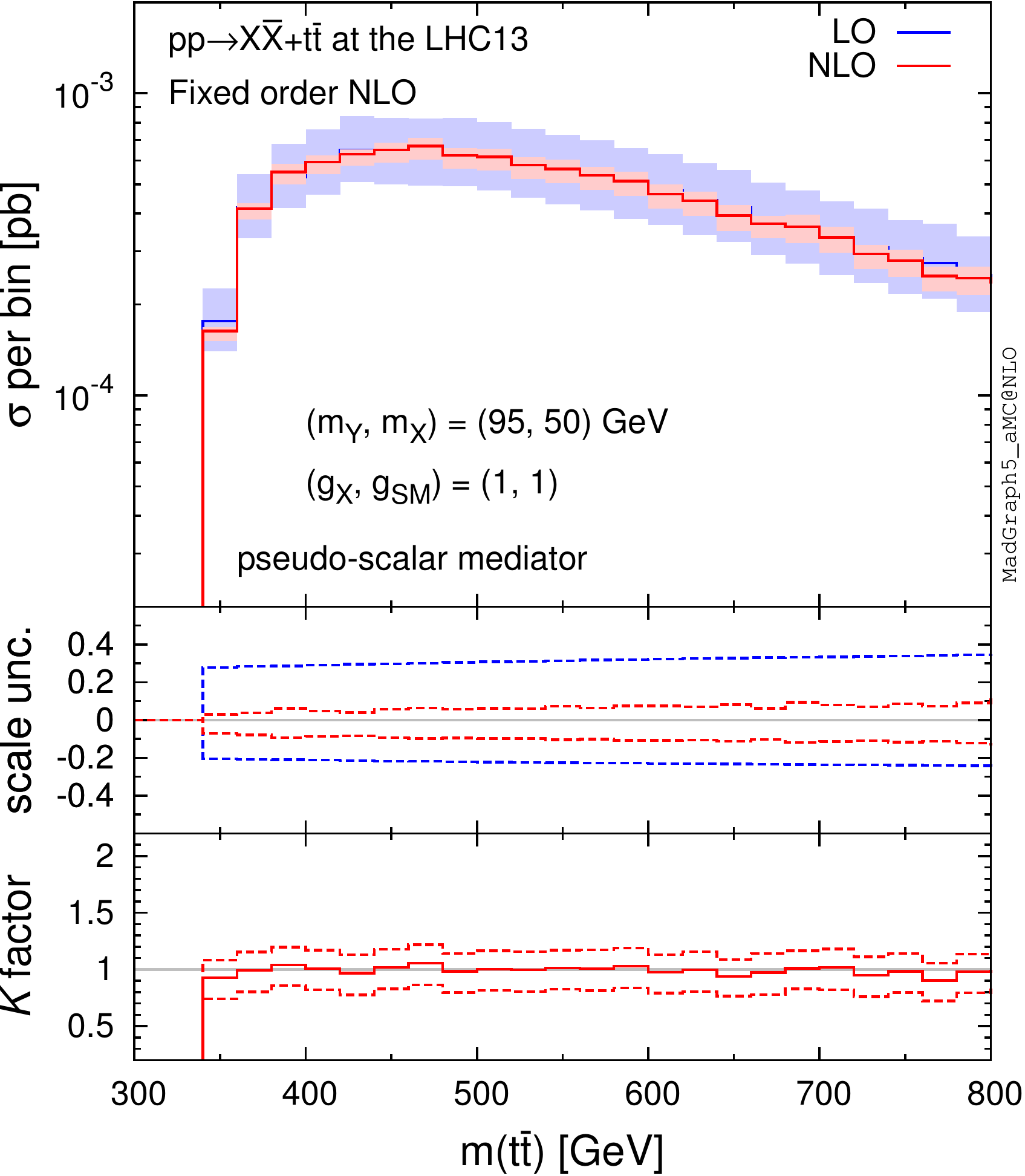}\quad
\includegraphics[width=0.32\textwidth]{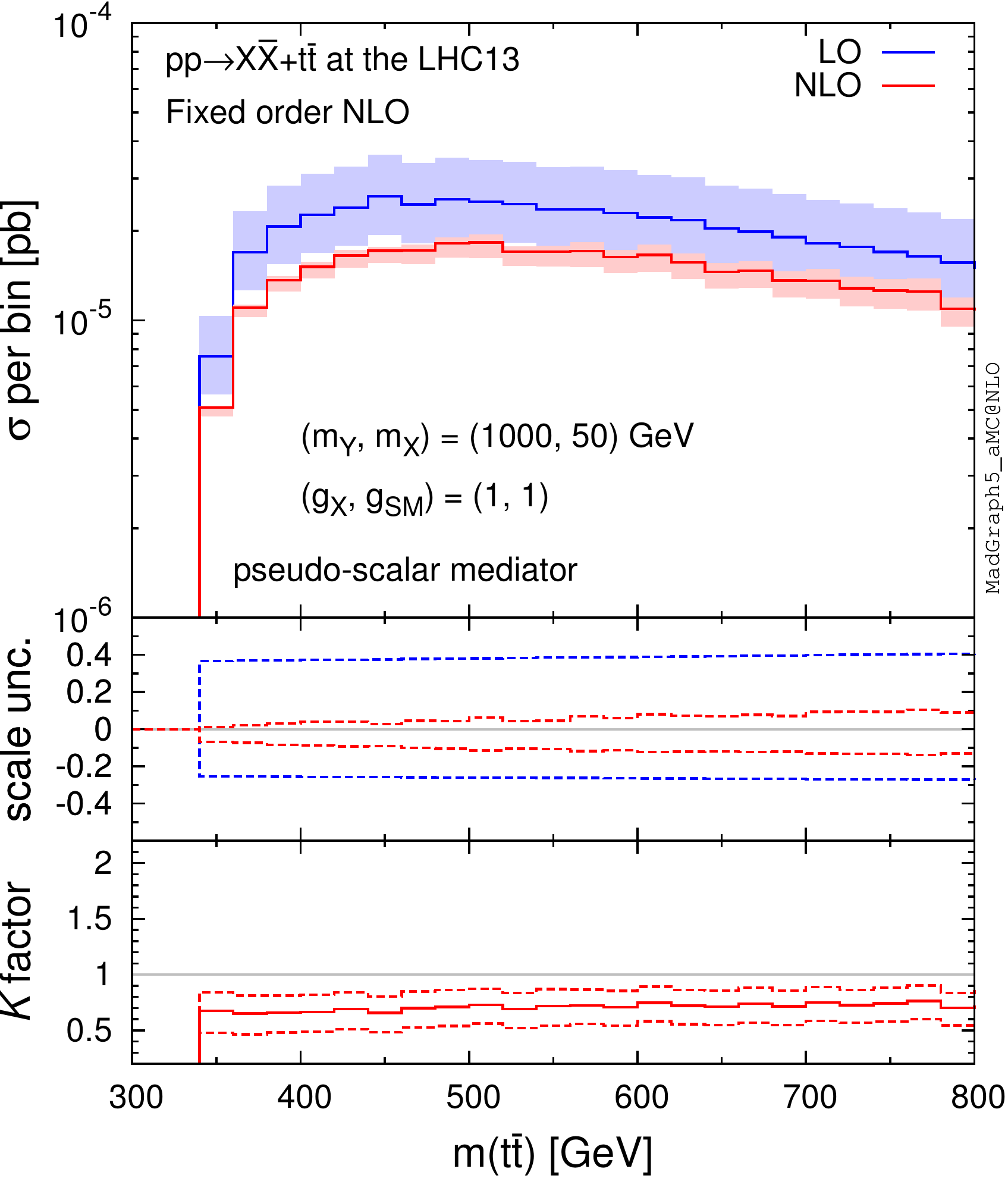}
\caption{Same as Fig.~\ref{fig:tt1}, but for the pseudo-scalar mediator
 scenario.}
\label{fig:tt2}
\end{figure*}

For the total production rates, the NLO effects are very mild for the
light mediator case, while they are significant for the heavy case.
The inclusion of
NLO corrections results in a drastic improvement of the scale uncertainties, 
from up to 40\,\% at LO to typically only about 10\,\% at NLO. Also, the PDF uncertainties are reduced by approximately a factor of two when going from LO to NLO. 

Table~\ref{tab:ttbarNLO} also shows clear differences between the
overall production rates in the cases of scalar and pseudo-scalar
mediators. For mediator mass of ${\cal O}(10)$~GeV we find that DM production cross section via scalar mediators is an order of magnitude larger compared to the production rate via the pseudo-scalar mediator with the same mass. The large difference occurs due to the fact that in case of $m_X < m_Y \ll m_t$, 
the production cross section is dominated by the $t \to t Y_0$
fragmentation. In case of the scalar mediator, the $t \to t Y_0$
fragmentation function contains terms with soft singularities of the
form $(1-x)/x$ -- where $x$ is the momentum fraction carried by the
mediator -- causing enhancements in the production
rate~\cite{Dawson:1997im}.
The soft-enhanced term is absent in the case of a pseudo-scalar mediator~\cite{Dittmaier:2000tc}, explaining the order of magnitude difference between the total rates of the scalar and pseudo-scalar mediators.

In cases where either DM or the mediator is produced close to threshold, we observe that the production cross section in the pseudo-scalar mediator case is larger.  
The effect can be attributed to the production rate originating mainly
from top fusion diagrams. The production of a DM (Dirac) pair via scalar
mediators  $t\bar{t} \rightarrow Y_0 \to X \bar X $ at threshold can
proceed only via a $P$-wave ($^3P_0$) and is hence suppressed by extra
two powers of $\beta = \sqrt{1 - 4m_t^2/s}$ \cite{Han:2014nja}. Conversely, production of DM pair via pseudo-scalar mediators can proceed via an $S$-wave ($^1S_0$) and hence does not suffer any kinematic suppression.

\subsection{Differential distributions}\label{sec:DM_top_distributions}

For the study of differential distributions, we consider the invariant
mass of the top-quark pair ($m(t\bar{t})$), without inclusion of a
parton shower.
Figures~\ref{fig:tt1} and \ref{fig:tt2} illustrate the scalar and
pseudo-scalar results, respectively, for different mediator masses
(off-shell, threshold, and on-shell) with the DM mass fixed at 50~GeV. 
In all cases the shape of $m(t\bar{t})$  is well modelled by the LO
calculation, and including a constant $K$ factor
reproduces the NLO results to an excellent degree, except in the threshold
region. 
However, the LO calculation suffers from significant scale uncertainties
which tend to increase with $m(t\bar{t})$, whereas the scale
uncertainties are under much better control at NLO. 

\begin{figure}
\center 
\includegraphics[width=1.\columnwidth]{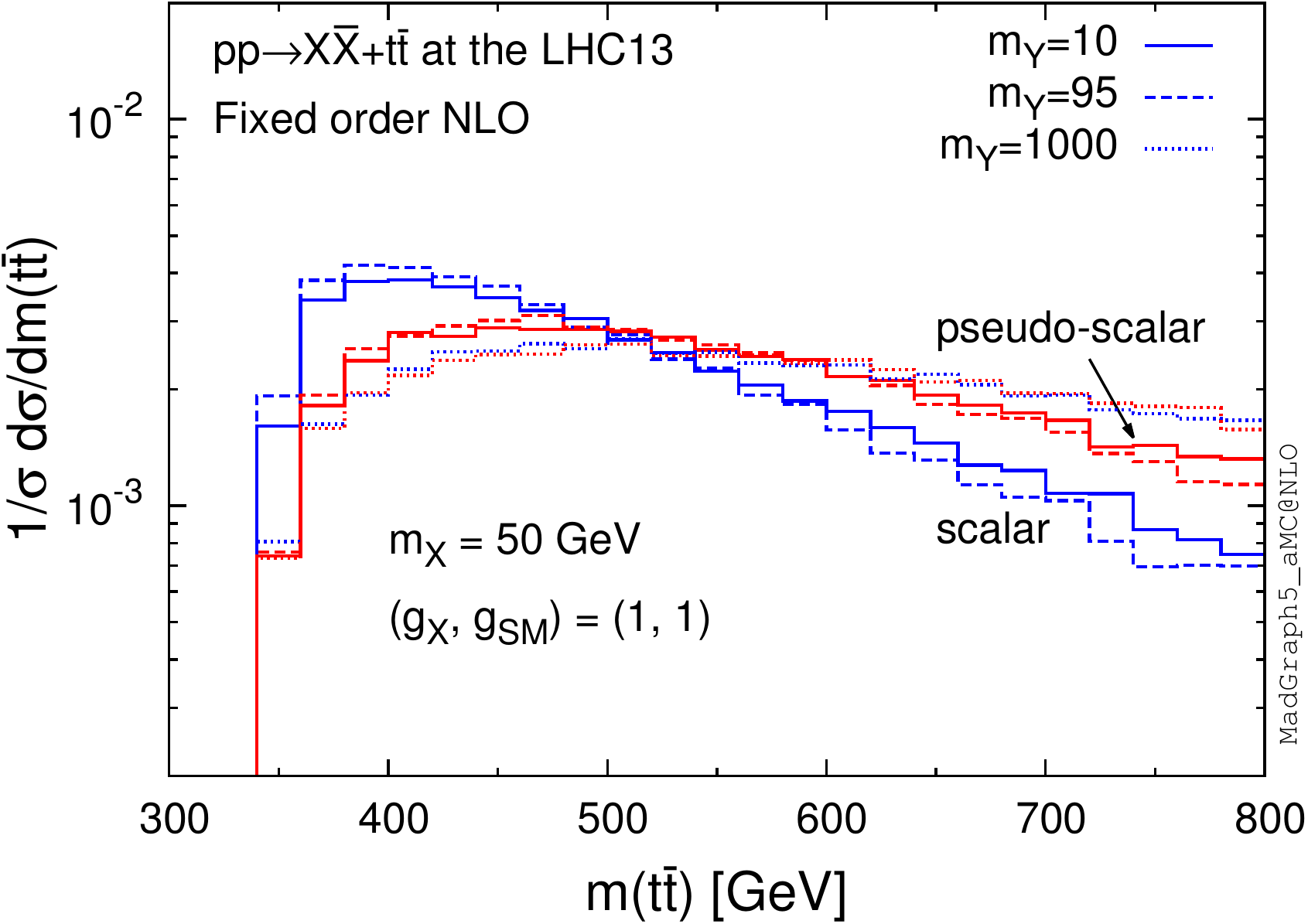}
\caption{Comparison of $t\bar t$ invariant mass distributions between
 the scalar (blue) and pseudo-scalar (red) mediator models for different
 mediator masses.}
\label{fig:mttSPS}
\end{figure}

Whether DM is produced via scalar or pseudo-scalar mediators can have a
dramatic effect on the shape of the $m(t\bar{t})$ distribution.
In Fig.~\ref{fig:mttSPS} we compare the NLO distributions in
Figs.~\ref{fig:tt1} and \ref{fig:tt2},
where we normalise the histograms to unit area to point out the shape
differences.
We observe that the shape of the distribution is particularly enhanced
for $m(t\bar{t}) \gtrsim 500 \GeV$ in the case of the pseudo-scalar
mediator, while the scalar mediator distribution displays a much more
prominent peak at lower $m(t\bar{t})$. 
The effect is severely damped in case of heavy mediators, where we find
no clear differences between the shapes of the scalar and pseudo-scalar
mediator distributions.  
Figure~\ref{fig:mttSPS} suggests that scalar and pseudo-scalar mediators
could be distinguished based on the shape of the $m(t\bar{t})$
distribution, as long as the mediator is sufficiently light and/or does
not decay highly off-shell.  An analogous observation has been made already in the case
of the study of the $CP$ properties of the Higgs boson~\cite{Demartin:2014fia}.

\section{Summary}\label{sec:conclusions}

Searches for DM are one of the main endeavours at the LHC Run~II. 
Accurate and precise predictions for production rates and distributions
are necessary to obtain robust constraints on DM models and characterise
possible DM signals. 
In this article we have provided a general implementation of the
simplified DM model approach into a calculation/simulation framework that
allows to systematically evaluate and include NLO QCD corrections to the
production of DM at the LHC. 
We have considered a class of simplified models where DM is a Dirac
fermion and couples to the SM via either spin-1 or spin-0 $s$-channel mediators,
making no restrictions on chiral couplings.
For the purpose of illustration, we analysed the NLO effects on the DM 
production via vector and axial-vector mediators in the context of
mono-jet signals.
In addition, we have presented detailed predictions of DM production in
association with a top-quark pair via
scalar and pseudo-scalar mediators. 
We presented our results for various DM and mediator masses to cover
benchmark points suggested by the ATLAS/CMS DM forum~\cite{Abercrombie:2015gea}.

For MET+jets in the spin-1 mediator model, our results show that higher-order corrections have
a significant effect both on the overall production rate as well as on the
shape of relevant differential distributions, with a sizeable reduction
of the scale and PDF uncertainties.
The NLO corrections to the LO production rates can be large, with $K$ factors of up to $K\lesssim 2$,  and typically occur in parts of the model parameter space where the
mass scale of DM and mediator is ${\cal O}(10-100)$~GeV.
For such scenarios, we also find large NLO
effects on the shape of differential distributions in $\MET$ and the transverse momentum of the associated
jets.
Simplified models with heavy ($e.g.$ ${\cal O}(1)$~TeV) mediators/DM do
not receive large NLO corrections, and we find that LO predictions
describe both total production rates and shapes of differential
distributions quite accurately.  
Distributions of the second hardest jet in the event are well modelled
by the parton shower for heavy mediator/DM cases. On the other hand, for
mediators/DM with masses of ${\cal O}(100)$~GeV, the inclusion of NLO effects is
essential for a proper description of $p_T(j_2)$ and $\eta (j_2)$
distributions, especially in the high-$p_T$ tails, where the NLO effects can
be an order of magnitude. 

So-called ``giant $K$ factors'' can occur in NLO computations of DM
production rates in the regions where $p_T^j\gg m_Y$.
Such effects can be extremely large when considering mono-jet production
rates, especially in phase-space regions with low $\MET$.
Imposing a sufficiently large $\MET$ cut and hence avoiding the
soft/collinear singularities associated with the mediator emissions from
high-$p_T$ jets efficiently mitigates the effect of giant $K$ factors.  

In our analysis we have gone beyond FO in perturbation theory
and studied the effects of jet multiplicity merging at NLO accuracy.
We found that FO calculations model the jet multiplicity and
other differential distributions adequately well, with no significant
effects on the shapes or overall rates coming from jet sample merging.  

Comparisons with the NLO predictions for the leading SM background channel
reveal that considerations of either inclusive or exclusive jet samples
beyond one jet could be beneficial for increasing the prospects for DM
detection.
The leading jet $p_T$ distributions in case of heavy mediators display
a milder decrease with the increase in $p_T$, suggesting that a
significant improvement in $S/B$ could be obtained by focusing on high-$p_T$ regions.
The second hardest jet $p_T$ distribution, on the other hand, was
characterised by a slope similar to the leading backgrounds, implying
that more inclusive cuts on the second jet should be used. 

For MET+$t\bar t$ in the spin-0 mediator model, our results show that 
the NLO corrections are very mild for the
light mediator case, while they are significant for the heavy case.
We observed a drastic improvement of the theory uncertainties
when going from LO to NLO. 
We have noted that the shape of the $m(t\bar{t})$ distribution can reveal
the chiral structure of the DM--SM interactions, as long as the mediator is relatively light 
($i.e.$ $\lesssim {\cal O}(100)$~GeV).  

The DM model we studied in this paper is publicly
available in the {\sc FeynRules} repository~\cite{FR-DMsimp:Online}. We emphasise that all results presented here have been
obtained in the {\sc FeynRules/MG5aMC} framework, and thus they can
be easily reproduced and used in DM searches at the LHC Run~II.

\section*{Acknowledgements}

We would like to thank C\'eline Degrande, Federico Demartin, Davide
Pagani  for useful discussions. We are thankful to Rikkert Frederix and
Stefano Frixione for useful comments regarding the FxFx merging.
We are particularly grateful to Olivier Mattelaer, Matthias Neubert,
Eleni Vryonidou, Jian Wang, Cen Zhang for the extensive collaboration on implementing and validating DM simplified models. 

This work has been performed in the framework of the ERC grant 291377
``LHCtheory: Theoretical predictions and analyses of LHC physics:
advancing the precision frontier'', the FP7 Marie Curie Initial
Training Network MCnetITN (PITN-GA-2012-315877), the research unit 
 ``New physics at the LHC'' of the German research foundation DFG and the Helmholtz Alliance for Astroparticle Physics.  
It is also supported in part by the Belgian Federal Science Policy
Office through the Interuniversity Attraction Pole P7/37.  
The work of MB, FM and AM is supported by the IISN ``MadGraph'' convention
4.4511.10 and the IISN ``Fundamental interactions'' convention 4.4517.08.
KM is supported in part by the Strategic Research Program ``High Energy
Physics'' and the Research Council of the Vrije Universiteit Brussel.

\bibliography{bibdm}

\providecommand{\href}[2]{#2}\begingroup\raggedright\begin{thebibliography}{10}

\bibitem{Drees:2012ji}
M.~Drees and G.~Gerbier, {\it {Mini-Review of Dark Matter: 2012}},
  \href{http://arxiv.org/abs/1204.2373}{{\tt arXiv:1204.2373}}.

\bibitem{Klasen:2015uma}
M.~Klasen, M.~Pohl, and G.~Sigl, {\it {Indirect and direct search for dark
  matter}},  {\em Prog. Part. Nucl. Phys.} {\bf 85} (2015) 1--32,
  [\href{http://arxiv.org/abs/1507.03800}{{\tt arXiv:1507.03800}}].

\bibitem{Aad:2015zva}
{\bf ATLAS} Collaboration, G.~Aad et~al., {\it {Search for new phenomena in
  final states with an energetic jet and large missing transverse momentum in
  pp collisions at $\sqrt{s}=8~$ TeV with the ATLAS detector}},  {\em Eur.
  Phys. J.} {\bf C75} (2015), no.~7 299,
  [\href{http://arxiv.org/abs/1502.01518}{{\tt arXiv:1502.01518}}].

\bibitem{CMS:2015jha}
{\bf CMS} Collaboration, {\it {Search for New Physics in the V-jet + MET final
  state}},  \href{http://arxiv.org/abs/CMS-PAS-EXO-12-055}{{\tt
  CMS-PAS-EXO-12-055}}.

\bibitem{Abercrombie:2015gea}
D.~Abercrombie, N.~Akchurin, E.~Akilli, J.~A. Maestre, B.~Allen, et~al., {\it
  {Dark Matter Benchmark Models for Early LHC Run-2 Searches: Report of the
  ATLAS/CMS Dark Matter Forum}},  \href{http://arxiv.org/abs/1507.00966}{{\tt
  arXiv:1507.00966}}.

\bibitem{Alves:2011wf}
{\bf LHC New Physics Working Group} Collaboration, D.~Alves, {\it {Simplified
  Models for LHC New Physics Searches}},  {\em J. Phys.} {\bf G39} (2012)
  105005, [\href{http://arxiv.org/abs/1105.2838}{{\tt arXiv:1105.2838}}].

\bibitem{Abdallah:2014hon}
J.~Abdallah et~al., {\it {Simplified Models for Dark Matter and Missing Energy
  Searches at the LHC}},  \href{http://arxiv.org/abs/1409.2893}{{\tt
  arXiv:1409.2893}}.

\bibitem{Malik:2014ggr}
S.~Malik et~al., {\it {Interplay and Characterization of Dark Matter Searches
  at Colliders and in Direct Detection Experiments}},  2014.
\newblock \href{http://arxiv.org/abs/1409.4075}{{\tt arXiv:1409.4075}}.

\bibitem{Goodman:2010yf}
J.~Goodman, M.~Ibe, A.~Rajaraman, W.~Shepherd, T.~M.~P. Tait, and H.-B. Yu,
  {\it {Constraints on Light Majorana dark Matter from Colliders}},  {\em Phys.
  Lett.} {\bf B695} (2011) 185--188,
  [\href{http://arxiv.org/abs/1005.1286}{{\tt arXiv:1005.1286}}].

\bibitem{Bai:2010hh}
Y.~Bai, P.~J. Fox, and R.~Harnik, {\it {The Tevatron at the Frontier of Dark
  Matter Direct Detection}},  {\em JHEP} {\bf 12} (2010) 048,
  [\href{http://arxiv.org/abs/1005.3797}{{\tt arXiv:1005.3797}}].

\bibitem{Fox:2011pm}
P.~J. Fox, R.~Harnik, J.~Kopp, and Y.~Tsai, {\it {Missing Energy Signatures of
  Dark Matter at the LHC}},  {\em Phys. Rev.} {\bf D85} (2012) 056011,
  [\href{http://arxiv.org/abs/1109.4398}{{\tt arXiv:1109.4398}}].

\bibitem{Shoemaker:2011vi}
I.~M. Shoemaker and L.~Vecchi, {\it {Unitarity and Monojet Bounds on Models for
  DAMA, CoGeNT, and CRESST-II}},  {\em Phys. Rev.} {\bf D86} (2012) 015023,
  [\href{http://arxiv.org/abs/1112.5457}{{\tt arXiv:1112.5457}}].

\bibitem{Busoni:2013lha}
G.~Busoni, A.~De~Simone, E.~Morgante, and A.~Riotto, {\it {On the Validity of
  the Effective Field Theory for Dark Matter Searches at the LHC}},  {\em Phys.
  Lett.} {\bf B728} (2014) 412--421,
  [\href{http://arxiv.org/abs/1307.2253}{{\tt arXiv:1307.2253}}].

\bibitem{Buchmueller:2013dya}
O.~Buchmueller, M.~J. Dolan, and C.~McCabe, {\it {Beyond Effective Field Theory
  for Dark Matter Searches at the LHC}},  {\em JHEP} {\bf 01} (2014) 025,
  [\href{http://arxiv.org/abs/1308.6799}{{\tt arXiv:1308.6799}}].

\bibitem{Wang:2011sx}
J.~Wang, C.~S. Li, D.~Y. Shao, and H.~Zhang, {\it {Next-to-leading order QCD
  predictions for the signal of Dark Matter and photon associated production at
  the LHC}},  {\em Phys.Rev.} {\bf D84} (2011) 075011,
  [\href{http://arxiv.org/abs/1107.2048}{{\tt arXiv:1107.2048}}].

\bibitem{Huang:2012hs}
F.~P. Huang, C.~S. Li, J.~Wang, and D.~Y. Shao, {\it {Searching for the signal
  of dark matter and photon associated production at the LHC beyond leading
  order}},  {\em Phys.Rev.} {\bf D87} (2013) 094018,
  [\href{http://arxiv.org/abs/1210.0195}{{\tt arXiv:1210.0195}}].

\bibitem{Fox:2012ru}
P.~J. Fox and C.~Williams, {\it {Next-to-Leading Order Predictions for Dark
  Matter Production at Hadron Colliders}},  {\em Phys.Rev.} {\bf D87} (2013),
  no.~5 054030, [\href{http://arxiv.org/abs/1211.6390}{{\tt arXiv:1211.6390}}].

\bibitem{Haisch:2013ata}
U.~Haisch, F.~Kahlhoefer, and E.~Re, {\it {QCD effects in mono-jet searches for
  dark matter}},  {\em JHEP} {\bf 1312} (2013) 007,
  [\href{http://arxiv.org/abs/1310.4491}{{\tt arXiv:1310.4491}}].

\bibitem{Mao:2014rga}
M.~Song, G.~Li, W.-G. Ma, R.-Y. Zhang, and J.-Y. Guo, {\it {Dark matter pair
  associated with a $W$ boson production at the LHC in next-to-leading order
  QCD}},  {\em JHEP} {\bf 1409} (2014) 069,
  [\href{http://arxiv.org/abs/1403.2142}{{\tt arXiv:1403.2142}}].

\bibitem{Frederix:2012ps}
R.~Frederix and S.~Frixione, {\it {Merging meets matching in MC@NLO}},  {\em
  JHEP} {\bf 12} (2012) 061, [\href{http://arxiv.org/abs/1209.6215}{{\tt
  arXiv:1209.6215}}].

\bibitem{Alloul:2013bka}
A.~Alloul, N.~D. Christensen, C.~Degrande, C.~Duhr, and B.~Fuks, {\it
  {FeynRules 2.0 - A complete toolbox for tree-level phenomenology}},  {\em
  Comput. Phys. Commun.} {\bf 185} (2014) 2250--2300,
  [\href{http://arxiv.org/abs/1310.1921}{{\tt arXiv:1310.1921}}].

\bibitem{Alwall:2014hca}
J.~Alwall, R.~Frederix, S.~Frixione, V.~Hirschi, F.~Maltoni, O.~Mattelaer,
  H.~S. Shao, T.~Stelzer, P.~Torrielli, and M.~Zaro, {\it {The automated
  computation of tree-level and next-to-leading order differential cross
  sections, and their matching to parton shower simulations}},  {\em JHEP} {\bf
  07} (2014) 079, [\href{http://arxiv.org/abs/1405.0301}{{\tt
  arXiv:1405.0301}}].

\bibitem{Degrande:2011ua}
C.~Degrande, C.~Duhr, B.~Fuks, D.~Grellscheid, O.~Mattelaer, and T.~Reiter,
  {\it {UFO - The Universal FeynRules Output}},  {\em Comput. Phys. Commun.}
  {\bf 183} (2012) 1201--1214, [\href{http://arxiv.org/abs/1108.2040}{{\tt
  arXiv:1108.2040}}].

\bibitem{deAquino:2011ub}
P.~de~Aquino, W.~Link, F.~Maltoni, O.~Mattelaer, and T.~Stelzer, {\it {ALOHA:
  Automatic Libraries Of Helicity Amplitudes for Feynman Diagram
  Computations}},  {\em Comput.Phys.Commun.} {\bf 183} (2012) 2254--2263,
  [\href{http://arxiv.org/abs/1108.2041}{{\tt arXiv:1108.2041}}].

\bibitem{Ossola:2008xq}
G.~Ossola, C.~G. Papadopoulos, and R.~Pittau, {\it {On the Rational Terms of
  the one-loop amplitudes}},  {\em JHEP} {\bf 05} (2008) 004,
  [\href{http://arxiv.org/abs/0802.1876}{{\tt arXiv:0802.1876}}].

\bibitem{Degrande:2014vpa}
C.~Degrande, {\it {Automatic evaluation of UV and R2 terms for beyond the
  Standard Model Lagrangians: a proof-of-principle}},  {\em Comput. Phys.
  Commun.} {\bf 197} (2015) 239--262,
  [\href{http://arxiv.org/abs/1406.3030}{{\tt arXiv:1406.3030}}].

\bibitem{Hahn:2000kx}
T.~Hahn, {\it {Generating Feynman diagrams and amplitudes with FeynArts 3}},
  {\em Comput. Phys. Commun.} {\bf 140} (2001) 418--431,
  [\href{http://arxiv.org/abs/hep-ph/0012260}{{\tt hep-ph/0012260}}].

\bibitem{FR-DMsimp:Online}
\url{http://feynrules.irmp.ucl.ac.be/wiki/DMsimp}.

\bibitem{Ossola:2007ax}
G.~Ossola, C.~G. Papadopoulos, and R.~Pittau, {\it {CutTools: A Program
  implementing the OPP reduction method to compute one-loop amplitudes}},  {\em
  JHEP} {\bf 0803} (2008) 042, [\href{http://arxiv.org/abs/0711.3596}{{\tt
  arXiv:0711.3596}}].

\bibitem{Hirschi:2011pa}
V.~Hirschi et~al., {\it {Automation of one-loop QCD corrections}},  {\em JHEP}
  {\bf 05} (2011) 044, [\href{http://arxiv.org/abs/1103.0621}{{\tt
  arXiv:1103.0621}}].

\bibitem{Cascioli:2011va}
F.~Cascioli, P.~Maierhofer, and S.~Pozzorini, {\it {Scattering Amplitudes with
  Open Loops}},  {\em Phys.Rev.Lett.} {\bf 108} (2012) 111601,
  [\href{http://arxiv.org/abs/1111.5206}{{\tt arXiv:1111.5206}}].

\bibitem{Frederix:2009yq}
R.~Frederix, S.~Frixione, F.~Maltoni, and T.~Stelzer, {\it {Automation of
  next-to-leading order computations in QCD: the FKS subtraction}},  {\em JHEP}
  {\bf 10} (2009) 003, [\href{http://arxiv.org/abs/0908.4272}{{\tt
  arXiv:0908.4272}}].

\bibitem{Frixione:2002ik}
S.~Frixione and B.~R. Webber, {\it {Matching NLO QCD computations and parton
  shower simulations}},  {\em JHEP} {\bf 0206} (2002) 029,
  [\href{http://arxiv.org/abs/hep-ph/0204244}{{\tt hep-ph/0204244}}].

\bibitem{Sjostrand:2006za}
T.~Sjostrand, S.~Mrenna, and P.~Z. Skands, {\it {PYTHIA 6.4 Physics and
  Manual}},  {\em JHEP} {\bf 0605} (2006) 026,
  [\href{http://arxiv.org/abs/hep-ph/0603175}{{\tt hep-ph/0603175}}].

\bibitem{Sjostrand:2007gs}
T.~Sjostrand, S.~Mrenna, and P.~Z. Skands, {\it {A Brief Introduction to PYTHIA
  8.1}},  {\em Comput. Phys. Commun.} {\bf 178} (2008) 852--867,
  [\href{http://arxiv.org/abs/0710.3820}{{\tt arXiv:0710.3820}}].

\bibitem{Corcella:2000bw}
G.~Corcella, I.~Knowles, G.~Marchesini, S.~Moretti, K.~Odagiri, et~al., {\it
  {HERWIG 6: An Event generator for hadron emission reactions with interfering
  gluons (including supersymmetric processes)}},  {\em JHEP} {\bf 0101} (2001)
  010, [\href{http://arxiv.org/abs/hep-ph/0011363}{{\tt hep-ph/0011363}}].

\bibitem{Bahr:2008pv}
M.~Bahr, S.~Gieseke, M.~Gigg, D.~Grellscheid, K.~Hamilton, et~al., {\it
  {Herwig++ Physics and Manual}},  {\em Eur.Phys.J.} {\bf C58} (2008) 639--707,
  [\href{http://arxiv.org/abs/0803.0883}{{\tt arXiv:0803.0883}}].

\bibitem{Belanger:2006is}
G.~Belanger, F.~Boudjema, A.~Pukhov, and A.~Semenov, {\it {MicrOMEGAs 2.0: A
  Program to calculate the relic density of dark matter in a generic model}},
  {\em Comput. Phys. Commun.} {\bf 176} (2007) 367--382,
  [\href{http://arxiv.org/abs/hep-ph/0607059}{{\tt hep-ph/0607059}}].

\bibitem{Belanger:2008sj}
G.~Belanger, F.~Boudjema, A.~Pukhov, and A.~Semenov, {\it {Dark matter direct
  detection rate in a generic model with micrOMEGAs 2.2}},  {\em Comput. Phys.
  Commun.} {\bf 180} (2009) 747--767,
  [\href{http://arxiv.org/abs/0803.2360}{{\tt arXiv:0803.2360}}].

\bibitem{Backovic:2013dpa}
M.~Backovic, K.~Kong, and M.~McCaskey, {\it {MadDM v.1.0: Computation of Dark
  Matter Relic Abundance Using MadGraph5}},  {\em Physics of the Dark Universe}
  {\bf 5-6} (2014) 18--28, [\href{http://arxiv.org/abs/1308.4955}{{\tt
  arXiv:1308.4955}}].

\bibitem{Backovic:2015cra}
M.~Backovic, K.~Kong, A.~Martini, O.~Mattelaer, and G.~Mohlabeng, {\it {Direct
  Detection of Dark Matter with MadDM v.2.0}},
  \href{http://arxiv.org/abs/1505.04190}{{\tt arXiv:1505.04190}}.

\bibitem{Neubert:2015fka}
M.~Neubert, J.~Wang, and C.~Zhang, {\it {Higher-Order QCD Predictions for Dark
  Matter Production in Mono-$Z$ Searches at the LHC}},
  \href{http://arxiv.org/abs/1509.05785}{{\tt arXiv:1509.05785}}.

\bibitem{Hirschi:2015iia}
V.~Hirschi and O.~Mattelaer, {\it {Automated event generation for loop-induced
  processes}},  \href{http://arxiv.org/abs/1507.00020}{{\tt arXiv:1507.00020}}.

\bibitem{Mattelaer:2015haa}
O.~Mattelaer and E.~Vryonidou, {\it {Dark matter production through
  loop-induced processes at the LHC: the s-channel mediator case}},  {\em Eur.
  Phys. J.} {\bf C75} (2015), no.~9 436,
  [\href{http://arxiv.org/abs/1508.00564}{{\tt arXiv:1508.00564}}].

\bibitem{Haisch:2012kf}
U.~Haisch, F.~Kahlhoefer, and J.~Unwin, {\it {The impact of heavy-quark loops
  on LHC dark matter searches}},  {\em JHEP} {\bf 07} (2013) 125,
  [\href{http://arxiv.org/abs/1208.4605}{{\tt arXiv:1208.4605}}].

\bibitem{Buckley:2014fba}
M.~R. Buckley, D.~Feld, and D.~Goncalves, {\it {Scalar Simplified Models for
  Dark Matter}},  {\em Phys.Rev.} {\bf D91} (2015), no.~1 015017,
  [\href{http://arxiv.org/abs/1410.6497}{{\tt arXiv:1410.6497}}].

\bibitem{Harris:2014hga}
P.~Harris, V.~V. Khoze, M.~Spannowsky, and C.~Williams, {\it {Constraining Dark
  Sectors at Colliders: Beyond the Effective Theory Approach}},  {\em
  Phys.Rev.} {\bf D91} (2015), no.~5 055009,
  [\href{http://arxiv.org/abs/1411.0535}{{\tt arXiv:1411.0535}}].

\bibitem{Haisch:2015ioa}
U.~Haisch and E.~Re, {\it {Simplified dark matter top-quark interactions at the
  LHC}},  {\em JHEP} {\bf 06} (2015) 078,
  [\href{http://arxiv.org/abs/1503.00691}{{\tt arXiv:1503.00691}}].

\bibitem{Alwall:2014bza}
J.~Alwall, C.~Duhr, B.~Fuks, O.~Mattelaer, D.~G. Özturk, and C.-H. Shen, {\it
  {Computing decay rates for new physics theories with FeynRules and
  MadGraph5\_aMC@NLO}},  {\em Comput. Phys. Commun.} {\bf 197} (2015) 312--323,
  [\href{http://arxiv.org/abs/1402.1178}{{\tt arXiv:1402.1178}}].

\bibitem{Botje:2011sn}
M.~Botje, J.~Butterworth, A.~Cooper-Sarkar, A.~de~Roeck, J.~Feltesse, et~al.,
  {\it {The PDF4LHC Working Group Interim Recommendations}},
  \href{http://arxiv.org/abs/1101.0538}{{\tt arXiv:1101.0538}}.

\bibitem{Whalley:2005nh}
M.~Whalley, D.~Bourilkov, and R.~Group, {\it {The Les Houches accord PDFs
  (LHAPDF) and LHAGLUE}},  \href{http://arxiv.org/abs/hep-ph/0508110}{{\tt
  hep-ph/0508110}}.

\bibitem{Frederix:2011ss}
R.~Frederix, S.~Frixione, V.~Hirschi, F.~Maltoni, R.~Pittau, et~al., {\it
  {Four-lepton production at hadron colliders: aMC@NLO predictions with
  theoretical uncertainties}},  {\em JHEP} {\bf 1202} (2012) 099,
  [\href{http://arxiv.org/abs/1110.4738}{{\tt arXiv:1110.4738}}].

\bibitem{Alekhin:2011sk}
S.~Alekhin, S.~Alioli, R.~D. Ball, V.~Bertone, J.~Blumlein, et~al., {\it {The
  PDF4LHC Working Group Interim Report}},
  \href{http://arxiv.org/abs/1101.0536}{{\tt arXiv:1101.0536}}.

\bibitem{Cacciari:2008gp}
M.~Cacciari, G.~P. Salam, and G.~Soyez, {\it {The Anti-k(t) jet clustering
  algorithm}},  {\em JHEP} {\bf 0804} (2008) 063,
  [\href{http://arxiv.org/abs/0802.1189}{{\tt arXiv:0802.1189}}].

\bibitem{Cacciari:2011ma}
M.~Cacciari, G.~P. Salam, and G.~Soyez, {\it {FastJet User Manual}},  {\em
  Eur.Phys.J.} {\bf C72} (2012) 1896,
  [\href{http://arxiv.org/abs/1111.6097}{{\tt arXiv:1111.6097}}].

\bibitem{Rubin:2010xp}
M.~Rubin, G.~P. Salam, and S.~Sapeta, {\it {Giant QCD K-factors beyond NLO}},
  {\em JHEP} {\bf 09} (2010) 084, [\href{http://arxiv.org/abs/1006.2144}{{\tt
  arXiv:1006.2144}}].

\bibitem{Cheung:2010zf}
K.~Cheung, K.~Mawatari, E.~Senaha, P.-Y. Tseng, and T.-C. Yuan, {\it {The Top
  Window for dark matter}},  {\em JHEP} {\bf 10} (2010) 081,
  [\href{http://arxiv.org/abs/1009.0618}{{\tt arXiv:1009.0618}}].

\bibitem{Lin:2013sca}
T.~Lin, E.~W. Kolb, and L.-T. Wang, {\it {Probing dark matter couplings to top
  and bottom quarks at the LHC}},  {\em Phys. Rev.} {\bf D88} (2013), no.~6
  063510, [\href{http://arxiv.org/abs/1303.6638}{{\tt arXiv:1303.6638}}].

\bibitem{Artoni:2013zba}
G.~Artoni, T.~Lin, B.~Penning, G.~Sciolla, and A.~Venturini, {\it {Prospects
  for collider searches for dark matter with heavy quarks}},  in {\em
  {Community Summer Study 2013: Snowmass on the Mississippi (CSS2013)
  Minneapolis, MN, USA, July 29-August 6, 2013}}, 2013.
\newblock \href{http://arxiv.org/abs/1307.7834}{{\tt arXiv:1307.7834}}.

\bibitem{Aad:2014vea}
{\bf ATLAS} Collaboration, G.~Aad et~al., {\it {Search for dark matter in
  events with heavy quarks and missing transverse momentum in $pp$ collisions
  with the ATLAS detector}},  {\em Eur. Phys. J.} {\bf C75} (2015), no.~2 92,
  [\href{http://arxiv.org/abs/1410.4031}{{\tt arXiv:1410.4031}}].

\bibitem{Khachatryan:2015nua}
{\bf CMS} Collaboration, V.~Khachatryan et~al., {\it {Search for the production
  of dark matter in association with top-quark pairs in the single-lepton final
  state in proton-proton collisions at sqrt(s) = 8 TeV}},  {\em JHEP} {\bf 06}
  (2015) 121, [\href{http://arxiv.org/abs/1504.03198}{{\tt arXiv:1504.03198}}].

\bibitem{Greiner:2014qna}
N.~Greiner, K.~Kong, J.-C. Park, S.~C. Park, and J.-C. Winter, {\it
  {Model-Independent Production of a Top-Philic Resonance at the LHC}},  {\em
  JHEP} {\bf 04} (2015) 029, [\href{http://arxiv.org/abs/1410.6099}{{\tt
  arXiv:1410.6099}}].

\bibitem{Artoisenet:2012st}
P.~Artoisenet, R.~Frederix, O.~Mattelaer, and R.~Rietkerk, {\it {Automatic
  spin-entangled decays of heavy resonances in Monte Carlo simulations}},  {\em
  JHEP} {\bf 1303} (2013) 015, [\href{http://arxiv.org/abs/1212.3460}{{\tt
  arXiv:1212.3460}}].

\bibitem{Dawson:1997im}
S.~Dawson and L.~Reina, {\it {QCD corrections to associated Higgs boson
  production}},  {\em Phys.Rev.} {\bf D57} (1998) 5851--5859,
  [\href{http://arxiv.org/abs/hep-ph/9712400}{{\tt hep-ph/9712400}}].

\bibitem{Dittmaier:2000tc}
S.~Dittmaier, M.~Kramer, Y.~Liao, M.~Spira, and P.~Zerwas, {\it {Higgs
  radiation off quarks in supersymmetric theories at $e^+ e^-$ colliders}},
  {\em Phys.Lett.} {\bf B478} (2000) 247--254,
  [\href{http://arxiv.org/abs/hep-ph/0002035}{{\tt hep-ph/0002035}}].

\bibitem{Han:2014nja}
T.~Han, J.~Sayre, and S.~Westhoff, {\it {Top-Quark Initiated Processes at
  High-Energy Hadron Colliders}},  {\em JHEP} {\bf 04} (2015) 145,
  [\href{http://arxiv.org/abs/1411.2588}{{\tt arXiv:1411.2588}}].

\bibitem{Demartin:2014fia}
F.~Demartin, F.~Maltoni, K.~Mawatari, B.~Page, and M.~Zaro, {\it {Higgs
  characterisation at NLO in QCD: CP properties of the top-quark Yukawa
  interaction}},  {\em Eur.Phys.J.} {\bf C74} (2014), no.~9 3065,
  [\href{http://arxiv.org/abs/1407.5089}{{\tt arXiv:1407.5089}}].

\end{thebibliography}\endgroup
\bibliographystyle{JHEP}

\end{document}